\input harvmac

\def\S{{\bf S}}

\def\pf{{\rm Pf}}

\def\F{{\widehat F}}

\def\N{{\cal N}}
\def\O{{\cal O}}
\def\T{{\bf T}}
\def\Z{{\bf Z}}
\def\R{{\bf R}}
\def\A{{\cal A}}

\def\K3{{\bf K3}}
\def\journal#1&#2(#3){\unskip, \sl #1\ \bf #2 \rm(19#3) }
\def\andjournal#1&#2(#3){\sl #1~\bf #2 \rm (19#3) }

\def\bar{\overline}
\def\hat{\widehat}

\def\tilde{\widetilde}

\def\frac#1#2{{#1\over#2}}

\def\half{\frac12}

\def\inbar{\,\vrule height1.5ex width.4pt depth0pt}
\def\IC{\relax\hbox{$\inbar\kern-.3em{\rm C}$}}
\def\IR{\relax{\rm I\kern-.18em R}}
\def\IP{\relax{\rm I\kern-.18em P}}
\def\Z{{\bf Z}}

%
%

%
\catcode`\@=11
\def\slash#1{\mathord{\mathpalette\c@ncel{#1}}}
\overfullrule=0pt

\def\TT{{\cal T}}

\def\underrel#1\over#2{\mathrel{\mathop{\kern\z@#1}\limits_{#2}}}

\catcode`\@=12


%

\def\det{{\rm det}}

\def\det{{\rm det}}
\def\exp{{\rm exp}}



\rightline{IASSNS-HEP-99/74}
\Title{
\rightline{hep-th/9908142}
} {\vbox{\centerline{String Theory and Noncommutative Geometry}}}
\medskip

\centerline{\it Nathan Seiberg and Edward Witten}
\bigskip
\centerline{School of Natural Sciences}
\centerline{Institute for Advanced Study}
\centerline{Olden Lane, Princeton, NJ 08540}

\smallskip

\vglue .3cm

\bigskip
\noindent
We extend earlier ideas about the appearance of noncommutative
geometry in string theory with a nonzero $B$-field.  We identify a
limit in which the entire string dynamics is described by a minimally
coupled (supersymmetric) gauge theory on a noncommutative space, and
discuss the corrections away from this limit.  Our analysis leads us
to an equivalence between ordinary gauge fields and noncommutative
gauge fields, which is realized by a change of variables that can be
described explicitly.  This change of variables is checked by
comparing the ordinary Dirac-Born-Infeld theory with its
noncommutative counterpart.  We obtain a new perspective on
noncommutative gauge theory on a torus, its $T$-duality, and Morita
equivalence.  We also discuss the $D0/D4$ system, the relation to
$M$-theory in DLCQ, and a possible noncommutative version of the
six-dimensional $(2,0)$ theory.

\Date{8/99}

\newsec{Introduction}

The idea that the spacetime coordinates do not commute is quite old 
\ref\snyder{H.S.~Snyder, ``Quantized Space-Time,''
Phys. Rev. {\bf 71} (1947) 38; ``The Electromagnetic Field In
Quantized Space-Time,'' Phys. Rev. {\bf 72} (1947) 68.}.  
It has been studied by many authors both from a mathematical
and a physical perspective.  The theory of operator algebras
has been suggested as a framework for physics in noncommutative
spacetime -- see 
\ref\connesb{A. Connes, ``Noncommutative Geometry,'' Academic Press
(1994).} for an exposition of the philosophy --
and Yang-Mills theory on a noncommutative torus has been
proposed as an example \ref\connesc{A. Connes and M. Rieffel,
``Yang-Mills For Noncommutative Two-Tori,'' in Operator Algebras
and Mathematical Physics (Iowa City, Iowa, 1985), pp. 237 Contemp.
Math. Oper. Alg. Math. Phys. 62, AMS 1987.}.
Though this example at first sight appears to be neither covariant
nor causal, it has proved to arise in string theory in a definite limit
\ref\connes{A. Connes, M. R. Douglas, and A. Schwarz, ``Noncommutative
Geometry and Matrix Theory: Compactification On Tori,'' JHEP {\bf
9802:003} (1998), hep-th/9711162.}, 
with the noncovariance arising from the expectation value of a
background field.
\nref\bfss{T.~Banks, W.~Fischler, S.H.~Shenker and L.~Susskind,
``M-Theory as a Matrix Model: A Conjecture,'' Phys. Rev. {\bf D55}
(1997) 5112, hep-th/9610043.}%
\nref\mmrev{T.~Banks, Nucl. Phys. Proc. Suppl. {\bf 67} (1998) 180
hep-th/9710231;  D.~Bigatti and L.~Susskind, ``Review of Matrix
Theory,'' hep-th/9712072.}%
This analysis involved toroidal compactification, in the
limit of small volume, with fixed and generic values of the worldsheet
theta angles.  This limit is fairly natural in the context of the
matrix model of $M$-theory \refs{\bfss,\mmrev}, and the original
discussion was made in this context.  Indeed, 
early work relating membranes to large matrices 
\ref\dwhn{B.~de Wit, J.~Hoppe and H.~Nicolai, ``On The Quantum
Mechanics Of Supermembranes,'' Nucl. Phys. {\bf B305} (1988) 545.},
has motivated in
\nref\hoppe{J.~Hoppe, ``Membranes And Integrable Systems,''
Phys. Lett. {\bf B250} (1990) 44.}%
\nref\fairlie{D.B. Fairlie, P. Fletcher and C.K. Zachos,
``Trigonometric Structure Constants For New Infinite-Dimensional
Algebras,'' Phys. Lett. {\bf 218} (1989) 203; ``Infinite
Dimensional Algebras and a Trigonometric Basis for the Classical Lie
Algebras,'' Journal of Math. Phys. {\bf 31} (1990) 1088.}%
\refs{\hoppe,\fairlie} constructions somewhat similar to \connesc.
For other thoughts about applications of noncommutative geometry
in physics, see e.g.\
\ref\froehlich{A. H. Chamseddine and J. Froehlich, ``Some Elements of
Connes' NC Geometry, and Space-Time Geometry,'' in: Chen Ning Yang, a
Great Physicist of the Twentieth Century, C. S. Liu and S.-T. Yau
(eds.), Intl. Press, Boston 1995; 
J. Froehlich and K. Gawedzki, ``Conformal Field Theory and Geometry of
Strings,'' in: Proc. of the Conference on Mathematical Quantum Theory
(Vancouver B.C. 1993), J. Feldman, R. Froese and L. Rosen (eds.),
Centre de Recherche Mathematiques-Proc. and Lecture Notes, vol. 7,
p. 57 (1994), AMS Publ;
A. H. Chamseddine and J. Froehlich, J. Math. Phys. 35 (1994) 5195;
J. Froehlich, O. Grandjean and A. Recknagel, Supersymmetry, ``NC
Geometry and Gravitation,'' Proc. of Les Houches Summer School 1995,
A. Connes, K. Gawedzki and J. Zinn-Justin (eds.), Elsevier 1998.}.
Noncommutative geometry has also been used as a framework for open
string field theory 
\ref\ewitten{E. Witten, ``Noncommutative Geometry And String Field
Theory,'' Nucl. Phys. {\bf B268} (1986) 253.}.

Part of the beauty of the
analysis in \connes\ was that $T$-duality acts within the
noncommutative Yang-Mills framework, rather than, as one might expect,
mixing the modes of noncommutative Yang-Mills theory with string
winding states and other stringy excitations.  This makes the
framework of noncommutative Yang-Mills theory seem very powerful.

\nref\douglas{M. R. Douglas and C. Hull, ``$D$-Branes And The
Noncommutative Torus,'' JHEP {\bf 9802:008,1998}, hep-th/9711165.}%
\nref\cheung{Y.-K. E. Cheung and M. Krogh, ``Noncommutative Geometry
{}From 0-Branes In A Background $B$ Field,'' Nucl. Phys. {\bf B528}
(1998) 185.}%
\nref\chuho{C.-S. Chu and P.-M. Ho, ``Noncommutative Open String
And $D$-Brane,'' Nucl. Phys. {\bf B550} (1999) 151, hep/th9812219;
``Constrained quantization of open string in background B field and
noncommutative $D$-brane,'' hep-th/9906192.} 
\nref\schomerus{V. Schomerus, ``$D$-Branes And Deformation Quantization,''
JHEP {\bf 9906:030} (1999), hep-th/9903205.}%
\nref\aash{F.~Ardalan, H.~Arfaei and M.M.~Sheikh-Jabbari,
``Mixed Branes and M(atrix) Theory on Noncommutative Torus,''
hep-th/9803067;  ``Noncommutative Geometry From Strings and Branes,''
JHEP {\bf 02}, 016 (1999) hep-th/9810072; ``Dirac Quantization of Open
Strings and Noncommutativity in Branes,'' hep-th/9906161.}%
\nref\schwarz{A. Schwarz, ``Morita Equivalence And Duality,''
Nucl. Phys. {\bf B534} (1998) 720, hep-th/9805034; M.A.~Rieffel and
A.~Schwarz, ``Morita Equivalence of Multidimensional Noncommutative
Tori,'' math.QA/9803057; M.A.~Rieffel, J. Diff. Geom. {\bf 31} (1990)
535;  quant-ph/9712009.}
\nref\astash{  A.~Astashkevich, N.~Nekrasov and A.~Schwarz,
``On Noncommutative Nahm Transform,'' hep-th/9810147.}
\nref\zumino{B.~Morariu and B.~Zumino, ``Super-Yang-Mills on the
Noncommutative Torus,'' hep-th/9807198;
D.~Brace and B.~Morariu, ``A Note on the BPS Spectrum of the Matrix
Model,'' JHEP {\bf 02}, 004 (1999) hep-th/9810185;
D. Brace, B. Morariu, and B. Zumino, ``Dualities Of The Matrix Model
{}From $T$ Duality Of The Type II String,'' Nucl. Phys. {\bf B545}
(1999) 192, hep-th/9810099;  
``$T$ Duality And Ramond-Ramond Backgrounds In The Matrix Model,''
Nucl. Phys. {\bf B549} (1999) 181, hep-th/9811213.}%
\nref\everl{C. Hofman and E. Verlinde, ``$U$ Duality Of Born-Infeld
On The Noncommutative Two Torus,'' JHEP {\bf 9812:010} (1998),
hep-th/9810116; ``Gauge Bundles And Born-Infeld On The Noncommutative
Torus,'' Nucl. Phys. {\bf B547} (1999) 157, hep-th/9810219.}%
\nref\piosch{B. Pioline and A. Schwarz, ``Morita Equivalence and
$T$-duality (or $B$ versus $\Theta$,'' hep-th/9908019.}%
\nref\konech{ A.~Konechny and
A.~Schwarz, ``BPS States on Noncommutative Tori and Duality,''
Nucl. Phys. {\bf B550} (1999) 561, hep-th/9811159; ``Supersymmetry
Algebra and BPS States of Super-Yang-Mills Theories on Noncommutative
Tori,'' Phys. Lett. {\bf B453} (1999) 23, hep-th/9901077; ``1/4 BPS
States on Noncommutative Tori,'' hep-th/9907008.}%
\nref\phoaw{P.~Ho and Y.~Wu, ``Noncommutative Geometry and $D$-branes,''
Phys. Lett. {\bf B398} (1997) 52, hep-th/9611233;  P.-M.~Ho, Y.-Y. Wu
and Y.-S.~Wu, ``Towards a Noncommutative Geometric Approach to Matrix
Compactification,'' Phys. Rev. {\bf D58} (1998) 026006,
hep-th/9712201;  P.~Ho, ``Twisted Bundle on Quantum Torus and BPS
States in Matrix Theory,'' Phys. Lett. {\bf B434} (1998) 41,
hep-th/9803166;  P.~Ho and Y.~Wu, ``Noncommutative Gauge Theories in
Matrix Theory,'' Phys. Rev. {\bf D58} (1998) 066003, hep-th/9801147}%
\nref\mli{M.~Li, ``Comments on Supersymmetric Yang-Mills Theory on a
Noncommutative Torus,'' hep-th/9802052.}%
\nref\kawano{T.~Kawano and K.~Okuyama, ``Matrix Theory on
Noncommutative Torus,'' Phys. Lett. {\bf B433} (1998) 29,
hep-th/9803044.}%
\nref\lizzi{F.~Lizzi and R.J.~Szabo, ``Noncommutative Geometry
and Space-time Gauge Symmetries of String Theory,'' hep-th/9712206;
G.~Landi, F.~Lizzi and R.J.~Szabo, ``String Geometry and the
Noncommutative Torus,'' hep-th/9806099;  F.~Lizzi and R.J.~Szabo,
``Noncommutative Geometry and String Duality,'' hep-th/9904064.}%
\nref\casal{R.~Casalbuoni, ``Algebraic Treatment of Compactification
on Noncommutative Tori,'' Phys. Lett. {\bf B431} (1998) 69,
hep-th/9801170.}%
\nref\katoku{M.~Kato and T.~Kuroki, ``World Volume Noncommutativity
Versus Target Space Noncommutativity,''  JHEP {\bf 03} (1999) 012
hep-th/9902004.}%
\nref\bigatti{D.~Bigatti, ``Noncommutative Geometry and
Super-Yang-Mills Theory,'' Phys. Lett. {\bf B451} (1999) 324,
hep-th/9804120.}%
\nref\bigsuss{D.~Bigatti and L.~Susskind, ``Magnetic Fields, Branes
and Noncommutative Geometry,'' hep-th/9908056.}%
\nref\hashitz{A.~Hashimoto and N.~Itzhaki, ``Noncommutative Yang-Mills
and the AdS / CFT correspondence,'' hep-th/9907166.}%
\nref\newmald{J.M. Maldacena and J.G. Russo, ``Large N Limit of
Non-Commutative Gauge Theories,'' hep-th/9908134.}%
\nref\cheunggan{Y.E.~Cheung, O.J.~Ganor, M.~Krogh and A.Y.~Mikhailov,
``Instantons on a Noncommutative ${\bf T}^4$ from Twisted (2,0) and Little
String Theories,'' hep-th/9812172.}%

Subsequent work has gone in several directions.  Additional arguments
have been presented extracting noncommutative Yang-Mills theory more
directly from open strings without recourse to matrix theory
\refs{\douglas-\aash}.  The role of Morita equivalence in establishing
$T$-duality has been understood more fully \refs{\schwarz, \astash}.
The modules and their $T$-dualities have been reconsidered in a more
elementary language \refs{\zumino-\piosch}, and the relation to the
Dirac-Born-Infeld Lagrangian has been explored \refs{\everl,\piosch}.
The BPS spectrum has been more fully understood \refs{\zumino, \everl,
\konech}.  Various related aspects of noncommutative gauge theories
have been discussed in \refs{\phoaw -\newmald}.  Finally, the authors
of \cheunggan\ suggested interesting relations between noncommutative
gauge theory and the little string theory
\ref\seilst{N.~Seiberg, Phys. Lett. {\bf B408} (1997) 98,
hep-th/9705221.}.

\bigskip\noindent{\it Large Instantons And The $\alpha'$ Expansion}

Our work has been particularly influenced by certain further developments,
including  the analysis of instantons on a noncommutative $\R^4$
\ref\nekrasov{N. Nekrasov and A. Schwarz, ``Instantons On
Noncommutative ${\bf R}^4$ And (2,0) Superconformal Field Theory,''
Commun. Math. Phys. {\bf 198} (1998) 689, hep-th/9802068.}. 
It was shown that instantons on a noncommutative $\R^4$ can be
described by adding a constant (a Fayet-Iliopoulos term)
to the ADHM equations.  This constant
had been argued, following
\ref\doumoo{M.R.~Douglas and G.~Moore, ``$D$-branes, Quivers, and ALE
Instantons,'' hep-th/9603167.},
to arise in the description of instantons on $D$-branes upon turning
on a constant $B$-field
\ref\abs{O. Aharony, M. Berkooz, and N. Seiberg, ``Light Cone
Description Of (2,0) Superconformal Theories In Six-Dimensions,''
Adv. Theor. Math. Phys. {\bf 2:119} (1998), hep-th/9712117.},
\foot{One must recall that in
the presence of a $D$-brane, a constant $B$-field cannot be gauged
away and can in fact be reinterpreted as a magnetic field on the brane.} 
so putting the two facts together it was proposed that instantons on
branes with a $B$-field should be described by noncommutative
\nref\berkooz{M.~Berkooz, ``Nonlocal field theories and the
noncommutative torus,'' Phys. Lett. {\bf B430} (1998) 237,
hep-th/9802069.}%
Yang-Mills theory \refs{\nekrasov,\berkooz}.  

\nref\ewitten{E. Witten, ``Small Instantons In String Theory,''
Nucl. Phys. {\bf B460} (1996) 541, hep-th/9511030.}
\nref\mdouglas{M. Douglas, ``Branes Within Branes,'' hep-th/9512077.}
Another very cogent argument for this  is as
follows.  Consider $N$ parallel threebranes of Type IIB.  They can
support supersymmetric configurations in the form of $U(N)$
instantons.  If the instantons are large, they can be described by the
classical self-dual Yang-Mills equations.  If the instantons are
small, the classical description of the instantons is no longer good.
However, it can be shown that, at $B=0$, the instanton moduli space
${\cal M}$ in string theory coincides precisely with the classical
instanton moduli space.  The argument for this is presented in section
2.3.  In particular, ${\cal M}$ has the small instanton singularities
that are familiar from classical Yang-Mills theory.  The significance
of these singularities in string theory is well known: they arise
because an instanton can shrink to a point and escape as a $-1$-brane
\refs{\ewitten,\mdouglas}.   Now if one turns on a $B$-field, the
argument that the stringy instanton moduli space coincides with the
classical instanton moduli space fails, as we will also see in section
2.3.  Indeed, the instanton moduli space must be corrected for
nonzero $B$.  The reason is that, at nonzero $B$ (unless $B$ is
anti-self-dual) a configuration of a threebrane and a separated
$-1$-brane is not BPS,\foot{This is shown in a footnote in section 
4.2; the configurations in question
are further studied in section 5.} 
so an instanton on the threebrane cannot shrink
to a point and escape.  The instanton moduli space must therefore be
modified, for non-zero $B$, to eliminate the small instanton
singularity.  Adding a constant to the ADHM equations resolves the
small instanton singularity
\nref\nakajima{H. Nakajima, ``Resolutions of Moduli Spaces of Ideal 
Instantons on $\IR^4$,'' in ``Topology, Geometry and Field Theory,''
 (World Scientific, 1994).}%
\refs{\nakajima}, and since going to noncommutative $\R^4$ does add
this constant \nekrasov, this strongly encourages us to believe that
instantons with the $B$-field should be described as instantons on a
noncommutative space.

This line of thought leads to an apparent paradox, however.
Instantons come in all sizes, and however else they can be described,
big instantons can surely be described by conventional Yang-Mills
theory, with the familiar stringy $\alpha'$ corrections that are of
higher dimension, but possess the standard Yang-Mills gauge
invariance.  The proposal in \nekrasov\ implies, however, that the
large instantons would be described by classical Yang-Mills equations
with corrections coming from the noncommutativity of spacetime.  For
these two viewpoints to agree means that noncommutative Yang-Mills
theory must be equivalent to ordinary Yang-Mills theory perturbed by
higher dimension, gauge-invariant operators.  To put it differently,
it must be possible (at least to all orders in a systematic asymptotic
expansion) to map noncommutative Yang-Mills fields to ordinary
Yang-Mills fields, by a transformation that maps one kind of gauge
invariance to the other and adds higher dimension terms to the
equations of motion.  This at first sight seems implausible, but we
will see in section 3  that it is true.

Applying noncommutative Yang-Mills theory to instantons on $\R^4$
leads to another puzzle.  The original application of noncommutative
Yang-Mills to string theory \connes\ involved toroidal
compactification in a small volume limit.  The physics of noncompact
$\R^4$ is the opposite of a small volume limit!  The small volume
limit is also puzzling even in the case of a torus; 
if the volume of the torus the strings
propagate on is taken to zero, how can we end up with a noncommutative
torus of finite size, as has been proposed?  
Therefore, a reappraisal of the range of
usefulness of noncommutative Yang-Mills theory seems called for.  For
this, it is desireable to have new ways of understanding the
description of $D$-brane phenomena in terms of physics on noncommuting
spacetime.  A suggestion in this direction is given by recent analyses
arguing for noncommutativity of string coordinates in the presence of
a $B$-field, in a Hamiltonian treatment \chuho\ and also in a
worldsheet treatment that makes the computations particularly simple
\schomerus.
\nref\tseytlin{E.S.~Fradkin and A.A.~Tseytlin,
``Nonlinear Electrodynamics From Quantized Strings,'' Phys. Lett. {\bf
163B} (1985) 123. }%
\nref\callan{ C. G. Callan, C. Lovelace, C. R. Nappi, S. A. Yost,
``String Loop Corrections To Beta Functions,'' Nucl. Phys.
{\bf B288} (1987) 525; A.~Abouelsaood, C. G.~Callan, C. R.~Nappi and S. 
A.~Yost,
``Open Strings In Background Gauge Fields,'' Nucl. Phys. {\bf B280}
(1987) 599.}%
In the latter paper, it was suggested that rather classical features
of the propagation of strings in a constant magnetic field
\refs{\tseytlin,\callan} can be reinterpreted in terms of
noncommutativity of spacetime. 

In the present paper, we will build upon these suggestions and
reexamine the quantization of open strings ending on $D$-branes in the
presence of a $B$-field.  We will show that noncommutative Yang-Mills
theory is valid for some purposes in the presence of any nonzero
constant $B$-field, and that there is a systematic and efficient
description of the physics in terms of noncommutative Yang-Mills
theory when $B$ is large.  The limit of a torus of small volume with
fixed theta angle (that is, fixed periods of $B$)
\refs{\connes,\douglas} is an example with large $B$, but it is also
possible to have large $B$ on $\R^n$ and thereby make contact with the
application of noncommutative Yang-Mills to instantons on $\R^4$.  An
important element in our analysis is a distinction between two
different metrics in the problem.  Distances measured with respect to
one metric are scaled to zero as in \refs{\connes,\douglas}.  However,
the noncommutative theory is on a space with a different metric with
respect to which all distances are nonzero.  This guarantees that both
on $\R^n$ and on $\T^n$ we end up with a theory with finite metric.

\bigskip\noindent{\it Organization Of The Paper}

This paper is organized as follows.  In section 2, we reexamine the
behavior of open strings in the presence of a constant $B$-field.  We
show that, if one introduces the right variables, the $B$ dependence
of the effective action is completely described by making spacetime
noncommutative.  In this description, however, there is still an
$\alpha'$ expansion with all of its usual complexity.  We further show
that by taking $B$ large or equivalently by taking $\alpha'\to 0$
holding the effective open string parameters fixed, one can get an
effective description of the physics in terms of noncommutative
Yang-Mills theory.  This analysis makes it clear that two different
descriptions, one by ordinary Yang-Mills fields and one by
noncommutative Yang-Mills fields, differ by the choice of
regularization for the world-sheet theory.  This means that (as we
argued in another way above) there must be a change of variables from
ordinary to noncommutative Yang-Mills fields.  Once one is convinced
that it exists, it is not too hard to find this transformation
explicitly: it is presented in section 3.  In section 4, we make a
detailed exploration of the two descriptions by ordinary and
noncommutative Yang-Mills fields, in the case of almost constant
fields where one can use the Born-Infeld action for the ordinary
Yang-Mills fields.  In section 5, we explore the behavior of
instantons at nonzero $B$ by quantization of the D0-D4 system. Other
aspects of instantons are studied in sections 2.3 and 4.2.   In
section 6, we consider the behavior of noncommutative Yang-Mills
theory on a torus and analyze the action of $T$-duality, showing how
the standard action of $T$-duality on the underlying closed string
parameters induces the action of $T$-duality on the noncommutative
Yang-Mills theory that has been described in the literature
\refs{\schwarz-\piosch}.  We also  show that many
mathematical statements about modules over a noncommutative torus and
their Morita equivalences -- used in analyzing $T$-duality
mathematically -- can be systematically derived by quantization of
open strings.  In the remainder of the paper, we reexamine the
relation of noncommutative Yang-Mills theory to DLCQ quantization of
$M$-theory, and we explore the possible noncommutative version of the
$(2,0)$ theory in six dimensions.

\bigskip\noindent{\it Conventions}

We conclude this introduction with a statement of our main
conventions about noncommutative gauge theory.

For $\R^n$ with  coordinates $x^i$ whose commutators are $c$-numbers,
we write 
\eqn\commnonc{[x^i,x^j]= i\theta^{ij}}
with real $\theta$.  
\nref\moyalb{A. Dhar, G. Mandal, and S. R. Wadia, ``Nonrelativistic
Fermions, Coadjoint Orbits of $W_\infty$, and String Field Theory
At $c=1$,'' hep-th/9207011, ``$W_\infty$ Coherent States And
Path-Integral Derivation Of Bosonization Of Non-Relativistic Fermions
In One Dimension,'' hep-th/9309028, ``String Field Theory Of Two
Dimensional QCD: A Realization Of $W_\infty$ Algebra,''
hep-th/9403050.}%
\nref\moyalc{D. B. Fairlie, T. Curtright, and C. K. Zachos, ``Integrable
Symplectic Trilinear Interactions For Matrix Membranes,'' Phys. Lett.
{\bf B405} (1997) 37, hep-th/9704037; D. Fairlie, ``Moyal Brackets In
$M$-Theory,'' Mod. Phys. Lett. {\bf A13} (1998) 263, hep-th/9707190;
``Matrix Membranes And Integrability,'' hep-th/9709042.}%
\nref\klimcik{C.~Klimcik, ``Gauge theories on the noncommutative
sphere,'' Commun. Math. Phys. {\bf 199} (1998) 257, hep-th/9710153.}%
Given such a Lie algebra, one seeks to deform the algebra of functions
on $\R^n$ to a noncommutative, associative algebra ${\cal A}$ such
that $f*g=fg+\half i\theta^{ij}\partial_if\partial_jg+{\cal
O}(\theta^2)$, with the coefficient of each power of $\theta$ being a
local differential expression bilinear in $f$ and $g$.  The
essentially unique solution of this problem (modulo redefinitions of
$f$ and $g$ that are local order by order in $\theta$) is given by the
explicit formula
\eqn\defprod{f(x)*g(x)=e^{{i\over 2}\theta^{ij} {\partial \over
\partial \xi^i} {\partial \over \partial
\zeta^j} }f(x+\xi)g(x+\zeta)\big|_{\xi=\zeta=0} = fg +
{i\over 2}\theta^{ij} \partial_i f \partial_j g
+\CO(\theta^2).}   
This formula defines what is often called the
 Moyal bracket of functions; it
has appeared in the physics literature in many contexts, including
applications to old and new matrix theories
\refs{\hoppe,\fairlie, \moyalb - \klimcik}.  We also consider the case of
$N\times N$ matrix-valued functions $f,g$.  In this case, we define
the $*$ product to be the tensor product of matrix multiplication with
the $*$ product of functions as just defined.  The extended $*$
product is still associative.

The $*$ product is compatible with integration in the sense that for
functions $f$, $g$ that vanish rapidly enough at infinity, so that one
can integrate by parts in evaluating the following integrals, one has
\eqn\ucuc{\int \Tr\ f*g=\int \Tr\ g*f.}
Here $\Tr$ is the ordinary trace of the $N\times N$ matrices, and
$\int$ is the ordinary integration of functions.

For ordinary Yang-Mills theory, we write the  gauge transformations and
field strength as
\eqn\ogauge{\eqalign{
&\delta_\lambda A_i=\partial_i \lambda +i [\lambda,A_i] \cr
&F_{ij}= \partial_i A_j - \partial_j A_i-i[A_i,A_j]\cr 
&\delta_\lambda F_{ij}= i[\lambda, F_{ij}],}}
where $A$ and $\lambda$ are $N\times N$ hermitian matrices.  The Wilson line is
\eqn\wilsonline{W(a,b)=Pe^{i\int_b^a A},}
where in the path ordering $A(b)$ is to the right.  Under the gauge
transformation \ogauge\
\eqn\delwilsonline{\delta W(a,b)=i\lambda(a)W(a,b)
-iW(a,b)\lambda(b).} 

\def\hat{\widehat}
\def\tilde{\widetilde}
For noncommutative gauge theory, one uses the same formulas for
the gauge transformation law and the field strength, except that
matrix multiplication is replaced by the $*$ product.  Thus, the
gauge parameter $\hat\lambda$ takes values in $\A$ tensored with $N\times N$
hermitian matrices, for some $N$, 
and the same is true for the components $\hat A_i$
of the gauge field $\hat A$.
The gauge transformations and field strength of  noncommutative
Yang-Mills theory are thus
\eqn\ogaugen{\eqalign{
&\hat \delta_{\hat \lambda} \hat A_i=\partial_i \hat \lambda
+ i\hat \lambda* \hat A_i - i\hat A_i* \hat \lambda \cr
&\hat F_{ij}= \partial_i \hat A_j - \partial_j
\hat A_i - i\hat A_i * \hat A_j + i\hat A_j * \hat A_i \cr
&\hat \delta_{\hat \lambda}F_{ij} = i\hat \lambda *\hat F_{ij} -i\hat
F_{ij}*\hat \lambda .}} 
The theory obtained this way reduces to conventional $U(N)$ Yang-Mills
theory for $\theta\to 0$.  Because of the way that the theory is
constructed from associative algebras, there seems to be no convenient
way to get other gauge groups.  The commutator of two infinitesimal
gauge transformations with generators $\hat\lambda_1$ and
$\hat\lambda_2$ is, rather as in ordinary Yang-Mills theory, a gauge
transformation generated by $i(\hat\lambda_1*\hat
\lambda_2-\hat\lambda_2*\hat\lambda_1)$.  Such commutators are
nontrivial even for the rank 1 case, that is $N=1$, though for
$\theta=0$ the rank 1 case is the Abelian $U(1)$ gauge theory.  For
rank 1, to first order in $\theta$, the above formulas for the gauge
transformations and field strength read
\eqn\ogaugea{\eqalign{
&\hat \delta_{\hat \lambda} \hat A _i =\partial _i\hat \lambda -
\theta^{kl} \partial_k \hat \lambda \partial_l \hat A_i +\CO(\theta^2)
\cr 
&\hat F_{ij}= \partial_i \hat A_j - \partial_j \hat A_i +
\theta^{kl}\partial_k \hat A_i \partial_l \hat A_j +\CO(\theta^2) \cr
&\hat \delta_{\hat \lambda} \hat F_{ij} = -\theta^{kl}\partial_k
\hat \lambda \partial_l \hat F_{ij} +\CO(\theta^2).} }

Finally, a matter of terminology: we will consider the opposite of a
``noncommutative'' Yang-Mills field to be an ``ordinary'' Yang-Mills
field, rather than a ``commutative'' one.  To speak of ordinary
Yang-Mills fields, which can have a nonabelian gauge group, as being
``commutative'' would be a likely cause of confusion.

\newsec{Open Strings In The Presence Of Constant $B$-Field}

\subsec{Bosonic Strings}

In this section, we will study strings in flat space, with metric
$g_{ij}$, in the presence of a constant Neveu-Schwarz $B$-field and
with $Dp$-branes.  The $B$-field is equivalent to a constant magnetic
field on the brane; the subject has a long history and the basic
formulas with which we will begin were obtained in the mid-80's
\refs{\tseytlin,\callan}.
 
We will denote the rank of the matrix $B_{ij}$ as $r$; $r$ is of
course even.  Since the components of $B$ not along the brane can be
gauged away, we can assume that $r\le p+1$.  When our target space has
Lorentzian signature, we will assume that $B_{0i}=0$, with ``0'' the
time direction.  With a Euclidean target space we will not impose such
a restriction.  Our discussion applies equally well if space is
$\R^{10}$ or if some directions are toroidally compactified with $x^i
\sim x^i + 2\pi r^i$.  (One could pick a coordinate system with
$g_{ij}=\delta_{ij}$, in which case the identification of the
compactified coordinates may not be simply $x^i \sim x^i + 2\pi r^i$,
but we will not do that.)  If our space is $\R^{10}$, we can
pick coordinates so that $B_{ij}$ is nonzero only for $i,j=1,\dots,r$
and that $g_{ij}$ vanishes for $i=1,\dots,r$, $j\not=1,\dots,r$.  If some
of the coordinates are on a torus, we cannot pick such coordinates
without affecting the identification $x^i \sim x^i + 2\pi r^i$.  For
simplicity, we will still consider the case $B_{ij}\not=0$ only for
$i,j=1,\dots,r$ and $g_{ij}=0$ for $i=1,\dots,r$, $j\not=1,\dots,r$.

The worldsheet action is 
\eqn\acti{\eqalign{
S= &{1\over 4\pi\alpha'}\int_\Sigma \left( g_{ij} \partial_a x^i
\partial^a x^j -2\pi i\alpha' B_{ij}\epsilon^{ab} \partial_a
x^i\partial_b x^j\right) \cr
& ={1\over 4\pi\alpha'}\int_\Sigma g_{ij}\partial_a x^i \partial^a x^j
- {i \over 2} \int_{\partial \Sigma} B_{ij}x^i\partial_t x^j,}}
where $\Sigma$ is the string worldsheet, which we take to be with
Euclidean signature.  (With Lorentz signature, one would omit the
``$i$'' multiplying $B$.) $\partial_t$ is a tangential derivative
along the worldsheet boundary $\partial\Sigma$.  The equations of
motion determine the boundary conditions.  For $i$ along the
$Dp$-branes they are
\eqn\boucon{g_{ij}\partial_n x^j + 2\pi i \alpha' B_{ij} \partial_t
x^j\big|_{\partial \Sigma}=0,}
where $\partial_n$ is a normal derivative to $\partial \Sigma $.
(These boundary conditions are not compatible with real $x$, though
with a Lorentzian worldsheet the analogous boundary conditions would
be real.  Nonetheless, the open string theory can be analyzed by
determining the propagator and computing the correlation functions
with these boundary conditions.  In fact, another approach to the open
string problem is to omit or not specify the boundary term with $B$ in
the action \acti\ and simply impose the boundary conditions \boucon.)

For $B=0$, the boundary conditions in \boucon\ are Neumann boundary
conditions.  When $B$ has rank $r=p$ and $B\to\infty$, or equivalently
$g_{ij} \rightarrow 0$ along the spatial directions of the brane, the
boundary conditions become Dirichlet; indeed, in this limit, the
second term in \boucon\ dominates, and, with $B$ being invertible,
\boucon\ reduces to $\partial_tx^j=0$. This interpolation from Neumann
to Dirichlet boundary conditions will be important, since we will
eventually take $B\to\infty$ or $g_{ij}\to 0$.  For $B$ very large or
$g$ very small, each boundary of the string worldsheet is attached to
a single point in the $Dp$-brane, as if the string is attached to a
zero-brane in the $Dp$-brane.  Intuitively, these zero-branes are
roughly the constituent zero-branes of the $Dp$-brane as in the matrix
model of $M$-theory \refs{\bfss,\mmrev}, an interpretation that is
supported by the fact that in the matrix model the construction of
$Dp$-branes requires a nonzero $B$-field.

Our main focus in most of this paper will be the case that $\Sigma$ is
a disc, corresponding to the classical approximation to open string
theory.  The disc can be conformally mapped to the upper half plane;
in this description, the boundary conditions \boucon\ are
\eqn\bouconu{g_{ij} (\partial -\bar \partial) x^j + 2\pi\alpha'
B_{ij}(\partial + \bar \partial ) x^j\big|_{z=\bar z}=0,}
where $\partial=\partial/\partial z$, $\bar\partial = \partial/\partial \bar 
z$,
and ${\rm Im}\,z\geq 0$. 
The propagator with these boundary conditions is
\refs{\tseytlin,\callan}  
\eqn\propa{\eqalign{
\langle x^i(z) x^j(z') \rangle =&-\alpha'\Big[g^{ij}\log|z-z'|
-g^{ij}\log|z-\bar z'| \cr
&+G^{ij} \log |z-\bar z'|^2 +{1 \over 2\pi \alpha'} \theta^{ij}
\log{z-\bar z' \over \bar z - z'} + D^{ij} \Big].\cr}}
Here
\eqn\gthed{\eqalign{
&G^{ij} = \left({1 \over g+2\pi \alpha' B}\right)^{ij}_S=\left({1\over
g+2\pi\alpha' B} g {1\over g-2\pi\alpha' B}\right)^{ij}, \cr
&G_{ij} =g_{ij}-(2\pi\alpha')^2 \big(B g^{-1} B\big)_{ij},
\cr 
&\theta^{ij}= 2\pi \alpha' \left({1 \over g+ 2\pi\alpha'
B}\right)^{ij}_A= -(2\pi \alpha')^2 \left({1 \over g+ 2\pi\alpha'
B} B {1 \over g- 2\pi\alpha'
B}\right)^{ij},}}
where $(~)_S$ and $(~)_A$ denote the symmetric and antisymmetric part
of the matrix.  The constants $D^{ij}$ in \propa\ can depend on $B$
but are independent of $z$ and $z'$; they play no essential role and
can be set to a convenient value.  The first three terms in \propa\
are manifestly single-valued.  The fourth term is single-valued, if
the branch cut of the logarithm is in the lower half plane.

In this paper, our focus will be almost entirely on the open string
vertex operators and interactions.  Open string vertex operators are
of course inserted on the boundary of $\Sigma$.  So to get the
relevant propagator, we restrict \propa\ to real $z$ and $z'$, which
we denote $\tau$ and $\tau'$.  Evaluated at boundary points, the
propagator is
\eqn\proparg{\langle x^i(\tau) x^j(\tau') \rangle =-\alpha'
G^{ij}\log (\tau-\tau')^2 +{i\over 2} \theta^{ij}
\epsilon(\tau-\tau'),} 
where we have set $D^{ij}$ to a convenient value.   $\epsilon(\tau)$
is the function that is $1$ or $-1$ for positive or negative $\tau$.

The object $G_{ij}$ has a very simple intuitive interpretation: it is
the effective metric seen by the open strings.  The short distance
behavior of the propagator between interior points on $\Sigma$ is
$\langle x^i(z) x^j(z') \rangle =-\alpha'g^{ij}\log|z-z'|$.  The
coefficient of the logarithm determines the anomalous dimensions of
closed string vertex operators, so that it appears in the mass shell
condition for closed string states.  Thus, we will refer to $g_{ij}$
as the closed string metric.  $G_{ij}$ plays exactly the analogous
role for open strings, since anomalous dimensions of open string
vertex operators are determined by the coefficient of
$\log(\tau-\tau')^2$ in \proparg, and in this coefficient $G^{ij}$
enters in exactly the way that $g^{ij}$ would enter at $\theta=0$.  We
will refer to $G_{ij}$ as the open string metric.

The coefficient $\theta^{ij}$ in the propagator also has a simple
intuitive interpretation, suggested in \schomerus.  In conformal field
theory, one can compute commutators of operators from the short
distance behavior of operator products by interpreting time ordering
as operator ordering.  Interpreting $\tau$ as time, we see that
\eqn\xcomm{[x^i(\tau),x^j(\tau)]=T\left(x^i(\tau) x^j(\tau^-)
-x^i(\tau)x^j(\tau^+)\right)=i\theta^{ij}.} 
That is, $x^i$ are coordinates on a noncommutative space with
noncommutativity parameter $\theta$.

Consider the product of tachyon vertex operators $e^{ip\cdot x}(\tau)$
and $e^{iq\cdot x}(\tau')$.  With $\tau>\tau'$, we get for the leading
short distance singularity
\eqn\dolfo{e^{ip\cdot x}(\tau)\cdot e^{iq\cdot x}(\tau')\sim
(\tau-\tau')^{2\alpha'G^{ij}p_iq_j}e^{-\half
i\theta^{ij}p_iq_j}e^{i(p+q)\cdot x}(\tau') +\dots.}  
If we could ignore the term $(\tau-\tau')^{2\alpha' p\cdot q}$, then
the formula for the operator product would reduce to a $*$ product; we
would get
\eqn\polfo{e^{ip\cdot x}(\tau)e^{iq\cdot x}(\tau')\sim e^{ip\cdot x}
*e^{iq\cdot x}(\tau').}
This is no coincidence.  If the dimensions of all operators were zero,
the leading terms of operator products ${\cal O}(\tau){\cal
O}'(\tau')$ would be independent of $\tau-\tau'$ for $\tau\to \tau'$,
and would give an ordinary associative product of multiplication of
operators.  This would have to be the $*$ product, since that product
is determined by associativity, translation invariance, and \xcomm\
(in the form $x^i*x^j-x^j*x^i= i\theta^{ij}$).

Of course, it is completely wrong in general to ignore the anomalous
dimensions; they determine the mass shell condition in string theory,
and are completely essential to the way that string theory works.
Only in the limit of $\alpha'\to 0$ or equivalently small momenta can
one ignore the anomalous dimensions.  When the dimensions are
nontrivial, the leading singularities of operator products ${\cal
O}(\tau){\cal O}'(\tau')$ depend on $\tau-\tau'$ and do not give an
associative algebra in the standard sense.  For precisely this reason,
in formulating open string field theory in the framework of
noncommutative geometry
\ewitten, 
instead of using the operator product expansion directly, it was
necessary to define the associative $*$ product by a somewhat messy
procedure of gluing strings.  For the same reason, most of the present
paper will be written in a limit with $\alpha'\to 0$ that enables us
to see the $*$ product directly as a product of vertex operators.

\bigskip\noindent{\it $B$ Dependence Of The Effective Action}

However, there are some important general features of the theory that
do not depend on taking a zero slope limit.  We will describe these
first.

Consider an operator on the boundary of the disc that is of the
general form $P(\partial x, \partial^2x,\dots)e^{ip\cdot x}$, where $P$
is a polynomial in derivatives of $x$, and $x$ are coordinates along
the $Dp$-brane (the transverse coordinates satisfy Dirichlet boundary
conditions).  Since the second term in the propagator \proparg\ is
proportional to $\epsilon(\tau-\tau')$, it does not contribute to
contractions of derivatives of $x$.  Therefore, the expectation value
of a product of $k$ such operators, of momenta $p^1,\dots,p^k$,
 satisfies
\eqn\typcoll{\eqalign{
&\left\langle \prod_{n=1}^k P_n(\partial x(\tau_n),\partial^2
x(\tau_n),\dots) e^{ip^n\cdot x(\tau_n)}\right\rangle_{G,\theta} \cr
&\qquad =e^{-{i\over 2} \sum_{n> m}p_i^n
\theta^{ij} p_j^m \epsilon(\tau_n-\tau_m)}\left\langle \prod_{n=1}^k
P_n(\partial x(\tau_n),\partial^2 x(\tau_n),\dots)  e^{ip^n\cdot
x(\tau_n)}\right\rangle_{G,\theta=0},}} 
where $\langle\dots\rangle_{G,\theta}$ is the expectation value with the
propagator \proparg\ parametrized by $G$ and $\theta$.  We see that when
the theory is described in terms of the open string parameters $G$ and
$\theta$, rather than in terms of $g$ and $B$, the $\theta$ dependence
of correlation functions is very simple.  Note that because of momentum
conservation ($\sum_m p^m=0$), the crucial factor
\eqn\muvn{\exp\left(-{i\over 2} \sum_{n> m}p_i^n
\theta^{ij} p_j^m \epsilon(\tau_n-\tau_m)\right)}
depends only on the cyclic ordering of the points $\tau_1,\dots,\tau_k$
around the circle.

The string theory $S$-matrix can be obtained from the conformal field
theory correlators by putting external fields on shell and integrating
over the $\tau$'s.  Therefore, it has a structure inherited from
\typcoll.  To be very precise, in a theory with $N\times N$ Chan-Paton
factors, consider a $k$ point function of particles with Chan-Paton
wave functions $W_i$, $i=1,\dots, k$, momenta $p_i$, and additional
labels such as polarizations or spins that we will generically call
$\epsilon_i$.  The contribution to the scattering amplitude in which
the particles are cyclically ordered around the disc in the order from
1 to $k$ depends on the Chan-Paton wave functions by a factor
$\Tr\,W_1W_2\dots W_k$.  We suppose, for simplicity, that $N$ is large
enough so that there are no identities between this factor and similar
factors with other orderings.  (It is trivial to relax this
assumption.)  By studying the behavior of the $S$-matrix of massless
particles of small momenta, one can extract order by order in
$\alpha'$ a low energy effective action for the theory.  If $\Phi_i$
is an $N\times N$ matrix-valued function in spacetime representing a
wavefunction for the $i^{th}$ field, then at $B=0$ a general term in
the effective action is a sum of expressions of the form
\eqn\tolgo{\int d^{p+1}x\sqrt{\det G} \Tr
\partial^{n_1}\Phi_1\partial^{n_2}\Phi_2\dots \partial^{n_k}\Phi_k.}
Here $\partial^{n_i}$ is, for each $i$, the product of $n_i$ partial
derivatives with respect to some of the spacetime coordinates; which
coordinates it is has not been specified in the notation.  The indices
on fields and derivatives are contracted with the metric $G$, though
this is not shown explicitly in the formula.

Now to incorporate the $B$-field, at fixed $G$, is very simple: if the
effective action is written in momentum space, we need only
incorporate the factor \muvn.  Including this factor is equivalent to
replacing the ordinary product of fields in \tolgo\ by a $*$ product.
(In this formulation, one can work in coordinate space rather than
momentum space.)  So the term corresponding to \tolgo\ in the
effective action is given by the same expression but with the wave
functions multiplied using the $*$ product:
\eqn\tpolgo{\int d^{p+1}x\sqrt{\det G} \Tr
\partial^{n_1}\Phi_1*\partial^{n_2}\Phi_2*\dots
*\partial^{n_k}\Phi_k.} 
It follows, then, that the $B$ dependence of the effective action for
fixed $G$ and constant $B$ can be obtained in the following very
simple fashion: replace ordinary multiplication by the $*$ product.
We will make presently an explicit calculation of an $S$-matrix
element to illustrate this statement, and we will make a detailed
check of a different kind in section 4 using almost constant fields
and the Dirac-Born-Infeld theory.

Though we have obtained a simple description of the $B$-dependence of
the effective action, the discussion also makes clear that going to
the noncommutative description does not in general enable us to
describe the effective action in closed form: it has an $\alpha'$
expansion that is just as complicated as the usual $\alpha'$ expansion
at $B=0$.  To get a simpler description, and increase the power of the
description by noncommutative Yang-Mills theory, we should take the
$\alpha'\to 0$ limit.

\bigskip\noindent{\it The $\alpha'\to 0$ Limit}

For reasons just stated, and to focus on the low energy behavior while
decoupling the string behavior, we would like to consider the zero
slope limit ($\alpha'\rightarrow 0$) of our open string system.
Clearly, since open strings are sensitive to $G$ and $\theta$, we
should take the limit $\alpha'\to 0$ keeping fixed these parameters
rather than the closed string parameters $g$ and $B$.

So we consider the limit
\eqn\limfi{\eqalign{
&\alpha'\sim \epsilon^\half \rightarrow 0\cr
&g_{ij}\sim \epsilon \rightarrow 0 \quad {\rm for}~i,j=1,\dots,r}}
with everything else, including the two-form $B$, held fixed.
Then \gthed\ become
\eqn\limmet{\eqalign{
&G^{ij}=\cases{
-{1 \over (2\pi\alpha')^2} \left({1 \over B}g {1 \over B}\right)^{ij}
& for $i,j=1,\dots,r$ \cr
g^{ij}& otherwise}\cr
&G_{ij}=\cases{
-(2\pi\alpha')^2 (Bg^{-1}B)_{ij} & for $i,j=1,\dots,r$ \cr
g_{ij}& otherwise}\cr
&\theta^{ij}=\cases{
\left({1 \over B}\right)^{ij} & for $i,j=1,\dots,r$ \cr
0& otherwise.}}}
Clearly, $G$ and $\theta$ are finite in the limit.  In this limit the
boundary propagator \proparg\ becomes
\eqn\limprop{\langle x^i(\tau) x^j(0)\rangle = {i\over 2}
\theta^{ij}\epsilon(\tau).} 

In this $\alpha'\to 0$ limit, the bulk kinetic term for the $x^i$ with
$i=1,\dots,r$ (the first term in \acti) vanishes.  Hence, their bulk
theory is topological.  The boundary degrees of freedom are governed
by the following action:
\eqn\bounlag{-{i \over 2} \int_{\partial\Sigma} B_{ij}x^i\partial_t
x^j.}
(A sigma model with only such a boundary interaction, plus gauge
fixing terms, is a special case of the theory used by Kontsevich in
studying deformation quantization 
\ref\kontsevich{M. Kontsevich, ``Deformation Quantization Of Poisson
Manifolds,'' q-alg/9709040.}, 
as has been subsequently elucidated 
\ref\catfel{A. S. Cataneo and G. Felder, ``A Path Integral Approach To
The Kontsevich Quantization Formula,'' math.QA/9902090.}.)  
If one regards \bounlag\ as a one-dimensional action (ignoring the
fact that $x^i(\tau)$ is the boundary value of a string), then it
describes the motion of electrons in the presence of a large magnetic
field, such that all the electrons are in the first Landau level.  In
this theory the spatial coordinates are canonically conjugate to each
other, and $[x^i,x^j]\not =0$.  As we will discuss in section 6.3, when
we construct the representations or modules for a noncommutative
torus, the fact that $x^i(\tau)$ is the boundary value of a string
changes the story in a subtle way, but the general picture that the
$x^i(\tau)$ are noncommuting operators remains valid.

With the propagator \limprop, normal ordered operators satisfy
\eqn\normal{:e^{ip_ix^i(\tau)} :~:e^{iq_ix^i(0)} : =e^{-{i\over 2}
\theta^{ij}p_iq_j\epsilon(\tau)}: e^{ipx(\tau)+iqx(0)}:,}
or more generally
\eqn\normalg{:f(x(\tau)) :~:g(x(0)) : = :e^{{i\over 2}\epsilon(\tau)
\theta^{ij}{\partial\over \partial x^i(\tau)} {\partial\over \partial
x^j(0)}} f(x(\tau)) g(x(0)) :,}
and
\eqn\normalgf{\lim_{\tau\rightarrow 0^+} :f(x(\tau)) :~:g(x(0)) : =
:f(x(0)) * g(x(0)) :,}
where
\eqn\defproda{f(x)*g(x)=e^{{i\over 2} \theta^{ij} {\partial \over
\partial \xi^i} {\partial \over \partial
\zeta^j} }f(x+\xi)g(x+\zeta)\big|_{\xi=\zeta=0}}
is the product of functions on a noncommutative space.  

As always in the zero slope limit, the propagator \limprop\ is not
singular as $\tau\rightarrow 0$.  
 This
lack of singularity ensures that the product of operators can be defined
without a subtraction and hence must be associative.
It is similar to a product of functions, but on a noncommutative space.

The correlation functions of exponential operators on the boundary of
a disc are
\eqn\corrfu{\left\langle \prod_n e^{ip_i^n x^i(\tau_n)}\right\rangle
= e^{-{i\over 2} \sum_{n> m}p_i^n \theta^{ij} p_j^m
\epsilon(\tau_n-\tau_m)}\delta\left(\sum p^n\right).} 
Because of the $\delta$ function and the antisymmetry of
$\theta^{ij}$, the correlation functions are unchanged under cyclic
permutation of $\tau_n$.  This means that the correlation functions
are well defined on the boundary of the disc.  More generally,
\eqn\corrfug{\left\langle \prod_n f_n(x(\tau_n))\right\rangle  = \int
dx f_1(x)*f_2(x)*\dots * f_n,}
which is invariant under cyclic permutations of the $f_n$'s.  As always
in the zero slope limit, the correlation functions \corrfu, \corrfug\ do
not exhibit singularities in $\tau$, and therefore there are no poles
associated with massive string states.

\bigskip\noindent{\it Adding Gauge Fields}

Background gauge fields couple to the string worldsheet by adding
\eqn\gaugecou{-i\int d\tau A_i(x)\partial_\tau x^i}
to the action \acti.  We assume for simplicity that there is only a
rank one gauge field; the extension to higher rank is straightforward.
Comparing \acti\ and \gaugecou, we see that a constant $B$-field can
be replaced by the gauge field $A_i= -\half B_{ij}x^j$, whose field
strength is $F=B$.  When we are working on $\R^n$, we are usually
interested in situations where $B$ and $F$ are constant at infinity,
and we fix the ambiguity be requiring that $F$ is zero at infinity.

Naively, \gaugecou\ is invariant under ordinary gauge transformations
\eqn\naiveg{\delta A_i = \partial_i\lambda}
because \gaugecou\ transforms by a total derivative 
\eqn\gaugetotal{\delta \int d\tau A_i(x)\partial_\tau x^i=
\int d\tau \partial_i\lambda \partial_\tau x^i= \int
d\tau \partial_\tau \lambda.}  
However, because of the infinities in quantum field theory, the theory
has to be regularized and we need to be more careful.  We will examine
a point splitting regularization, where different operators are never
at the same point.  

Then expanding the exponential of the action in powers of $A$ and
using the transformation law \naiveg, we find that the functional
integral transforms by
\eqn\noggau{-\int d\tau A_i(x)\partial_\tau x^i\cdot \int d\tau'
\partial_{\tau'}\lambda}
plus terms of higher order in $A$.  The product of operators in \noggau\
can be regularized in a variety of ways.  We will make a point-splitting
regularization in which we cut out the region $|\tau-\tau'|<\delta$
and take the limit $\delta\to 0$.  Though the integrand is a total
derivative, the $\tau'$ integral contributes
surface terms at $\tau-\tau'=\pm \delta$. In the limit $\delta\to 0$,
the surface terms contribute 
\eqn\correctgau{\eqalign{
-\int d\tau & :A_i(x(\tau))\partial_\tau x^i(\tau):~
:\left(\lambda(x(\tau^-)) -\lambda(x(\tau^+))\right):\cr
& = -\int d\tau :\left(A_i(x)*\lambda -\lambda*A_i(x)
\right)\partial_\tau x^i:}}
Here we have used the relation of the operator product to the $*$ product,
and the fact that with the propagator \limprop\ there is no contraction between
$\partial_\tau x$ and $x$. To cancel this term, we must add another
term to the variation of the gauge field; the theory is invariant not
under \naiveg, but under
\eqn\newg{\hat \delta \hat A_i= \partial_i\lambda +i\lambda* \hat A_i-
i \hat A_i*\lambda.}
This is the gauge invariance of noncommutative Yang-Mills theory,
and in recognition of that fact we henceforth denote the gauge field
in the theory defined with point splitting regularization as $\hat A$.
A sigma model expansion with Pauli-Villars regularization would have
preserved the standard gauge invariance of open string gauge field, so
whether we get ordinary or noncommutative gauge fields depends on the
choice of regulator.

We have made this derivation to lowest order in $\hat A$, but it is
straightforward to go to higher orders.  
At the $n$-th order in $\hat A$, the variation is
\eqn\nthord{\eqalign{
{i^{n+1}\over n!}\int &\hat A(x(t_1))\dots\hat A(x(t_n))\partial_t
\lambda(x(t))\cr 
&+{i^{n+1}\over (n-1)!}\int \hat A(x(t_1))\dots\hat A(x(t_{n-1} ))
\left(\lambda*\hat A(x(t_n)) - \hat A* \lambda(x(t_n))\right),}}
where the integration region excludes points where some $t$'s
coincide.  The first term in \nthord\ arises by using the naive gauge
transformation \naiveg, and expanding the action to $n$-th order in
$\hat A$ and to first order in $\lambda$.  The second term arises from
using the correction to the gauge transformation in \newg\ and
expanding the action to the same order in $ \hat A$ and $\lambda$.
The first term can be written as
\eqn\nthordm{\eqalign{
&{i^{n+1}\over n!}\sum_j\int
\hat A(x(t_1))\dots\hat A(x(t_{j-1}))\hat A(x(t_{j+1}))\dots
\hat A(x(t_n))\left(\hat A*\lambda(x(t_j))- \lambda*\hat A(x(t_j))
\right)\cr  
&\qquad = {i^{n+1}\over (n-1)!}\int \hat A(x(t_1))\dots\hat
A(x(t_{n-1})) 
\left(\hat A*\lambda(x(t_n)) - \lambda*\hat A(x(t_n))\right),}}
making it clear that \nthord\ vanishes.  Therefore, there is no need
to modify the gauge transformation law \newg\ at higher orders in
$\hat A$. 

Let us return to the original theory before taking the zero slope limit
\limfi, and examine the correlation functions of the physical vertex
operators of gauge fields 
\eqn\phiso{V=\int \xi\cdot \partial x e^{ip\cdot x}}
These operators are physical when
\eqn\phycon{\xi\cdot p=p\cdot p=0,}
where the dot product is with the open string metric $G$ \gthed.  We
will do an explicit calculation to illustrate the statement that the
$B$ dependence of the $S$-matrix, for fixed $G$, consists of replacing
ordinary products with $*$ products.  Using the conditions \phycon\
and momentum conservation, the three point function is
\eqn\corrfun{\eqalign{
\big\langle \xi^1\cdot \partial x& e^{ip^1 \cdot x(\tau
_1)}~\xi^2\cdot \partial x e^{ip^2 \cdot x(\tau _2)}~\xi^3\cdot
\partial x e^{ip^3 \cdot x(\tau _3)} \big\rangle \sim {1 \over (\tau
_1-\tau _2)(\tau _2-\tau _3)(\tau _3-\tau _1)} \cr 
&\cdot \left( \xi^1 \cdot\xi^2 p^2 \cdot \xi^3 +  \xi^1 \cdot \xi^3p^1
\cdot \xi^2 +  \xi^2 \cdot \xi^3 p^3 \cdot \xi^1 + 2\alpha' p^3 \cdot
\xi^1 p^1\cdot \xi^2 p^2 \cdot \xi^3 \right) \cr
&\cdot e^{-{i\over 2}\left(p^1_i\theta^{ij}p^2_j \epsilon (\tau
_1-\tau _2) +p^2_i\theta^{ij}p^3_j\epsilon (\tau _2-\tau _3) 
+p^3_i\theta^{ij}p^1_j\epsilon (\tau _3-\tau _1) \right)}.}} 
This expression should be multiplied by the Chan-Paton matrices.  The
order of these matrices is correlated with the order of $\tau_n$.
Therefore, for a given order of these matrices we should not sum over
different orders of $\tau_n$.  Generically, the vertex operators
\phiso\ should be integrated over $\tau_n$, but in the case of the
three point function on the disc, the gauge fixing of the $SL(2;{\bf
R})$ conformal group cancels the integral over the $\tau$'s.  All we
need to do is to remove the denominator $(\tau _1-\tau _2)(\tau
_2-\tau _3)(\tau _3-\tau _1)$.  This leads to the amplitude
\eqn\threeamp{\left( \xi^1 \cdot\xi^2 p^2 \cdot \xi^3 + \xi^1 \cdot
\xi^3p^1 \cdot \xi^2 +  \xi^2 \cdot \xi^3 p^3 \cdot \xi^1 + 2\alpha'
p^3 \cdot \xi^1 p^1\cdot \xi^2 p^2 \cdot \xi^3 \right)\cdot
e^{-{i\over 2} p^1_i\theta^{ij}p^2_j }.}

The first three terms are the same as the three point function
evaluated with the action
\eqn\effact{{(\alpha')^{3-p\over 2}\over 4(2\pi)^{p-2}G_s } \int
\sqrt{\det G} G^{ii'}G^{jj'}\Tr\ \hat F_{ij}*\hat F_{i'j'},}  
where $G_s$ is the string coupling and
\eqn\noncmfn{\hat F_{ij}=\partial_i\hat A_j -
\partial_j \hat A_i-i\hat A_i*\hat A_j+i\hat A_j*\hat A_i} 
is the noncommutative field strength.  The normalization is the
standard normalization in open string theory. The effective open
string coupling constant $G_s$ in \effact\ can differ from the closed
string coupling constant $g_s$.  We will determine the relation
between them shortly.  The last term in \threeamp\ arises from
the $(\partial \hat A)^3$ part of a term $\alpha' \hat F^3$ in the
effective action.  This term vanishes for $\alpha'\to 0$ (and in any
event is absent for superstrings).

Gauge invariance of \effact\ is slightly more subtle than in ordinary
Yang-Mills theory.  Since under gauge transformations $\hat \delta
\hat F= i\lambda *\hat F -i\hat F*\lambda$, the gauge variation of
$\hat F * \hat F$ is not zero.  But this gauge variation is
$\lambda*(i\hat F *\hat F)-(i\hat F*\hat F)*\lambda$, and the integral
of this vanishes by virtue of \ucuc.  Notice that, because the scaling
in \limfi\ keeps all components of $G$ fixed as $\epsilon\to 0$,
\effact\ is uniformly valid whether the rank of $B$ is $p+1$ or
smaller. 

The three point function \corrfun\ can easily be generalized to any
number of gauge fields.  Using \typcoll
\eqn\nptint{\left\langle \prod_n \xi^n\cdot \partial x e^{ip^n\cdot
x(\tau_n)}\right\rangle_{G,\theta} =e^{-{i\over 2} \sum_{n> m}p_i^n
\theta^{ij} p_j^m \epsilon(\tau_n-\tau_m)}\left\langle \prod_n \xi^n
\cdot  \partial x e^{ip^n\cdot x(\tau_n)}\right\rangle_{G,\theta=0}.}
This illustrates the claim that when the effective action is expressed
in terms of the open string variables $G$, $\theta$ and $G_s$ (as
opposed to $g$, $B$ and $g_s$), $\theta$ appears only in the $*$
product.

The construction of the effective Lagrangian from the $S$-matrix
elements is always subject to a well-known ambiguity.  The $S$-matrix
is unchanged under field redefinitions in the effective Lagrangian.
Therefore, there is no canonical choice of fields.  The vertex operators
determine the linearized gauge symmetry, but field redefinitions $A_i
\rightarrow A_i + f_i(A_j)$ can modify the nonlinear terms.  It is
conventional in string theory to define an effective action for ordinary
gauge fields with ordinary gauge invariances that generates the $S$-matrix.
In this formulation, the $B$-dependence of the effective action
is very simple: it is described by everywhere replacing $F$ by $F+B$.
(This is manifest in the sigma model approach that we mention presently.)

We now see that it is also natural to generate the $S$-matrix from an
effective action written for noncommutative Yang-Mills fields. In
this description, the $B$-dependence is again simple, though different.
 For fixed $G$ and
$G_s$, $B$ affects only $\theta$, which determines the $*$ product.
Being able to describe the same $S$-matrix with the two kinds of
fields means that there must be a field redefinition of the form $A_i
\rightarrow A_i + f_i(A_j)$, which relates them.

This freedom to write the effective action in terms of different
fields has a counterpart 
in the sigma model description of string theory.  Here we can use
different regularization schemes.  With Pauli-Villars regularization
(such as the regularization we use in section 2.3),
the theory has ordinary gauge symmetry, as the total derivative in
\gaugetotal\ integrates to zero.  
Additionally, with such a regularization, the effective action can depend
on $B$ and $F$ only in the combination $F+B$, since there is a symmetry
$A\to A+\Lambda$, $B\to B-d\Lambda$, for any one-form $\Lambda$.
With point-splitting regularization,
we have found noncommutative gauge symmetry, and a different description
of the $B$-dependence.  

The difference between
different regularizations is always in a choice of contact terms; theories
defined with different regularizations
are related by coupling constant redefinition.  Since the
coupling constants in the worldsheet Lagrangian are the spacetime
fields, the two descriptions must be related by a field redefinition.
The transformation from ordinary to noncommutative Yang-Mills fields
that we will describe in section 3 is thus an example of a
transformation of coupling parameters that is required to compare two
different regularizations of the same  quantum field
theory.

In the $\alpha'\to 0$ limit \limfi, the amplitudes and the effective
action are simplified.  For example, 
the $\alpha' \hat F^3$ term coming from 
the last term in the amplitude
\threeamp\ is negligible in this limit.  
More generally, using dimensional analysis and the
fact that the $\theta$ dependence is only in the definition of the $*$
product, it is clear that all higher dimension operators involve more
powers of $\alpha'$.  Therefore they can be neglected, and the $\hat
F^2$ action \effact\ becomes exact for $\alpha'\to 0$.

The lack of higher order corrections to \effact\ can also be
understood as follows.  In the limit \limfi, there are no on-shell
vertex operators with more derivatives of $x$, which would correspond
to massive string modes.  Since there are no massive string modes,
there cannot be corrections to \effact.  As a consistency check, note
that there are no poles associated with such operators in
\corrfu\ or in \nptint\ in our limit.

All this is standard in the zero slope limit, and the fact that the
action for $\alpha'\to 0$ reduces to $\hat F^2$ is quite analogous to
the standard reduction of open string theory to ordinary Yang-Mills
theory for $\alpha'\to 0$.  The only novelty in our discussion is the
fact that for $B\not=0$, we have to take $\alpha'\to 0$ keeping fixed
$G$ rather than $g$.  Even before taking the $\alpha'\to 0$ limit, the
effective action, as we have seen, can be written in terms of the
noncommutative variables.  The role of the zero slope limit is just to
remove the higher order corrections to $\hat F^2$ from the effective
action.

It remains to determine the relation between the effective open string
coupling $G_s$ which appears in \effact\ and the closed string
variables $g$, $B$ and $g_s$.  For this, we examine the constant term
in the effective Lagrangian.  For slowly varying fields, the effective
Lagrangian is the Dirac-Born-Infeld Lagrangian (for a recent review of
the DBI theory see
\ref\tsyrev{ A.A. Tseytlin, ``Born-Infeld Action, Supersymmetry And
String Theory,'' to appear in the Yuri Golfand memorial volume,
ed. M. Shifman, World Scientific (2000), hep-th/9908105.} 
and references therein)
\eqn\biagma{\CL_{DBI}={1 \over g_s (2\pi)^p(\alpha')^{p+1\over
2}}\sqrt{\det (g+2\pi\alpha'(B+F))}.}
The coefficient is determined by the $Dp$-brane tension which for $B=0$
is 
\eqn\dpten{T_p(B=0)={1 \over g_s(2\pi)^p(\alpha')^{p+1\over 2}}.}
Therefore
\eqn\biagmam{\CL(F=0)={1 \over g_s
(2\pi)^p(\alpha')^{p+1\over 2}}\sqrt{\det (g+2\pi\alpha'B)}.}
Above we argued that when the effective action is expressed in terms
of noncommutative gauge fields and 
the open string variables $G$, $\theta$ and $G_s$, the $\theta$
dependence is entirely in the $*$ product.  In this description,
the analog of \biagma\  is
\eqn\bibagmama{\CL(\widehat F)={1 \over G_s (2\pi)^p(\alpha')^{p+1\over
2}}\sqrt{\det G+2\pi \alpha'\widehat F},}
and  the constant term
in the effective Lagrangian is
\eqn\biagmama{\CL(\widehat F=0)={1 \over G_s (2\pi)^p(\alpha')^{p+1\over
2}}\sqrt{\det G}.}
Therefore, 
\eqn\limpropg{G_s=g_s\left({\det G\over \det{(g+2\pi
\alpha'B)}}\right)^{1\over 2}=
g_s\left({\det G \over \det g}\right)^{1\over 4} =
g_s\left({\det (g+2\pi \alpha' B) \over \det g}\right)^\half,}
where the definition \gthed\ of $G$ has been used.
As a (rather trivial) consistency check, note that when $B=0$ we have
$G_s=g_s$.  In the zero slope limit \limfi\ it becomes
\eqn\limpropgz{G_s= g_s\det' (2\pi \alpha' Bg^{-1})^\half,}
where $\det'$ denotes a determinant in the $r\times r$ block with
nonzero $B$.

The effective Yang-Mills coupling is determined from the $\widehat F^2$
term in \bibagmama\ and is
\eqn\ymcoup{{1 \over g_{YM}^2} = {(\alpha')^{3-p\over 2}\over
(2\pi)^{p-2}G_s }= {(\alpha')^{3-p\over 2}\over
(2\pi)^{p-2}g_s}\left({\det (g+2\pi \alpha' B) \over \det
G}\right)^\half.} 
Using \limpropgz\ we see that in order to keep it finite in our limit
such that we end up with a 
quantum theory, we should scale
\eqn\scalegs{\eqalign{
&G_s \sim \epsilon^{3-p\over 4}\cr
&g_s \sim \epsilon^{3-p+r\over 4}.}}
Note that the scaling of $g_s$ depends on the rank $r$ of the $B$
field, while the scaling of $G_s$ is independent of $B$.  The scaling
of $G_s$ just compensates for the dimension of the Yang-Mills
coupling, which is proportional to $p-3$ as the Yang-Mills theory on a
brane is scale-invariant precisely for threebranes.

If several $D$-branes are present, we should scale $g_s$ such that all
gauge couplings of all branes are finite.  For example, if there are
some $D0$-branes, we should scale $g_s \sim\epsilon^{3\over 4} $
($p=r=0$ in \scalegs).  In this case, all branes for which $p>r$ can be
treated classically, and branes with $p=r$ are quantum.

If we are on a torus, then the limit \limfi\ with $g_{ij}\to 0$ and
$B_{ij}$ fixed is essentially the limit used in \connes.  This limit
takes the volume to zero while keeping fixed the periods of $B$.  On
the other hand, if we are on $\R^n$, then by rescaling the
coordinates, instead of taking $g_{ij}\to 0$ with $B_{ij}$ fixed, one
could equivalently keep $g_{ij}$ fixed and take $B_{ij}\to\infty$.
(Scaling the coordinates on $\T^n$ changes the periodicity, and
therefore it is more natural to scale the metric in this case.) In
this sense, the $\alpha'\to 0$ limit can, on $\R^n$, be interpreted as
a large $B$ limit.

It is crucial that $g_{ij}$ is taken to zero with fixed $G_{ij}$.  The
latter is the metric appearing in the effective Lagrangian.
Therefore, either on $\R^n$ or on a torus, all distances measured with
the metric $g$ scale to zero, but the noncommutative theory is
sensitive to the metric $G$, and with respect to this metric the
distances are fixed.  This is the reason that we end up with finite
distances even though the closed string metric $g$ is taken to zero.

\subsec{Worldsheet Supersymmetry}

We now add fermions to the theory and consider worldsheet
supersymmetry.  Without background gauge fields we have to add to the
action \acti\
\eqn\fermiont{ {i\over 4\pi\alpha'}\int_\Sigma \left(g_{ij}\psi^i \bar
\partial\psi^j + g_{ij}\bar \psi^i \partial \bar \psi^j\right)}
and the boundary conditions are
\eqn\bouconf{g_{ij}(\psi^j -\bar \psi^j) +2\pi\alpha' B_{ij}
(\psi^j+\bar\psi^j) \big|_{z=\bar z }=0}
($\bar \psi$ is not the complex conjugate of $\psi$).  The action and
the boundary conditions respect the supersymmetry transformations
\eqn\susytran{\eqalign{
&\delta x^i=-i\eta (\psi^i +\bar \psi^i)\cr
&\delta \psi^i=\eta \partial x^i\cr
&\delta \bar \psi^i=\eta \bar \partial x^i,\cr}}

In studying sigma models, the boundary
 interaction \gaugecou\ is typically extended to
\eqn\gaugecous{L_{A}=-i\int d\tau \left(A_i(x)\partial_\tau x^i -i
F_{ij}\Psi^i\Psi^j\right)} 
with $F_{ij}=\partial_i A_j-\partial_j A_i$ and 
\eqn\psiide{\Psi^i=\half(\psi^i
+\bar\psi^i)= \left({1\over g-2\pi\alpha' B} g\right)^i_j \psi^j.}

The expression \gaugecous\ seems to be invariant under \susytran\
because its variation is a total derivative
\eqn\susytotal{\delta\int d\tau \left(A_i(x)\partial_\tau x^i -i
F_{ij}\Psi^i\Psi^j \right)= -2i\eta\int d\tau
\partial_\tau(A_i\Psi^i).} 
However, as in the derivation of \correctgau, with point splitting
regularization, a total derivative such as the one in \susytotal\ can
contribute a surface term.  In this case, the surface term is obtained
by expanding the $\exp(-L_{A})$ term in the path integral in powers of
$A$.  The variation of the path integral coming from \susytotal\
reads, to first order in $L_{A}$,
\eqn\nubuo{i\int d\tau\int d\tau'\left(A_i\partial_\tau x^i(\tau)-iF_{ij}
\Psi^i\Psi^j(\tau)\right)\left(-2i\eta\partial_{\tau'} 
A_k\Psi^k(\tau')\right).}
With point splitting regularization, one picks up surface terms as
$\tau'\to\tau^+$ and $\tau'\to \tau^-$, similar to those in
\correctgau.  The surface terms can be canceled by the supersymmetric
variation of an additional interaction term $\int d\tau
A_i*A_j\Psi^i\Psi^j(\tau)$, and the conclusion is that with
point-splitting regularization, \gaugecous\ should be corrected to
\eqn\gaugecousn{-i\int d\tau \left(A_i(x)\partial_\tau x^i -i \hat
F_{ij}\Psi^i\Psi^j\right)} 
with $\hat F$ the noncommutative field strength \noncmfn.

Once again, if supersymmetric Pauli-Villars regularization were used
(an example of an explicit regularization procedure will be given
presently in discussing instantons), the more naive boundary coupling
\gaugecous\ would be supersymmetric.  Whether ``ordinary'' or
``noncommutative'' gauge fields and symmetries appear in the formalism
depends on the regularization used, so there must be a transformation
between them.

\subsec{Instantons On Noncommutative ${\bf R}^4$}

As we mentioned in the introduction, one of the most fascinating
applications of noncommutative Yang-Mills theory has been to
instantons on $\R^4$.  Given a system of $N$ parallel $D$-branes with
worldvolume $\R^4$, one can study supersymmetric configurations in the
$U(N)$ gauge theory.  (Actually, most of the following discussion
applies just as well if $\R^4$ is replaced by $\T^n\times \R^{4-n}$
for some $n$.)  In classical Yang-Mills theory, such a configuration
is an instanton, that is a solution of $F^+=0$.  (For any two-form on
$\R^4$ such as the Yang-Mills curvature $F$, we write $F^+$ and $F^-$
for the self-dual and anti-self-dual projections.)  So the objects we
want are a stringy generalization of instantons.  {\it A priori} one
would expect that classical instantons would be a good approximation
to stringy instantons only when the instanton scale size is very large
compared to $\sqrt{\alpha'}$.  However, we will now argue that with a
suitable regularization of the worldsheet theory, the classical or
field theory instanton equation is exact if $B=0$.  This implies that
with any regularization, the stringy and field theory instanton
moduli spaces are the same. The argument, which is similar to an argument
about sigma models with K3 target 
\ref\bankseiberg{T. Banks and N. Seiberg, ``Nonperturbative
Infinities,'' Nucl. Phys. {\bf B273} (1986) 157.}, 
also suggests that for $B\not= 0$, the classical instanton equations
and moduli space are {\it not} exact.  We have given some arguments
for this assertion in the introduction, and will give more arguments
below and in the rest of the paper.

At $B=0$, the free worldsheet theory in bulk
\eqn\ugug{S={1\over 4\pi \alpha'}\int_\Sigma\left(g_{ij}\partial_ax^i
\partial^ax^i+ig_{ij}\psi^i\bar\partial\psi^j+
ig_{ij}\bar\psi^i\partial\bar\psi^j \right)}
actually has a $(4,4)$ worldsheet supersymmetry.  This is a
consequence of the $\N=1$ worldsheet supersymmetry described in
\susytran\ plus an $R$ symmetry group.  In fact, we have a
symmetry group $SO(4)_L$ acting on the $\psi^i$ and another $SO(4)_R$
acting on $\bar\psi^i$.  We can decompose $SO(4)_L =SU(2)_{L,+}\times
SU(2)_{L,-}$, and likewise $SO(4)_R=SU(2)_{R,+}\times
\times SU(2)_{R,-}$.  $SU(2)_{R,+}$, together with the $\N=1$
supersymmetry in \susytran, generates an $\N=4$ supersymmetry of the
right-movers, and $SU(2)_{L,+}$, together with \susytran, likewise
generates an $\N=4$ supersymmetry of left-movers.  So altogether in
bulk we get an $\N=(4,4)$ free superconformal model.  Of course, we
could replace $SU(2)_{R,+}$ by $SU(2)_{R,-}$ or $SU(2)_{L,+}$ by
$SU(2)_{L,-}$, so altogether the free theory has (at least) four
$\N=(4,4)$ superconformal symmetries.  But for the instanton problem,
we will want to focus on just one of these extended superconformal
algebras.

Now consider the case that $\Sigma$ has a boundary, but with $B=0$ and
no gauge fields coupled to the boundary.  The boundary conditions on
the fermions are, from \bouconf, $\psi^j=\bar\psi^j$.  This breaks
$SO(4)_L\times SO(4)_R$ down to a diagonal subgroup
$SO(4)_D=SU(2)_{D,+}\times SU(2)_{D,-}$ (here $SU(2)_{D,+}$ is a
diagonal subgroup of $SU(2)_{L,+}\times SU(2)_{R,+}$, and likewise for
$SU(2)_{D,-}$).  We can define an $\N=4$ superconformal algebra in
which the $R$-symmetry is $SU(2)_{D,+}$ (and another one with
$R$-symmetry $SU(2)_{D,-}$).  As is usual for open superstrings, the
currents of this $\N=4$ algebra are mixtures of left and right
currents from the underlying $\N=(4,4)$ symmetry in bulk.

Now let us include a boundary interaction as in \gaugecous:
\eqn\tumulo{L_A=-i\int d\tau\left(A_i(x)\partial_\tau x^i-
iF_{ij}\Psi^i\Psi^j \right).}
The condition that the boundary interaction preserves some spacetime
supersymmetry is that the theory with this interaction is still an
$\N=4$ theory.  This condition is easy to implement, at the classical
level.  The $\Psi^i$ transform as $(1/2,1/2)$ under $SU(2)_{D,+}\times
SU(2)_{D,-}$.  The $F_{ij}\Psi^i\Psi^j$ coupling in $L_A$ transforms
as the antisymmetric tensor product of this representation with
itself, or $(1,0)\oplus (0,1)$, where the two pieces multiply,
respectively, $F^+$ and $F^-$, the self-dual and anti-self-dual
parts of $F$.  Hence, the condition that $L_A$ be invariant under
$SU(2)_{D,+}$ is that $F^+=0$, in other words that the gauge field
should be an instanton.  For invariance under $SU(2)_{D,-}$ we need
$F^-=0$, an anti-instanton.  Thus, at the classical level, an
instanton or anti-instanton gives an $\N=4$ superconformal
theory,\foot{Our notation is not well adapted to nonabelian gauge
theory.  In this case, the factor $e^{-L_A}$ in the path integral must
be reinterpreted as a trace $\Tr\,P\exp\oint_{\partial\Sigma}
\left(iA_i\partial_\tau x^i+F_{ij}\Psi^i\Psi^j\right)$ where the exponent
is Lie algebra valued.  This preserves $SU(2)_{D,\pm}$ if
$F^{\pm}=0$.} and hence a supersymmetric or BPS configuration.

To show that this conclusion is valid quantum mechanically, we need 
a regularization that preserves (global) $\N=1$ supersymmetry and
also the $SO(4)_D$ symmetry.  This can readily be provided by
Pauli-Villars regularization.  First of all, the fields  $x^i,\psi^i,
\bar\psi^i$, together with auxiliary fields $F^i$, can be interpreted
in the standard way as components of $\N=1$ superfields $\Phi^i$, 
$i=1,\dots,4$.

To carry out Pauli-Villars regularization, we introduce two sets of
superfields $C^i$ and $E^i$, where $E^i$ are real-valued and $C^i$
takes values in the same space (${\bf R}^4$ or more generally ${\bf
T}^n\times {\bf R}^{4-n}$) that $\Phi^i$ does, and we write
$\Phi^i=C^i-E^i$.  For $C^i$ and $E^i$, we consider the following
Lagrangian:
\eqn\tuggo{L=\int d^2xd^2\theta\left(\epsilon^{\alpha\beta}
D_\alpha C^iD_\beta C^i\right)-\int
d^2xd^2\theta\left(\epsilon^{\alpha
\beta}D_\alpha E^iD_\beta E^i+M^2(E^i)^2\right).}
This regularization of the bulk theory is manifestly invariant under
global ${\cal N}=1$ supersymmetry.  But since it preserves an
$SO(4)_D$ (which under which all left and right fermions in $C$ or $E$
transform as $(1/2,1/2)$), it actually preserves a global $\N=4$
supersymmetry.

This symmetry can be preserved in the presence of boundaries.  We
simply consider free boundary conditions for both $C^i$ and $E^i$.
The usual short distance singularity is absent in the $\Phi^i$
propagator (as it cancels between $C^i$ and $E^i$).  Now, include a
boundary coupling to gauge fields by the obvious superspace version of
\gaugecous:
\eqn\hunny{L_A=-i\int d\tau d\theta \,A_i(\Phi)D\Phi^i=
-i\int d\tau\left(A_i(x)\partial_\tau x^i-iF_{ij}\Psi^i\Psi^j
\right).}
Classically (as is clear from the second form, which arises upon doing
the $\theta$ integral), this coupling preserves $SU(2)_{D,+}$ if
$F^+=0$, or $SU(2)_{D,-}$ if $F^-=0$.  Because of the absence of a
short distance singularity in the $\Phi$ propagator, all Feynman
diagrams are regularized.\foot{In most applications, Pauli-Villars
regularization fails to regularize the one-loop diagrams, because it
makes the vertices worse while making the propagators better.  The
present problem has the unusual feature that Pauli-Villars
regularization eliminates the short distance problems even from the
one-loop diagrams.}  Hence, for every classical instanton, we get a
two-dimensional quantum field theory with global $\N=4$ supersymmetry.

If this theory flows in the infrared to a conformal field theory, this
theory is $\N=4$ superconformal and hence describes a configuration
with spacetime supersymmetry.  On the other hand, the global $\N=4$
supersymmetry, which holds precisely if $F^+=0$, means that any
renormalization group flow that occurs as $M\to\infty$ would be a flow
on classical instanton moduli space.  Such a flow would mean that
stringy corrections generate a potential on instanton moduli space.
But there is too much supersymmetry for this, and therefore there is
no flow on the space; i.e. different classical instantons lead to
distinct conformal field theories.  We conclude that, with this
regularization, every classical instanton corresponds in a natural way
to a supersymmetric configuration in string theory or in other words
to a stringy instanton.  Thus, with this regularization, the stringy
instanton equation is just $F^+=0$.  Since the moduli space of
conformal field theories is independent of the regularization, it also
follows that with any regularization, the stringy instanton moduli
space coincides with the classical one.

\bigskip\noindent{\it Turning On $B$}

Now, let us reexamine this issue in the presence of a constant $B$
field.  The boundary condition required by supersymmetry was given in
\bouconf:
\eqn\kidni{(g_{ij}+2\pi \alpha' B_{ij})\psi^j=(g_{ij}-2\pi \alpha
B_{ij}) \bar\psi^j.}

To preserve \kidni, if one rotates $\psi^i$ by an $SO(4)$ matrix $h$,
one must rotate $\bar\psi^i$ with a {\it different} $SO(4)$ matrix
$\bar h$.  The details of the relation between $h$ and $\bar h$ will
be explored below, in the context of point-splitting regularization.
At any rate, \kidni\ does preserve a diagonal subgroup $SO(4)_{D,B}$
of $SO(4)_L\times SO(4)_R$, but as the notation suggests, which
diagonal subgroup it is depends on $B$.

The Pauli-Villars regularization introduced above preserves $SO(4)_D$,
which for $B\not= 0$ does not coincide with $SO(4)_{D,B}$.  The
problem arises because the left and right chiral fermions in the
regulator superfields $E^i$ are coupled by the mass term in a way that
breaks $SO(4)_L\times SO(4)_R$ down to $SO(4)_D$, but they are coupled
by the boundary condition in a way that breaks $SO(4)_L\times SO(4)_R$
down to $SO(4)_{D,B}$.  Thus, the argument that showed that classical
instanton moduli space is exact for $B=0$ fails for $B\not= 0$.

This discussion raises the question of whether a different
regularization would enable us to prove the exactness of classical
instantons for $B\not= 0$.  However, a very simple argument mentioned
in the introduction shows that one must expect stringy corrections to
instanton moduli space when $B\not= 0$.  In fact, if $B^+\not= 0$, a
configuration containing a threebrane and a separated $-1$-brane is
not BPS (we will explore it in section 5), so the small instanton
singularity that is familiar from classical Yang-Mills theory should
be absent when $B^+\not= 0$.

It has been proposed \refs{\nekrasov,\berkooz} that the stringy
instantons at $B^+\not= 0$ are the instantons of noncommutative
Yang-Mills theory, that is the solutions of $\hat F^+=0$ with a
suitable $*$ product.  We can now make this precise in the $\alpha'\to
0$ limit.  In this limit, the effective action is, as we have seen,
$\hat F^2$, with the indices in $\hat F$ contracted by the open string
metric $G$.  In this theory, the condition for a gauge field to leave
unbroken half of the linearly realized supersymmetry on the branes is
$\hat F^+=0$, where the projection of $\hat F$ to selfdual and
antiselfdual parts is made with respect to the open string metric $G$,
rather than the closed string metric $g$.  Hence, at least in the
$\alpha'=0$ limit, BPS configurations are described by noncommutative
instantons, as has been suggested in \refs{\nekrasov,\berkooz}.  If we
are on $\R^4$, then, as shown in \nekrasov, deforming the classical
instanton equation $\F^+=0$ to the noncommutative instanton equation
$\hat F^+=0$ has the effect of adding a Fayet-Iliopoulos (FI) constant
term to the ADHM equations, removing the small instanton
singularity\foot{Actually, it was assumed in \nekrasov\ that $\theta$
is self-dual.  The general situation, as we will show at the end of
section 5, is that the small instanton singularity is removed
precisely if $B^+\not= 0$, or equivalently $\theta^+\not= 0$.}.  The
ADHM equations with the FI term have a natural interpretation in terms
of the DLCQ description of the six-dimensional $(2,0)$ theory
\abs, and have been studied mathematically in \nakajima.

What happens if $B\not= 0$ but we do not take the $\alpha'\to 0$
limit?  In this case, the stringy instanton moduli space must be a
hyper-Kahler deformation of the classical instanton moduli space, with
the small instanton singularities eliminated if $B^+\not= 0$, and
reducing to the classical instanton moduli space for instantons of
large scale size if we are on $\R^4$.  We expect that the most general
hyper-Kahler manifold meeting these conditions is the moduli space of
noncommutative instantons, with some $\theta$ parameter and with some
effective metric on spacetime $G$.\foot{The effective metric on spacetime
must be hyper-Kahler for supersymmetry, so it is a flat metric if we
are on $\R^4$ or $\T^n\times \R^{4-n}$, or a hyper-Kahler metric if we
are bold enough to extrapolate the discussion to a K3 manifold or a
Taub-NUT or ALE space.}

\bigskip\noindent{\it Details For Instanton Number One}

Though we do not know how to prove this in general, one can readily
prove it by hand for the case of instantons of instanton number one on
$\R^4$.  The ADHM construction for such instantons, with gauge group
$U(N)$, expresses the moduli space as the moduli space of vacua of a
$U(1)$ gauge theory with $N$ hypermultiplets $H^a$ of unit charge
(times a copy of $\R^4$ for the instanton position).  In the
$\alpha'\to 0 $ limit with non-zero $B$, there is a FI term.  If we
write the hypermultiplets $H^a$, in a notation that makes manifest
only half the supersymmetry, as a pair of chiral superfields
$A^a,B_a$, with respective charges $1,-1$, then the ADHM equations
read
\eqn\nuffy{\eqalign{\sum_a A^aB_a & = \zeta_c.\cr
                    \sum_a|A^a|^2-\sum_a|B_a|^2& = \zeta.\cr}}
One must divide by $A^a\to e^{i\alpha}A^a$, $B_a\to e^{-i\alpha}B_a$.
Here $\zeta_c$ is a complex constant, and $\zeta$ a real constant.
$\zeta_c$ and $\zeta$ are the FI parameters.  The real and imaginary
part of $\zeta_c$, together with $\zeta$, transform as a triplet of an
$SU(2)$ $R$-symmetry group, which is broken to $U(1)$ (rotations of
$\zeta_c$) by our choice of writing the equations in terms of chiral
superfields.  To determine the topology of the moduli space ${\cal
M}$, we make an $SU(2)_R$ transformation (or a judicious choice of
$A^a$ and $B_a$) to set $\zeta_c=0$ and $\zeta>0$.  Then, if we set
$B_a=0$, the $A^a$, modulo the action of $U(1)$, determine a point in
${\bf CP}^N$; the equation $\sum_aA^aB_a=0$ means that the $B_a$
determine a cotangent vector of ${\bf CP}^N$, so ${\cal M}$ is the
cotangent bundle $T^*{\bf CP}^N$.

The second homology group of ${\cal M}$ is of rank one, being
generated by a two-cycle in ${\bf CP}^N$.  Moduli space of
hyper-Kahler metrics is parametrized by the periods of the three
covariantly constant two-forms $I,J,K$.  As there is only one period,
there are precisely three real moduli, namely $\zeta, \,\,{\rm
Re}\,\zeta_c,$ and ${\rm Im}\,\zeta_c$.

Hence, at least for instanton number one, the stringy instanton moduli
space on $\R^4$, for any $B$, must be given by the solutions of $\hat
F^+=0$, with some effective metric on spacetime and some effective
theta parameter.  It is tempting to believe that these may be the
metric and theta parameter found in \gthed\ from the open string
propagator.

\bigskip\noindent
{\it Noncommutative Instantons And ${\cal N}=4$ Supersymmetry}

We now return to the question of what symmetries are preserved by the
boundary condition \kidni.  We work in the $\alpha'\to 0$ limit, so
that we know the boundary couplings and the gauge invariances
precisely.  The goal is to show, by analogy with what happened for
$B=0$, that noncommutative gauge fields that are self-dual with
respect to the open string metric lead to $\N=4$ worldsheet
superconformal symmetry.

It is convenient to introduce a vierbein $e^i_a$ for the closed string
metric  Thus $g^{-1}=e e^t$ ($e^t$ is the transpose of $e$) or
$g^{ij}=\sum_ae^i_ae_a^j$.  Then, we express the fermions in terms of
the local Lorentz frame in spacetime
\eqn\jugbo{\psi^i=e^i_a\chi^a, ~~\bar \psi^i=
e^i_a\bar\chi^a.}  
The $SO(4)_L\times SO(4)_R$ automorphism group of the supersymmetry
algebra rotates these four fermions by $\chi\rightarrow h\chi$ and
$\bar \chi \rightarrow \bar h \bar \chi$.  The boundary conditions
\kidni\ breaks $SO(4)_L\times SO(4)_R$ to a diagonal subgroup
$SO(4)_{D,B}$ defined by
\eqn\oneso{\bar h = e^{-1}{1\over g-2\pi\alpha'B} (g +2\pi\alpha' B)e
h e^{-1} {1\over g+2\pi\alpha'B} (g -2\pi\alpha' B) e.} 

In terms of $\chi$ and $\bar\chi$, the boundary coupling of the
original fermions \gaugecousn\ becomes
\eqn\bouncoupl{\chi^t e^t g {1 \over g+2\pi\alpha'B}\hat F {1
\over g-2\pi\alpha'B}g e \chi.}
We have used \psiide\ to express $\Psi$ in terms of $\psi$, and
\jugbo\ to express $\psi$ in terms of $\chi$.  Under $SO(4)_{D,B}$, 
this coupling transforms as
\eqn\bountrans{\chi^t e^t g {1 \over g+2\pi\alpha'B}\hat F {1
\over g-2\pi\alpha'B}g e \chi \rightarrow \chi^t h^t e^t g {1 \over
g+2\pi\alpha'B}\hat F {1 \over g-2\pi\alpha'B}g e h \chi ,}
and the theory is invariant under the subgroup of $SO(4)_{D,B}$ for which 
\eqn\bountranssu{e^t g {1 \over g+2\pi\alpha'B}\hat F {1 \over
g-2\pi\alpha'B}g e =  h^t e^t g {1 \over 
g+2\pi\alpha'B}\hat F {1 \over g-2\pi\alpha'B}g e h.}
In order to analyze the consequences of this equation, we define
a vierbein for the open string metric by the following very
convenient formula:
\eqn\vierde{E=  {1 \over g-2\pi\alpha'B}g e.}
To verify that this is a vierbein, we compute
\eqn\newviermet{E E^t = {1 \over g-2\pi\alpha'B}g {1 \over
g+2\pi\alpha'B}= {1 \over g+2\pi\alpha'B}g {1 \over
g-2\pi\alpha'B}=G^{-1}.} 
In terms of $E$, \bountranssu\ reads
\eqn\bountranssu{E^t \hat F E =  h^t E^t \hat F E h.}
For this equation to hold for $h$ in an $SU(2)$ subgroup of
$SO(4)_{D,B}$, $E^t\hat FE^t$ must be selfdual, or anti-selfdual, with
respect to the trivial metric of the local Lorentz frame.  This is
equivalent to $\hat F$ being selfdual or anti-selfdual with respect to
the open string metric $G$.  Thus, we have shown that the boundary
interaction preserves an $SU(2)$ $R$ symmetry, and hence an ${\cal
N}=4$ superconformal symmetry, if $\hat F^+=0$ or $\hat F^-=0$ with
respect to the open string metric.

\newsec{Noncommutative Gauge Symmetry vs. Ordinary Gauge Symmetry}

We have by now seen that ordinary and noncommutative Yang-Mills fields
arise from the same two-dimensional field theory regularized in
different ways.  Consequently, there must be a transformation from
ordinary to noncommutative Yang-Mills fields that maps the standard
Yang-Mills gauge invariance to the gauge invariance of noncommutative
Yang-Mills theory.  Moreover, this transformation must be local in the
sense that to any finite order in perturbation theory (in $\theta$)
the noncommutative gauge fields and gauge parameters are given by
local differential expressions in the ordinary fields and 
parameters.

At first sight, it seems that we want a local field redefinition $\hat
A=\hat A(A,\partial A,\partial^2A,\dots;\theta)$ of the gauge fields,
and a simultaneous reparametrization $\hat
\lambda=\hat\lambda(\lambda,\partial \lambda,\partial^2\lambda,\dots;
\theta)$ of the gauge parameters that maps one gauge invariance to the
other.  However, this must be relaxed.  If there were such a
map intertwining with the gauge
invariances, it would follow that the gauge group of ordinary
Yang-Mills theory is isomorphic to the gauge group of noncommutative
Yang-Mills theory.  This is not the case.  For example, for rank one,
the ordinary gauge group, which acts by
\eqn\juggo{\delta A_i=\partial_i\lambda,}
is Abelian, while the noncommutative gauge invariance, which acts by
\eqn\floog{\delta A_i=\partial_i\lambda+i\lambda *A_i-iA_i*\lambda,}
is nonabelian.  An Abelian group cannot be isomorphic to a nonabelian
group, so no redefinition of the gauge parameter can map the ordinary
gauge parameter to the noncommutative one while intertwining with the
gauge symmetries.

What we actually need is less than an identification between the two
gauge groups.  To do physics with gauge fields, we only need to know
when two gauge fields $A$ and $A'$ should be considered
gauge-equivalent.  We do not need to select a particular set of
generators of the gauge equivalence relation -- a gauge group that
generates the equivalence relation\foot{Fadde'ev-Popov quantization
of gauge theories is formulated in terms of the gauge group, but in
the more general Batalin-Vilkovisky approach to quantization, the
emphasis is on the equivalence relation generated by the gauge
transformations.  For a review of this approach, see
\ref\henneaux{M. Henneaux, ``Lectures On The Antifield-BRST Formalism
For Gauge Theories,'' Nucl. Phys. {\bf B} (Proc. Suppl.)  {\bf 18A}
(1990) 47; M. Henneaux and C. Teitelboim, {\it Quantization Of Gauge
Systems} (Princeton University Press, 1992).}.}.
In the problem at hand, it turns out that we can map $A$ to $\hat A$
in a way that preserves the gauge equivalence relation, even though
the two gauge groups are different.

What this means in practice is as follows.  We will find a mapping
{}from ordinary gauge fields $A$ to noncommutative gauge fields $\hat
A$ which is local to any finite order in $\theta$ and has the
following further property.  Suppose that two ordinary gauge fields
$A$ and $A'$ are equivalent by an ordinary gauge transformation by
$U=\exp(i\lambda)$.  Then, the corresponding noncommutative gauge
fields $\hat A$ and $\hat A'$ will also be gauge-equivalent, by a
noncommutative gauge transformation by $\hat U=\exp(i\hat\lambda)$.
However, $\hat\lambda$ will depend on both $\lambda $ and $A$.  If
$\hat\lambda$ were a function of $\lambda$ only, the ordinary and
noncommutative gauge groups would be the same; since $\hat\lambda$ is
a function of $A$ as well as $\lambda$, we do not get any well-defined
mapping between the gauge groups, and we get an identification only of
the gauge equivalence relations.

Note that the situation that we are considering here is the opposite
of a gauge theory in which the gauge group has field-dependent
structure constants or only closes on shell.  This means (see
\henneaux\ for a fuller explanation) that one has a well-defined gauge
equivalence relation, but the equivalence classes are not the orbits
of any useful group, or are such orbits only on shell.  In the
situation that we are considering, there is more than one group that
generates the gauge equivalence relation; one can use either the
ordinary gauge group or (with one's favorite choice of $\theta$) the
gauge group of noncommutative Yang-Mills theory.

Finally, we point out in advance a limitation of the discussion.  The
arguments in section 2 (which involved, for example, comparing two
different ways of constructing an $\alpha'$ expansion of the string
theory effective action) show only that ordinary and noncommutative
Yang-Mills theory must be equivalent to all finite orders in a long
wavelength expansion.  By dimensional analysis, this means that they
must be equivalent to all finite orders in $\theta$.  However, it is
not clear that the transformation between $A$ and $\hat A$ should
always work nonperturbatively.  Indeed, the small instanton problem
discussed in section 2.3 seems to give a situation in which the
transformation between $\hat A$ and $A$ breaks down, presumably
because the perturbative series that we will construct does not
converge.

\subsec{\it The Change Of Variables}

Once one is convinced that a transformation of the type described
above exists, it is not too hard to find it.  We take the gauge fields
to be of arbitrary rank $N$, so that all fields and gauge parameters
are $N\times N$ matrices (with entries in the ordinary ring of
functions or the noncommutative algebra defined by the $*$ product of
functions, as the case may be).  We look for a mapping $\hat A(A)$ and
$\hat \lambda(\lambda,A)$ such that
\eqn\redcondi{\hat A(A)+ \hat \delta_{\hat \lambda} \hat A(A)= \hat
A(A+\delta_\lambda A),}
with infinitesimal $\lambda$ and $\hat \lambda$.  This will ensure
that an ordinary gauge transformation of $A$ by $\lambda$ is
equivalent to a noncommutative gauge transformation of $\hat A$ by
$\hat\lambda$, so that ordinary gauge fields that are gauge-equivalent
are mapped to noncommutative gauge fields that are likewise
gauge-equivalent.  The gauge transformation laws $\delta_\lambda$ and
$\hat\delta_{\hat\lambda}$ were defined at the end of the
introduction.  We first work to first order in $\theta$.  We write
$\hat A= A+A'(A)$ and $\hat \lambda(\lambda, A)= \lambda +
\lambda'(\lambda,A) $, with $A'$ and $\lambda'$ local function of
$\lambda$ and $A$ of order $\theta$.  Expanding \redcondi\ in powers
of $\theta$, we find that we need
\eqn\leadthecon{A'_i (A+\delta_\lambda A) -A'_i(A) -\partial_i \lambda' -
i[\lambda',A_i]-i[\lambda, A'_i]= -{1\over 2}\theta^{kl} (\partial_k
\lambda \partial_l 
A_i + \partial_l A_i \partial_k \lambda) +\CO(\theta^2) .} 
In arriving at this formula, we have used the expansion $f*g=fg+\half
i \theta^{ij}\partial_if\partial_jg+{\cal O}(\theta^2)$, and have
written the ${\cal O}(\theta)$ part of the $*$ product explicitly on
the right hand side.  All products in \leadthecon\ are therefore
ordinary matrix products, for example
$[\lambda',A_i]=\lambda'A_i-A_i\lambda'$, where (as $\lambda'$ is of
order $\theta$), the multiplication on the right hand side should be
interpreted as ordinary matrix multiplication at $\theta=0$.

Equation \leadthecon\ is solved by
\eqn\firstvar{\eqalign{
&\hat A_i(A)=A_i+A_i'(A) = A_i -{1 \over 4}\theta^{kl} \{A_k,
\partial_l A_i +F_{li}\} +\CO(\theta^2) \cr  
&\hat \lambda(\lambda, A)=\lambda +\lambda'(\lambda,A) = \lambda +{1
\over 4} \theta^{ij} \{\partial_i \lambda, A_j\}+\CO(\theta^2) \cr}}
where again the products on the right hand side, such as
$\{A_k,\partial_l A_i\}=A_k\cdot \partial_lA_i+\partial_lA_i \cdot
A_k$ are ordinary matrix products.  From the formula for $\hat A$, it
follows that
\eqn\changefi{\hat F_{ij} = F_{ij} +{1\over 4}\theta^{kl} \left( 2
\{F_{ik}, F_{j l} \} -\{A_k, D_l F_{ij} +\partial_l
F_{ij}\}\right)+\CO(\theta^2) .} 
These formulas exhibit the desired change of variables to first
nontrivial order in $\theta$.  

By reinterpreting the above formulas, it is a rather short step to
write down a differential equation that generates the desired change
of variables to all finite orders in $\theta$.  Consider the problem
of mapping noncommutative gauge fields $\hat A(\theta)$ defined with
respect to the $*$ product with one choice of $\theta$, to
noncommutative gauge fields $\hat A(\theta + \delta \theta)$, defined
for a nearby choice of $\theta$.  To first order in $\delta\theta$,
the problem of converting from $\hat A(\theta)$ to $\hat
A(\theta+\delta \theta)$ is equivalent to what we have just solved.
Indeed, apart from associativity, the only property of the $*$ product
that one needs to verify that \firstvar\ obeys \redcondi\ to first
order in $\theta$ is that for any variation $\delta \theta^{ij}$ of
$\theta$,
\eqn\nifgo{\delta \theta^{ij}{\partial\over \partial\theta^{ij}}
\left(f*g\right)={i \over 2}\delta\theta^{ij}{\partial f\over \partial
x^i}* {\partial g\over \partial x^j}}
at $\theta=0$.  But this is true for any value of $\theta$, as one can
verify with a short perusal of the explicit formula for the $*$
product in \defprod.  Hence, adapting the above formulas, we can write
down a differential equation that describes how $\hat A(\theta)$ and
$\hat\lambda(\theta)$ should change when $\theta$ is varied, to
describe equivalent physics:
\eqn\firstvaranya{\eqalign{
&\delta \hat A_i(\theta ) =
\delta\theta^{kl}{\partial\over\partial\theta^{kl}} 
\hat A_i(\theta )=  -{1 \over 4}\delta\theta^{kl} \big[\hat A_k*
(\partial_l \hat A_i + \hat F_{li}) + (\partial_l \hat A_i +\hat
F_{li})* \hat A_k \big]  \cr 
&\delta\hat\lambda(\theta)=\delta\theta^{kl}{\partial\over
\partial\theta^{kl}}\hat\lambda(\theta)= 
{1\over 4}\delta \theta^{kl}\left(\partial_k\lambda* A_l+
A_l*\partial_k\lambda\right) \cr
&\delta \hat F_{ij}(\theta) =  \delta
\theta^{kl}{\partial\over\partial\theta^{kl}} \hat F_{ij}(\theta) =
{1\over 4}\delta \theta^{kl}  \biggl[ 2 \hat F_{ik}* \hat F_{j l} + 2
\hat F_{j l} *\hat F_{ik}  - \hat A_k * \left(\hat  D_l \hat F_{ij}
+\partial_l \hat F_{ij}\right)\biggr. \cr 
&\qquad\qquad\qquad\qquad\qquad\qquad - \biggl.\left(\hat  D_l \hat
F_{ij} +\partial_l \hat F_{ij}\right)* \hat A_k \biggr].} }
On the right hand side, the $*$ product is meant in the generalized
sense explained in the introduction: the tensor product of matrix
multiplication with the $*$ product of functions.  This differential
equation generates the promised change of variables to all finite
orders in $\theta$. To what extent the series in $\theta$ generates by
this equation converges is a more delicate question, beyond the scope
of the present paper.  The equation is invariant under a scaling
operation in which $\theta$ has degree $-2$ and $A$ and
$\partial/\partial x$ have degree one, so one can view the expansion
it generates as an expansion in powers of $\theta$ for any $A$, which
is how we have derived it, or as an expansion in powers of $A$ and
$\partial/\partial x$ for any $\theta$.

The differential equation \firstvaranya\ can be solved explicitly for
the important case of a rank one gauge field with constant $\hat F$.
In this case, the equation can be written
\eqn\rankonedel{\delta \hat F= -\hat F \delta \theta\hat F}
(the Lorentz indices are contracted as in matrix multiplication).  Its
solution with the boundary condition $\hat F(\theta=0)=F$ is
\eqn\rankonesol{\hat F={1 \over 1+ F\theta}F.}
{}From \rankonesol\ we find $F$ in terms of $\hat F$
\eqn\rankonesolo{F=\hat F {1 \over 1 -\theta \hat F}.}
We can also write these relations as
\eqn\anotherw{\hat F-{1\over \theta}=-{1 \over \theta({1 \over
\theta}+F)\theta}.} 
We see that when $F=-\theta^{-1}$ we cannot use the noncommutative
description because $\hat F$ has a pole.  Conversely,
$F$ is singular when $\hat
F=\theta^{-1}$, so in that case, the commutative description does not exist.
Using our identification in the zero slope limit (or in a natural
regularization scheme which will be discussed below) $\theta={1 \over
B}$, equations \rankonesol, \rankonesolo\ and \anotherw\ become
\eqn\rankonesolb{\hat F=B{1 \over B+F}F,}
\eqn\rankonesolbo{F=\hat F{1 \over B- \hat F}B}
and
\eqn\rankonesolba{\hat F-B=-B{1 \over B+F}B.}
So an ordinary Abelian gauge field with constant curvature $F$ and
Neveu-Schwarz two-form field $B$ is equivalent to a noncommutative
gauge field with $\theta=1/B$ and the value of $\hat F$ as in
\rankonesolb.  When $B+F=0$ we cannot use the noncommutative
description.  It is natural that this criterion depends only on $B+F$,
since in the description by ordinary Abelian gauge theory, $B$ and $F$
are mixed by a gauge symmetry, with only the combination $B+F$ being
gauge-invariant.

\bigskip\noindent{\it Application To Instantons}

Another interesting application is to instantons in four dimensions.
We have argued in section 2.3 (following \refs{\nekrasov,\berkooz})
that a stringy instanton is a solution of the noncommutative instanton
equation
\eqn\insteq{\hat F^+_{ij}=0.}
We can evaluate this equation to first nontrivial order in $\theta$
using \changefi.  Since
$\theta^{kl}\{A_k,D_lF_{ij}+\partial_lF_{ij}\}^+=0$ if $F_{ij}^+=0$,
to evaluate the ${\cal O}(\theta)$ deviation of \insteq\ from the
classical instanton equation $F_{ij}^+=0$, we can drop those
non-gauge-invariant terms in \changefi.  We find that to first order
in $\theta$, the noncommutative instanton equation can be written in
any of the following equivalent forms:
\eqn\modins{\eqalign{
0=&F^+_{ij} + {1\over 2}\left(\theta^{kl} \{F_{ik}, F_{jl}\}
\right)^+ +\CO(\theta^2) \cr
=& F^+_{ij}-{1 \over 8}{1 \over \sqrt {\det G}} \epsilon^{rstu}F_{rs}
F_{tu} G_{ik} G_{jl}(\theta^+)^{kl} +\CO(\theta^2)\cr =&F^+_{ij}-{1
\over 4} (F\tilde F) \theta^+_{ij} +\CO(\theta^2).}}  
Here $G$ is the open string metric, which is used to determine the
self-dual parts of $F$ and $\theta$.  In \modins, we used the facts
that $F^-=\CO(1)$ and $F^+=\CO(\theta)$, along with various identities
of $SO(4)$ group theory.  For example, in evaluating
$(\theta^{kl}\{F_{ik},F_{jl}\})^+$ to order $\theta$, one can replace
$F$ by $F^-$. According to $SO(4)$ group theory, a product of any
number of anti-selfdual tensors can never make a selfdual tensor, so
we can likewise replace $\theta$ by $\theta^+$.  $SO(4)$ group theory
also implies that there is only one self-dual tensor linear in
$\theta^+$ and quadratic in $F^-$, namely $\theta^+(F^-)^2$, so the
${\cal O}(\theta)$ term in the equation is a multiple of this.  To
first order in $\theta$, we can replace $(F^-)^2$ by
$(F^-)^2-(F^+)^2$, which is a multiple of $F\tilde F={1\over
2\sqrt{\det G}}\epsilon^{rstu}F_{rs}F_{tu}$; this accounts for the
other ways of writing the equation given in \modins.

In \modins, we see that to first order, the corrections to the
instanton equation depend only on $\theta^+$ and not $\theta^-$; in
section 5, we explore the extent to which this is true to all orders.

\bigskip\noindent{\it More Freedom In The Description}

What we have learned is considerably more than was needed to account
for the results of section 2.  In section 2, we found that, using
a point-splitting regularization,  string theory with given closed
string parameters $g$ and $B$ can be described, in the open string
sector, by a noncommutative Yang-Mills theory with $\theta$ given in eqn.
\gthed.  There must, therefore, exist a transformation from commutative
Yang-Mills to noncommutative Yang-Mills with that value of $\theta$.

In our present discussion, however, we have obtained a mapping
{}from ordinary Yang-Mills to non-commutative Yang-Mills that is
completely independent of $g$ and $B$ and hence allows us to express
the  open string sector in terms of a noncommutative Yang-Mills theory
with an arbitrary value of $\theta$.
It is plausible that this type of description would arise if one
uses a suitable regularization that somehow interpolates between
Pauli-Villars and point-splitting.

\def\hat{\widehat}
What would the resulting description look like?  In the description
by ordinary Yang-Mills fields, the effective action is a function of
$F+B$, and is written using ordinary multiplication of functions.
In the description obtained with point-splitting regularization,
the effective action is a function of $\hat F$, but the multiplication is
the $*$ product with $\theta$ in \gthed.  If one wishes a description
with an arbitrary $\theta$, the variable in the action will have to somehow
interpolate from $F+B$ in the description by ordinary Yang-Mills fields
to $\hat F$ in the description with the canonical value of $\theta$ in \gthed.
The most optimistic hypothesis is that there is some two-form $\Phi$,
which depends on $B$, $g$, and $\theta$, such that the $\theta$ dependence
of the effective action is completely captured by replacing $\hat F$ by
\eqn\hatfphi{\hat F+\Phi,}
using the appropriate $\theta$-dependent $*$ product, and using
an appropriate effective metric $G$ and string coupling $G_s$.

We propose that this is so, with $G$, $G_s$, and $\Phi$ determined
in terms of $g, $ $B$, and $\theta$ by the following formulas, whose
main justification will be given in section 4.1:
\eqn\changevarg{\eqalign{
&{1 \over G+ 2\pi\alpha' \Phi}= -{\theta \over 2\pi \alpha'} +{1 \over
g+ 2\pi\alpha' B}\cr  
&G_s=g_s \left({\det (G+ 2\pi\alpha'\Phi) \over \det (g+ 2\pi\alpha'
B)}\right)^\half = g_s {1 \over \det\left[({1 \over g+
2\pi\alpha' B}-{\theta \over 2\pi \alpha'})  (g+ 2\pi\alpha'
B)\right]^\half}.}}
In the first equation, $G$ and $\Phi$ are determined because they
are symmetric and antisymmetric respectively.  The second equation is
motivated, as in \limpropg, by demanding that for $F=\hat F=0$ the
constant terms in the Lagrangians using the two set of variables are
the same. 

We will show in section 4 that for slowly varying fields -- governed
by the Dirac-Born-Infeld action -- such a general description, depending
on an arbitrary $\theta$, does exist.  The first equation in \changevarg\
has been determined because it is the unique formula compatible with
the analysis in section 4.1.  We will also see in section 3.2 that
a special case of the transformation in \changevarg\ has a natural
microscopic explanation in noncommutative Yang-Mills theory.
We do not have a general proof of the existence of a description
with the properties proposed in \hatfphi\ and \changevarg.  Such
a proof might be obtained by finding a regularization that suitably
generalizes point-splitting and Pauli-Villars and leads to these
formulas.

A few special cases of \changevarg\ are particularly interesting:
\item{(1)}  $\theta=0$.  Here we recover the commutative description,
where $G=g$, $G_s=g_s$ and $\Phi=B$.  
\item{(2)}  $\Phi=0$.  This is  the description we studied in section 2. 
\item{(3)}  In the zero slope limit with fixed $G$, $B$ and $\Phi$, we
take $g=\epsilon g^{(0)} +\CO(\epsilon^2)$, $B=B^{(0)} +\epsilon
B^{(1)} +\CO(\epsilon^2)$ and $\alpha' = \CO(\epsilon^\half)$ (we
assume for simplicity that the rank of $B$ is maximal, i.e.\
$r=p+1$).  Expanding the first expression in \changevarg\ in powers of
$\epsilon$ we find
\eqn\expandch{\eqalign{
{1\over G }- 2\pi\alpha' {1 \over G}\Phi {1 \over G} &+
\CO(\epsilon) = -{\theta \over 2\pi \alpha'} +{1\over 2\pi
\alpha' B^{(0)}} -{\epsilon \over (2\pi\alpha')^2 } {1\over B^{(0)}} 
g^{(0)} {1\over B^{(0)}} \cr
&+ {\epsilon^2 \over (2\pi\alpha')^3 } {1\over
B^{(0)}} g^{(0)} {1\over B^{(0)}} g^{(0)} {1\over B^{(0)}}  -
{\epsilon \over 2\pi \alpha'} {1 \over  B^{(0)} }B^{(1)}{1 \over
B^{(0)} }  + \CO(\epsilon) .}}
Equating the different orders in $\epsilon$ we have
\eqn\zeroslog{\eqalign{
&\theta={1 \over B^{(0)}} +\CO(\epsilon) \cr
& G=-{(2\pi\alpha')^2\over\epsilon  } B^{(0)} { 1\over
g^{(0)}}B^{(0)}+\CO(\epsilon)\cr
& \Phi= - B^{(0)}  + { (2\pi \alpha')^2 \over \epsilon}B^{(0)} {
1\over g^{(0)}} B^{(1)}{ 1\over g^{(0)}}B^{(0)}  +\CO(\epsilon) \cr
&G_s=g_s\det\left({2\pi \alpha' \over \epsilon} B^{(0)} {1\over
g^{(0)}}\right)^\half\left(1+ \CO(\epsilon)\right),}}
which agree with our zero slope values \limmet, \limpropgz\ (except
the new value of $\Phi$).  The freedom in our description is the
freedom in the way we take the zero slope limit; i.e.\ in the value of
$B^{(1)}$.  It affects only the value of $\Phi$.  For example, for
$B^{(1)} =0$ we have $\Phi= - B^{(0)}$, and for $B^{(1)}= {\epsilon
\over (2\pi \alpha')^2}g^{(0)}{1\over B^{(0)}} g^{(0)}$ we have
$\Phi=0$, as in the discussion in section 2.  The fact that there is
freedom in the value of $\Phi$ in the zero slope limit has a simple
explanation.  In this limit the effective Lagrangian is proportional
to $\Tr (\hat F+\Phi)^2= \Tr (\hat F^2+2\Phi\hat F +\Phi^2)$.  The
$\Phi$ dependence affects only a total derivative term and a constant
shift of the Lagrangian. Such terms are neglected when the effective
Lagrangian is derived, as in perturbative string theory, from the
equations of motion or the S-matrix elements.
\item{(4)}  We can extend the leading order expressions in the zero
slope limit \zeroslog\ (again in the case of maximal rank $r=p+1$)
with $B^{(1)}=0$ to arbitrary value of $\epsilon$, away from the zero
slope limit, and find 
\eqn\zerosloge{\eqalign{
&\theta={1 \over B} \cr
& G=-(2\pi\alpha')^2 B { 1\over g}B\cr
& \Phi= - B ,}}
which satisfy \changevarg.  With this choice of $\theta$ the string
coupling \changevarg\ keeps its zero slope limit value \limpropgz\  
\eqn\scsf{G_s=g_s\det(2\pi \alpha' Bg^{-1})^\half.}
In the next subsection, we will see that the existence of a description
with these values of the parameters is closely related
to background independence of noncommutative Yang-Mills theory.  
These are also the values for which the pole in $\hat
F(F)$, given in \rankonesol, occurs at $F+B=0$ as in \rankonesolb.

\subsec{Background Independence Of Noncommutative Yang-Mills
On ${\bf R}^n$}

In the language of ordinary Yang-Mills theory, the gauge-invariant
combination of $B$ and $F$ is $M=2\pi\alpha'(B+F)$. (The $2\pi\alpha'$ is
for later convenience.) The same gauge-invariant field $M$
can be split in different ways as $2\pi\alpha'(B+F)$ or $2\pi\alpha'(
B'+F')$ where $B$
and $B'$ are constant two-forms.  Given such a splitting, we
incorporate the background $B$ or $B'$ as a boundary condition in an
exactly soluble conformal field theory, as described in section 2.
Then we treat the rest of $M$ by a boundary interaction.  As we have
seen in section 2 and above, the boundary interaction can be
regularized either by Pauli-Villars, leading to ordinary Yang-Mills
theory, or by point splitting, leading to noncommutative Yang-Mills.

In the present discussion, we will focus on noncommutative Yang-Mills,
and look at the background dependence.  Thus, by taking the background
to be $B$ or $B'$, we should get a noncommutative description with
appropriate $\theta$ or $\theta'$, and different $\widehat F$'s.
Note the contrast with the discussion in sections 2 and 3.1: here we
are sticking with point-splitting regularization, and changing the
background from $B$ to $B'$, while in our previous analysis, we kept
the background fixed at $B$, but changed the regularization.

We make the following remarks:
\item{(1)}  If we are on a torus, a shift in background from $B$ to
$B'$ must be such that the difference $B-B'$ obeys Dirac quantization
(the periods of $B-B'$ are integer multiples of $2\pi$) because the
ordinary gauge fields with curvatures $F$ and $F'$ each obey Dirac
quantization, so their difference $F-F'$ does also.  Such quantized
shifts in $B$ are elements of the $T$-duality group.

\item{(2)}   Even if we are on $\R^n$, there can be at most one value
of $B$ for which the noncommutative curvature vanishes at infinity.
Thus, if we are going to investigate background independence in the
form proposed above, we have to be willing to consider noncommutative
gauge fields whose curvature measured at infinity is constant.
\item{(3)}  This  has a further consequence.  Since the condition for
$\hat F$ to vanish at infinity will not be background independent,
there is no hope for the noncommutative action as we have written it
so far, namely,
\eqn\icovo{{1\over g_{YM}^2}\int d^nx \sqrt G G^{ik}G^{jl}
\Tr \hat F_{ij}*\hat F_{kl}}
to be background independent.  Even the condition that this action
converges will not be background independent.  We will find it
necessary to extend the action to
\eqn\nicovo{\hat L= {1\over g_{YM}^2 }\int d^nx \sqrt G
G^{ik}G^{jl}(\hat F_{ij} -\theta^{-1}_{ij})(\hat
F_{kl}-\theta^{-1}_{kl}),} 
which will be background independent.  The constant we added
corresponds to $\Phi=-\theta^{-1}$ in \hatfphi.  It is easy to see
that with this value of $\Phi$ equation \changevarg\ determines
$\theta=B^{-1}$, $G= -(2\pi\alpha')^2 B g^{-1}B$ as in \zerosloge.  It
is important that even though the expressions for $\theta$ and $G$ are
as in the zero slope limit \limmet, in fact they are exact even away
{}from this limit as they satisfy \changevarg.

Note that these remarks apply to background independence, and not to
behavior under change in regularization.  (Change in regularization is
not particularly restricted by being on a torus, since for instance
\firstvaranya\ makes perfect sense on a torus; leaves fixed the
condition that the curvature vanishes at infinity; and does not leave
fixed any particular Lagrangian.)  Note also that in the description
of open strings by ordinary gauge theory, the symmetry of shift in $B$
(keeping fixed $B+F$) is made at fixed closed string metric $g$, so we
want to understand background independence of noncommutative
Yang-Mills at fixed $g$.

Remark (1) above makes it clear that Morita equivalence must be an
adequate tool for proving background independence, since more
generally \refs{\connes,\schwarz-\piosch}, Morita equivalence is an
effective tool for analyzing $T$-duality of noncommutative Yang-Mills
theory.  We will here consider only background independence for
noncommutative Yang-Mills on $\R^n$, and not surprisingly in this case
the discussion reduces to something very concrete that we can write
down naively without introducing the full machinery of Morita
equivalence.

We will consider first the case that the rank $r$ of $\theta$ equals the
dimension $n$ of the space, so that  $\theta$ is invertible.  (The
generalization is straightforward and is briefly indicated below.)
The gauge fields are described by covariant derivatives
\eqn\humgo{D_i={\partial\over \partial x^i}-iA_i,}
where the $A_i$ are elements of the algebra ${\cal A}$ generated by
the $x_i$ (tensored with $N\times N$ matrices if the gauge field has
rank $N$).  We recall that ${\cal A}$ is defined by the relations
$[x^i,x^j]=i\theta^{ij}$.

The $\partial/\partial x^i$ do not commute with the $x$'s that appear
in $A_i$, and this is responsible for the usual complexities of gauge
theory.  The surprising simplification of noncommutative Yang-Mills
theory is that this complexity can be eliminated by a simple change of
variables.  We write
\eqn\tumbo{D_i=\partial_i'-iC_i,}
where 
\eqn\upumbo{\partial_i'={\partial\over\partial x^i}+iB_{ij}x^j}
and
\eqn\ubumbo{C_i=A_i+B_{ij}x^j.}
The point of this is that the $\partial_i'$ commute with the $x^i$.
Hence, the curvature $\widehat F_{ij}=i[D_i,D_j]$ is simply
\eqn\plumbo{\widehat F_{ij}=i[\partial_i',\partial_j']-i[C_i,C_j]}
or more explicitly
\eqn\ugumbo{\widehat F_{ij} =B_{ij} -i[C_i,C_j].}

Now we are almost ready to explain what background independence means.
The $C_i$ are given as functions of the $x^i$, and as such they are
elements of an algebra ${\cal A}$ that depends on $\theta$.  However,
as an abstract algebra, ${\cal A}$ only depends on the rank of
$\theta$, and because no derivatives of $C$ appear in the formula for
the curvature, we can treat the $C_i$ as elements of an abstract
algebra.  For example, we can take any fixed algebra $[y^a,y^b]=i
t^{ab}$, $a,b=1,\dots,n$ with $t^{ab}$ being any
invertible antisymmetric tensor.  Then, picking a ``vierbein''
$f^k_a$, such that $\theta^{jk}=f^j_af^k_bt^{ab}$, we write
the $x^k$ that appear in the argument of $C_i$ as $x^k =f^k_ay^a$,
and regard the $C_i$ as functions of $y^a$.   We make no such
transformation of the $x^i$ that appear in the definition of
$\partial_i'$.  Thus, the covariant derivatives are
\eqn\ikkle{D_i=\partial_i'-iC_i(y^a)={\partial\over \partial x^i}+
iB_{ij}x^j-iC_i(y_a).}
Because $[\partial_i',x^j]=0$, we can make this change of variables
for the ``internal'' $x$'s that appear as arguments of $C_i$, without
touching the $x$'s that appear explicitly on the right hand side of
\ikkle, and without changing the formula for the curvature.  There is
no analog of this manipulation in ordinary Yang-Mills theory.

One is now tempted to define background independence by varying
$\theta^{ij}$, and its inverse $B_{ij}$, while keeping fixed
$C_i(y_a)$.  Then, writing \ugumbo\ in the form $\widehat
F_{ij}=\theta^{-1}_{ij}- i[C_i,C_j]$, we see that under this operation
\eqn\umib{N_{ij}= \widehat F_{ij}-\theta^{-1}_{ij}=-i[C_i,C_j]} 
is invariant.  However, this operation, taken with fixed open string
metric $G$, does not leave fixed the action \nicovo, since $G$ depends
on $\theta$: $G=-(2\pi\alpha')^2Bg^{-1}B=
-(2\pi\alpha')^2\theta^{-1}g^{-1}\theta^{-1}$.  Instead, we want to
vary $\theta$ while keeping fixed the components of $C_i$ in a fixed
local Lorentz frame.  If $e_i^a$ is a vierbein for the closed string
metric $g$ (so $g_{ij}=\sum_a e_i^ae_j^a$), then a vierbein for $G$ is
$E_i^a=2\pi\alpha' B_{ij}e^{j}{}_a$.  We write $C_i=E_i^aC_a$.  Now we
can formulate background independence: it is an operation in which one
varies $\theta$, keeping fixed $g$ and $C_a$.

It is easy to see now that
\eqn\icuv{G^{ik}G^{jl}\Tr\,(\hat F_{ij}-\theta^{-1}_{ij})*(\hat
F_{kl}- \theta^{-1}_{kl})}
is invariant under this operation.  With $G^{il}=-(2\pi\alpha')^{-2}
\theta^{ij}g_{jk}\theta^{kl}$, this follows from \umib\ and the fact
that $\theta^{ij}C_j$ is background independent.  Background
independence of \icuv\ is equivalent to the form of $G^{ij}$ and the
fact that the quantity defined by
$Q^{il}=-i\theta^{ij}[C_j,C_k]\theta^{kl}$, or more simply 
\eqn\ocx{Q=\theta\hat F\theta -\theta,}
is background independent.  In the rank one case, for constant $\hat
F$, we can via \rankonesolba\ express $Q$ in terms of the equivalent
ordinary Abelian gauge field that could be used in an alternative
description of the same physics.  We find simply
\eqn\juniper{Q=-{1\over B+F}.}
This is a satisfying result; it says that in this case the background
independent object $Q$ defined with point-splitting regularization and
noncommutative Yang-Mills theory is a function of the background
independent object $B+F$ found with Pauli-Villars regularization and
ordinary Yang-Mills theory.  It also shows that we should not try to
use the noncommutative description if $B+F=0$ where $Q$ is infinite,
and we should not try to expand around $Q=0$, where $B+F$ is infinite.

For background independence of the action \nicovo, background
independence of \icuv\ is not quite enough.  We need, in addition,
that the measure
\eqn\picuv{{d^nx \sqrt G\over g_{YM}^2 }}
should be background independent.  This will tell how $g_{YM}^2$ must
transform under the change of background.  Since the action density is
most naturally written as a function of the $y$'s, we should convert
the integration measure to an integral over $y$.  From $x^k=f^k_ay^a$
and $f^k_af^l_bt^{ab}=\theta^{kl}$, we get $d^nx = d^ny
\,\det(f) =d^ny\,\sqrt{\det\theta}/\sqrt{\det \,t}.$ We also have
$G=-(2\pi\alpha')^2 \theta^{-1}g\theta^{-1}$ so $\det
G=(2\pi\alpha')^{2n}(\det \theta)^{-2} \det g$.  So the measure is 
\eqn\kiggu{{d^ny(2\pi\alpha')^{n} \sqrt {\det g}\over 
      g_{YM}^2 \sqrt{\det \theta} \sqrt{\det \,t}}.}

So $g_{YM}$ must transform under a change in $\theta$ in such a way
that $g_{YM}^2\sqrt{\det\theta}$ is invariant.  Since $g_{YM}^2 \sim
G_s$ this means that $G_s /\sqrt {\det B}$ is invariant.  This is
clearly the case for the value of $G_s=g_s\det(2\pi \alpha'
Bg^{-1})^\half$ as determined in \scsf.  This means that under a shift
in the $ B$-field,
\eqn\ologo{B\to B'=B+b}
(with $b$ a constant antisymmetric tensor), which induces
\eqn\nologo{\theta\to\theta'= \theta{1\over 1+\theta b},}
we require
\eqn\prologo{g_{YM}\to g_{YM}'=g_{YM}\left(\det(1+\theta
b)\right)^{1\over 4}.}
If we are on a torus, we require $b$ to have periods that are integer
multiples of $2\pi$, and then \ologo\ is a special case of a
$T$-duality transformation.  The transformation \prologo\ is in this
situation a special case of the $T$-duality transformations of
noncommutative Yang-Mills, as analyzed in
\refs{\connes,\schwarz-\piosch}; in section 6 (see
equation (6.14)), we derive this formula from the standard $T$-duality
transformations of closed strings and the mapping from closed string
to open string parameters.

Finally, let us consider the more general case that $\theta$ might
have rank $r<n$.  The algebra then has a center generated by $n-r$
coordinates, which we can call $x^1,\dots, x^{n-r}$.  $\theta$ is only
invertible in the space of $x^{n-r+1},\dots, x^n$.  We let $b_{ij}$ be
a partial inverse of $\theta^{ij}$, with $b_{ij}$ zero unless
$i,j>n-r$, and $b_{ij}\theta^{jk}=\delta_i{}^k$ for $i,r>n-r$.  A
construction just as above, defining $\partial_i'$ and $C_i$ by the
same formulas, gives now invariance under change of background, as
long as one preserves the center of the algebra and the rank of
$\theta$.  But otherwise, one can change $\theta$ as one pleases.

\newsec{Slowly Varying Fields}

The purpose of the present section is to do some explicit calculations
verifying and illustrating our theoretical claims.  We have argued
that the same open string theory effective action can be expressed in
terms of either ordinary Yang-Mills theory or noncommutative
Yang-Mills theory.  In the description by ordinary Yang-Mills, the
$B$-dependence is described by replacing everywhere $F$ by $B+F$.  In
the description by noncommutative Yang-Mills, the $B$-dependence is
entirely contained in the dependence on $B$ of the open string metric
$G$, the open string effective coupling $G_s$, and the $\theta$
dependence of the $*$ product.  We have also argued in section 3.1 that
there must exist a continuous interpolation between these two
descriptions with arbitrary $\theta$.

To compare these different descriptions, we need a situation in which
we can compute in all of them.  For this we will take the limit of
slowly varying, but not necessarily small, gauge fields of rank one.
The effective action is \refs{\tseytlin,\callan} the Dirac-Born-Infeld
(DBI) action.  We will compare the ordinary DBI action as a function
of the closed string metric and coupling and $B+F$, to its
noncommutative counterpart, as a function of the open string metric
and coupling and $\hat F+\Phi$.\foot{The comparison cannot be made
just using the formula \rankonesolb\ for constant $F$, since as we
will see, terms in \firstvaranya\ of the general form $A\partial F$
contribute in the analysis.  It is necessary to integrate by parts in
comparing the DBI actions, and one cannot naively treat $F$ as a
constant.}  After proving the equivalence between them and exploring
the zero slope limit in section 4.1, we specialize in section 4.2 to
the case of four dimensions, and compare the respective BPS conditions
-- the stringy generalization of the instanton equation.

\subsec{Dirac-Born-Infeld Action}

For slowly varying fields on a single $Dp$-brane, the effective
Lagrangian is the Dirac-Born-Infeld Lagrangian
\eqn\biagm{\CL_{DBI}={1 \over g_s (2\pi)^p(\alpha')^{p+1\over
2}}\sqrt{\det (g+2\pi\alpha'(B+F))}.}
We discussed in section 2 the  normalization and the fact that $B+F$
is the gauge invariant combination.  

We have argued in section 2 that the effective Lagrangian of the noncommutative
gauge fields $\hat A$ must be such that when expressed in terms of the
open string variables $G$, $\theta$, and $G_s$ given in \gthed\ and
\limpropgz, the dependence on
$\theta$ is only in the $*$ product.  Therefore, for slowly varying
$\hat F$ it is
\eqn\biagmnt{\hat \CL_{DBI}={1 \over G_s (2\pi)^p(\alpha')^{p+1\over
2}}\sqrt{\det (G+2\pi\alpha' \hat F)}.}
We also proposed in section 3.1 that there is a more general
description with an arbitrary $\theta$ and parameters $G, \Phi$, $G_s$
determined in \changevarg
\eqn\changevarga{\eqalign{
&{1 \over G+ 2\pi\alpha' \Phi}= -{\theta \over 2\pi \alpha'} +{1 \over
g+ 2\pi\alpha' B}\cr  
&G_s=g_s \left({\det (G+ 2\pi\alpha'\Phi) \over \det (g+ 2\pi\alpha'
B)}\right)^\half = g_s {1 \over \det\left[({1 \over g+
2\pi\alpha' B}-{\theta \over 2\pi \alpha'})  (g+ 2\pi\alpha'
B)\right]^\half}.}}
Under the assumption of \hatfphi, the effective action in these
variables should be 
\eqn\biagmn{\hat \CL_{DBI}={1 \over G_s (2\pi)^p(\alpha')^{p+1\over
2}}\sqrt{\det (G+2\pi\alpha' (\hat F+\Phi))}}
We will here demonstrate that in the limit of slowly varying fields,
\biagm, \biagmnt, and \biagmn\ are all equivalent.  In particular,
this will  verify the equivalence of \biagm\ and \biagmnt, for which
we gave an {\it a priori} explanation in section 2, it will 
give (along with the special case considered in section 3.2) our main
evidence that there exists a description with an arbitrary $\theta$ and
$\hat F$ shifted as in \hatfphi, and it will show that in such a description
$G,\Phi$, and $G_s$ must be given by the formulas in \changevarga.

In all of the above  formulas, we can expand $\hat
\CL_{DBI}$ in powers of $\hat F$ and all the resulting products
can be regarded as $*$ products.  If instead we treat them as ordinary
products, our answer will differ by terms including derivatives of
$\hat F$.  Since the DBI Lagrangian is obtained in string theory after
dropping such terms, there is no reason to keep some of them but not
others.  Therefore, we will ignore all such derivatives of $\hat F$
and regard the products in the expansion of \biagmn\ as ordinary
products, i.e.\ the $\theta$ dependence will be only in the definition
of $\hat F$. 

We want to show
that using the change of variables \changevarga\ and the
transformations of the fields $\hat A(A)$ given in  \changefi, the two
Lagrangians are related as
\eqn\mainre{\CL_{DBI}=\hat \CL_{DBI}  + {\rm total~ derivative} +
\CO(\partial F) .}  
The difference in total derivative arises from the fact that the
action is derived in string theory by using the equations of motion or
the $S$-matrix elements, which are not sensitive to such total
derivatives.  Furthermore, the effective Lagrangian in terms of $\hat
A$ is gauge invariant only up to total derivatives, and we will permit
ourselves to integrate by parts and discard total derivatives.  The
$\CO(\partial F)$ term in \mainre\ is possible because these two
Lagrangians are derived in string theory in the approximation of
neglecting derivatives of $F$, and therefore they can differ by such
terms.

For $\theta=0$, the
change of variables \changevarga\ is trivial and so is \mainre.
Therefore, in order to prove \mainre\ it is enough to prove its
derivative with respect to $\theta$ holding fixed the closed string
parameters $g$, $B$, $g_s$ and the commutative gauge field $A$.
In other words, we will show that this variation of the right hand
side of \mainre\ vanishes.

In order to keep the equations simple we will set $2\pi \alpha'=1$;
the $\alpha'$ dependence can be easily restored on dimensional
grounds.  In preparation for the calculation we differentiate
\changevarga\ holding $g$, $B$ and $g_s$ fixed, and express the
variation of $G$, $\Phi$ and $G_s$ in terms of the variation of
$\theta$:
\eqn\vchangevarg{\eqalign{
&\delta G+\delta \Phi= (G+ \Phi) \delta\theta (G+\Phi)\cr 
&\delta G_s={1\over 2} G_s \Tr{1 \over G+\Phi}(\delta G+
\delta \Phi)={1\over 2} G_s \Tr(G+\Phi)\delta \theta={1\over 2} G_s
\Tr\Phi\delta \theta, }} 
where $\delta G$ and $\delta \Phi$ are symmetric and antisymmetric
respectively.  We also
need the variation of $\hat F$ with respect to $\theta$.  Equation
\firstvaranya\ can be written as
\eqn\firstranko{\delta \hat F_{ij}(\theta) =
\delta \theta^{kl}  \biggl[  \hat F_{ik} \hat F_{j l} - {1 \over2}
\hat A_k(\partial_l\hat F_{ij}  +\hat D_l\hat F_{ij}) \biggr]+
\CO(\partial \hat F).} 
Since we are going to ignore derivatives of $\hat F$, we replaced the
$*$ products with ordinary products.  We kept, however, the $*$
products in the definitions of $\hat F$ and in $\hat D_l \hat F_{ij}$.
Similarly, we kept the explicit derivative and covariant derivative of
$\hat F$ since they multiply $\hat A_k$, and can become terms without
derivatives of $\hat F$ after integration by parts.

We are now ready to vary $\hat \CL_{DBI}$ \biagmn:
\eqn\ftaca{\eqalign{
&\delta \left[{1 \over G_s} \det(G+ \hat F+ \Phi)^\half
\right] ={\det(G+ \hat F+ \Phi)^\half \over G_s}
\left[-{\delta G_s \over G_s} +{1\over 2} \Tr {1 \over G+
\hat F+ \Phi} (\delta G+ \delta \hat F+ \delta
\Phi)\right] \cr
&\qquad\qquad={1 \over 2} {\det(G+ \hat F+ \Phi)^\half
\over G_s}\left[ -\Tr \delta\theta (G+\Phi) +\Tr {1 \over G+ \hat
F+ \Phi}(G+ \Phi) \delta \theta (G+\Phi)\right.\cr
&\qquad\qquad\quad\left. +  \left({1 \over G+
\hat F + \Phi }\right)_{ji}\delta \theta^{kl}\left( \hat F_{ik} \hat
F_{jl} - {1\over 2} \hat A_k (\partial_l\hat F_{ij}  +\hat D_l\hat
F_{ij}) \right)\right] +\CO(\partial \hat F)\cr
&\qquad\qquad={1 \over 2} {\det(G+ \hat F+ \Phi)^\half
\over G_s}\left[ - \Tr {1 \over G+ \hat
F+ \Phi}\hat F \delta \theta(G+ \Phi)  \right. \cr
&\qquad\qquad\quad \left.- \Tr {1 \over G+
\hat F +\Phi} \hat F\delta \theta \hat F + \Tr \delta \theta
\hat F\right]  +\CO(\partial \hat F)+ {\rm total~ derivative}\cr 
&\qquad\qquad=\CO(\partial \hat F) + {\rm total~ derivative},}}
where in the third step we used
\eqn\usedeq{\eqalign{
&\partial_l\det(G+ \hat F + \Phi)^\half = {1\over 2}
\det(G+ \hat F +\Phi)^\half  \left({1\over
G+ \hat F+ \Phi}\right)_{ji}\partial_l \hat F_{ij},\cr
&\hat D_l\det(G+ \hat F + \Phi)^\half = {1\over 2}
\det(G+ \hat F +\Phi)^\half  \left({1\over
G+ \hat F+ \Phi}\right)_{ji}\hat D_l F_{ij}+\CO(\partial_l\hat F \hat
D_l\hat F),\cr} }
to integrate by parts, and then we used
\eqn\findfhat{\delta\theta^{kl}(\partial_l\hat A_k +\hat D_l \hat A_k) =
\delta\theta^{kl} \hat F_{lk}.}
This completes the proof of \mainre\ and of our change of variables
\changevarg.  

In fact, we first found the change of variables
\changevarg\ by demanding that \mainre\ is satisfied.  It is, however,
quite nontrivial that there exists a change of
variables like \changevarg\ for which \mainre\ is satisfied.  This
depends on our relation between the commutative and the noncommutative
gauge fields \firstranko\ and on the particular form of the DBI
Lagrangian.  One could actually use this computation to motivate the
DBI Lagrangian as the only Lagrangian for which \mainre\ is true.

Even though we proved \mainre\ for every value of the parameters, it
is instructive to examine it in various limits, comparing the
commutative and the noncommutative sides of \mainre\ explicitly.  Some
of the technical aspects of these comparisons are similar to the
calculation in \ftaca, but they are conceptually different.  Rather
than varying the description by changing $\theta$ for a fixed
background (fixed $g$, $B$ and $g_s$), here we will use a description
with $\Phi=0$ and will vary the background.  We will perform two
computations.  The first will be for small $\alpha' B$ and the second
will be in the zero slope limit.  The reader who is not interested
in the details could jump to section 4.2.

\bigskip\noindent{\it Comparison For Small $B$}

First, we consider the comparison for small $B$.  In this regime, the
open string variables are
\eqn\chavarblessb{\eqalign{
&G= g-(2\pi\alpha')^2 Bg^{-1}B \cr 
&\theta=  -(2\pi \alpha')^2 g^{-1} B g^{-1} +\CO(B^3) \cr
&G_s=g_s\left(1- (\pi \alpha' )^2 \Tr (g^{-1}B)^2 +\CO(B^4)\right).}}

Since $\theta$ is small for small $B$, we begin by expanding \biagmnt\
in powers of $\theta$.  The change of variables in the rank one case
\eqn\changran{\hat F_{ij} = F_{ij} + \theta^{kl} \left(F_{ik} F_{j l}
-A_k \partial_l F_{ij}\right)+\CO(\theta^2)}
leads to
\eqn\ftaca{\eqalign{
&\det(G+2\pi \alpha' \hat F)^\half=
\det\left[G_{ij}+2\pi \alpha' \left(F_{ij} + \theta^{kl}
\left( F_{ik} F_{j l} -A_k \partial_l F_{ij}\right)+\CO(\theta^2)
\right) \right]^\half \cr & =
\quad\det(G+2\pi \alpha' F)^\half\left[1+ \pi \alpha' \left({1
\over G+2\pi \alpha' F}\right)_{ji}\theta^{kl}\left( F_{ik} F_{j l} -A_k
\partial_l F_{ij} \right)+\CO(\theta^2)\right] \cr & =
\quad  \det(G+2\pi \alpha' F)^\half\left[1- \pi\alpha' \Tr {1
\over G+2\pi \alpha' F} F\theta F + \half \Tr \theta F
+\CO(\theta^2)\right] + {\rm total~ derivative}\cr & =
\quad  \det(G+2\pi \alpha' F)^\half\left[1- {1\over 4\pi\alpha'} \Tr
{1 \over G+2\pi \alpha' F} G\theta G +\CO(\theta^2)\right] + {\rm
total~ derivative},}}
where in the second step we used \usedeq\ and then integrated by
parts, and in the third step we used $\Tr\, G\theta=0$.  Using
\chavarblessb, to first  order in $B$, we find as expected
\eqn\biagmns{\eqalign{
\hat \CL_{DBI}=&{1 \over g_s (2\pi)^p(\alpha')^{p+1\over
2}}\det(g+2\pi \alpha' F)^\half\left(1+ \pi \alpha' \Tr {1 \over
g+2\pi \alpha' F}B  \right)\cr 
&\qquad +\CO(B^2) + {\rm total~ derivative} =\cr
=& \CL_{DBI} +\CO(B^2) + {\rm total~ derivative}  .}}

To extend this comparison to order $B^2$, we would need to use
\firstvaranya\ to determine the order $\theta^2$ terms in \changran.
However, since the order $\theta^2$ terms in \changran\ involve three
factors of $F$ or terms which become three factors of $F$ after
integration by parts, we can compare the $B^2F^2$ terms in the two
Lagrangians without needing the corrections to \changran.

We use the identity for antisymmetric $M$
\eqn\usefid{\det(1+M)^\half = 1-{1 \over 4}\Tr M^2 - {1 \over 8} \Tr
M^4+{1\over 32}\left(\Tr M^2\right)^2 + \CO(M^6)}
to write the DBI Lagrangian density as 
\eqn\biagme{\eqalign{
\CL_{DBI}=&{\sqrt{\det g } \over g_s (2\pi)^p(\alpha')^{p+1\over
2}}\big[ 1-(\pi\alpha')^2\Tr (g^{-1}(B+F))^2  -2(\pi\alpha')^4\Tr
(g^{-1}(B+F))^4\cr 
&+{(\pi\alpha')^4\over 2} \left(\Tr (g^{-1}(B+F))^2\right)^2
+\CO((B+F)^6) \big]\cr 
=&{(\pi\alpha')^2\sqrt{\det g } \over g_s (2\pi)^p(\alpha')^{p+1\over
2}}\big[-\Tr (g^{-1}F)^2 -2(\pi\alpha')^2\Tr
(g^{-1}F)^4+{(\pi\alpha')^2\over 2} \left(\Tr
(g^{-1}F)^2\right)^2\cr 
& -8(\pi\alpha')^2\Tr g^{-1}B(g^{-1}F)^3+2(\pi\alpha')^2
\Tr (g^{-1}F)^2\Tr g^{-1}Fg^{-1}B \cr
& -8(\pi\alpha')^2\Tr (g^{-1}B)^2(g^{-1}F)^2+(\pi\alpha')^2 \Tr
(g^{-1}B)^2\Tr (g^{-1}F)^2 \cr 
&+{\rm constant}+{\rm total~derivative}+\CO(B^3, BF^5,B^2F^4)
\big],}} 
where in the last step we used the fact that
\eqn\usetot{2\Tr (g^{-1}Bg^{-1}F)^2 - \left(\Tr
g^{-1}Bg^{-1}F\right)^2 = {\rm total~derivative}.}
Similarly,
\eqn\biagmeh{\eqalign{
\hat\CL_{DBI}=&{\sqrt{\det G } \over G_s (2\pi)^p(\alpha')^{p+1\over
2}}\big[ 1-(\pi\alpha')^2\Tr (G^{-1}\hat F )^2  -2(\pi\alpha')^4\Tr
(G^{-1}\hat F)^4\cr 
&+{(\pi\alpha')^4\over 2} \left(\Tr (G^{-1}\hat F)^2\right)^2
+\CO(\hat F^6) \big]\cr 
=&{(\pi\alpha')^2\sqrt{\det g } \over g_s (2\pi)^p(\alpha')^{p+1\over
2}}\big[-\Tr (g^{-1}F)^2 -2(\pi\alpha')^2\Tr (g^{-1}F)^4
+{(\pi\alpha')^2\over 2} \left(\Tr (g^{-1}F)^2\right)^2 \cr 
& -8(\pi\alpha')^2\Tr g^{-1}B(g^{-1}F)^3+2(\pi\alpha')^2
\Tr (g^{-1}F)^2\Tr g^{-1}Fg^{-1}B \cr
& -8(\pi\alpha')^2\Tr (g^{-1}B)^2(g^{-1}F)^2+(\pi\alpha')^2 \Tr
(g^{-1}B)^2\Tr (g^{-1}F)^2 \cr 
&+{\rm constant}+{\rm total~derivative}+\CO(B^3, BF^5,B^2F^4)
\big]\cr 
=&\CL_{DBI}+ {\rm total~derivative}+\CO(B^3, BF^5,B^2F^4) ,}}
where we have used \changran\ and \chavarblessb.

This demonstrates explicitly that
\eqn\matchsmb{\hat\CL_{DBI}= \CL_{DBI} +{\rm total~derivative}+
\CO(B^3F^3,B^2 F^4).} 

\bigskip\noindent{\it Comparison In The Zero Slope Limit}

We now turn to another interesting limit -- our zero slope limit.  In
this limit the entire string effective Lagrangian becomes quadratic in
$\hat F$.  The same is true for the DBI Lagrangian
\eqn\dbizer{\eqalign{
\hat \CL_{DBI} = &{(\alpha')^{3-p\over 2}\over 
4(2\pi)^{p-2}G_s} \sqrt{\det G} G^{im}G^{jn}\hat F_{ij}*\hat F_{mn} \cr
&+ {\rm ~total~ derivative ~+  ~constant~ + ~higher~ powers ~ of~ }
\alpha'.} }
Ignoring the total derivative and the constant we can set $\Phi=0$.

If we take the zero slope limit, and $F$ is slowly varying, then
we can get a description either using ordinary gauge fields and the
DBI Lagrangian $\CL_{DBI} $ , or using noncommutative gauge fields
using the $\hat F^2$ Lagrangian \dbizer.

We work on a single Euclidean $p$-brane with $B$ of rank $r=p+1$, and
for simplicity we consider the metric $g_{ij}=\epsilon \delta_{ij}$.
We are interested in the zero slope limit, i.e.\ $\alpha'\sim
\epsilon^\half \to 0$.  Expanding the DBI action density in powers of
$\epsilon$, we find
\eqn\biadm{\eqalign{
\CL_{DBI}=&{1\over (2\pi)^p(\alpha')^{p+1\over
2}g_s}\left(|\pf(M)|+{\epsilon\over 2}|\pf(M)|\Tr{1\over
M}+{\epsilon^2\over 8}|\pf(M)|\left(\Tr{1\over M}\right)^2 \right.\cr
&\left. -{\epsilon^2\over 4}|\pf(M)|\Tr{1\over M^2}
+\CO(\epsilon^3)\right),\cr}} 
where 
\eqn\mdefa{M=2\pi \alpha'(B+F).}
The absolute value sign arises from the branch of the square root in
\biagm.  Since $M$ is antisymmetric, the second and third terms vanish.
The first term is a constant plus a total derivative in spacetime,
which we ignore in this discussion.  In the limit
$\epsilon \rightarrow 0$, the leading term is
\eqn\biacm{\eqalign{
\CL_{DBI}=&-{\epsilon^2 \over 4(2\pi)^{p+3\over
2}(\alpha')^2g_s }|\pf (B+F)|\Tr {1\over (B+F)^2} \cr
&+ {\rm ~total~ derivative ~+  ~constant~ + ~higher~ powers ~ of~ }
\alpha'.}}

Our general discussion above shows that the Lagrangian \biacm\ must be
the same as the $\hat F^2$ Lagrangian \dbizer.  We now verify this
explicitly in a power series in $F$.

We define three auxiliary functions
\eqn\defhe{\eqalign{
&f(B,F)=|\pf (B+F)|\Tr {1\over (B+F)^2}- |\pf B|\Tr {1\over B^2} \cr
&q(B,F,\eta)= \left|\pf \left(B+F +\eta {1\over B}\right)\right|\cr
&h(B,F) = 4{\partial q \over \partial \eta}(\eta=0)
-q(\eta=0)\Tr {1 \over B^2}.}}
$q$ and therefore also all its derivatives with respect to $\eta$ are
total derivatives in spacetime.  In particular, $h(B,F) $ is a total
derivatives in spacetime. 
Expanding $f(B,F)$ in powers of $F$ we find
\eqn\biacme{\eqalign{ 
f(B,F)=&h(B,F)+|\pf B| \Tr \left({1 \over B^2 }F{1 \over B^2 }F\right)
+ {1\over 2} |\pf B| \Tr \left({1 \over B }F \right)\Tr \left({1 \over
B^2 }F{1 \over B^2 }F\right) \cr
&-2|\pf B|\Tr \left({1\over B}F{1 \over B^2 }F{1 \over B^2
}F\right) +\CO(F^4).}} 

If $\hat F$ is small, the $\hat F^2$ action can be expressed in terms
of ordinary gauge fields.  In the rank one case \changefi\ becomes
\eqn\phangran{\hat F_{ij} = F_{ij} + \theta^{kl} \left(F_{ik} F_{j l}
-A_k \partial_l F_{ij}\right)+\CO(F^3).}
Here in asserting that the corrections are $\CO(F^3)$, we consider two
derivatives or two powers of $A$ (or one of each) to be equivalent to
one power of $F$.  
Substituting \phangran\ in \dbizer\ we find
\eqn\ftac{\eqalign{
{(\alpha')^{3-p\over 2}\over 4(2\pi)^{p-2}G_s }&\sqrt{\det G}
G^{im}G^{jn}\hat F_{ij}*\hat F_{mn} = {(\alpha')^{3-p\over 2}\over
4(2\pi)^{p-2}G_s } \sqrt{\det G} \left[G^{im}G^{jn}F_{ij}
F_{mn}\right. \cr
& \qquad \left. +2G^{im} G^{jn}\theta^{kl} \left( F_{ik} F_{j l} -A_k
\partial_l F_{ij}\right)F_{mn} \right] + {\rm
total~derivatives}+\CO(F^4)\cr 
&={(\alpha')^{3-p\over 2}\over 4(2\pi)^{p-2}G_s } \sqrt{\det G}\left[-
\Tr\left( G^{-1}F G^{-1}F\right) +2\Tr\left(F\theta  FG^{-1} 
FG^{-1}\right)\right.\cr
&\qquad \left.  -{1\over 2} \Tr \left(F\theta
\right)\left(FG^{-1}FG^{-1}\right) \right] + {\rm
total~derivatives}+\CO(F^4) .}}  

For $g_{ij}=\epsilon \delta_{ij}$ with $B$ of rank $r=p+1$, we find
{}from \limmet\ that $G=-\epsilon^{-1}(2\pi\alpha')^2B^2$ (which is
finite as $\epsilon \rightarrow 0$), $\theta = {1 \over B}$ and $G_s=
g_s \left({2 \pi \alpha' \over \epsilon }\right)^{p+1\over 2}|\pf B|$.
Using these formulas, with \biacme\ and \ftac, we get
\eqn\matcht{\CL_{DBI}={(\alpha')^{3-p\over 2}\over 4(2\pi)^{p-2}G_s }
\sqrt{\det G} G^{im}G^{jn}\hat F_{ij}*\hat F_{mn}+ {\rm
total~derivatives} +{\rm constant} +\CO(F^4).}

We conclude that the zero slope limit of $\CL_{DBI}$ is the
nonpolynomial action \biacm.  It conicides with the zero slope limit
of $\hat \CL_{DBI}$, which is simply $\hat F^2$ (we have checked it
explicitly only up to terms of order $F^4$).

\subsec{Supersymmetric Configurations}

Now we will specialize to four dimensions (though some of the
introductory remarks are more general) and analyze supersymmetric
configurations, the stringy instantons.

We recall first that, in general, a $Dp$-brane preserves only half of
the supersymmetry of Type II superstring theory.  In an interpretation
\ref\hp{J. Hughes and J. Polchinski, ``Partially Broken Global
Supersymmetry And The Superstring,'' Nucl. Phys. {\bf B278} (1986)
147.} 
that actually predates the $D$-brane era, this means the theory along
the brane has spontaneously broken (or ``nonlinearly realized'')
supersymmetry along with its unbroken (or ``linearly realized'')
supersymmetry.

For example, in the extreme low energy limit, the theory along a
threebrane is the $F^2$ theory.  Its minimal supersymmetric extension
in four dimensions is obtained by adding a positive chirality
``photino'' field $\lambda_\alpha$,\foot{We recall that $SO(4)$
decomposes as $SU(2)_+\times SU(2)_-$.  A positive chirality spinor
transforms as $(1/2,0)$, while a negative chirality spinor transforms
as $(0,1/2)$.  Our conventions are such that a selfdual antisymmetric
tensor transforms as $(1,0)$ and an anti-selfdual one as $(0,1)$.  The
full low energy $D3$-brane action has additional fields and
supersymmetries beyond those discussed here, but we do not expect them
to affect the particular issues we will address.}  with its CPT
conjugate $\bar\lambda_{\dot\alpha}$ of opposite chirality.  The
linearly realized or unbroken supersymmetry acts by the standard
formulas 
\eqn\linsusy{ \delta \lambda_\alpha = {1 \over 2\pi \alpha'}
M^+_{ij} \sigma^{ij\beta}_{\alpha } \eta_\beta }
\eqn\linsusyo{\delta \bar \lambda_{\dot \alpha} = {1 \over 2\pi
\alpha'}  M^-_{ij} \bar \sigma^{ij\dot \beta}_{\dot \alpha } \bar
\eta_{\dot\beta},}
where we have included a $B$-field, and written the standard formula
in terms of $M$ (defined in \mdefa) rather than $F$.  As usual,
$\sigma^{ij}=\half\left[\Gamma^i,\Gamma^j\right]$, while $\alpha,\
\beta$ and $\dot\alpha,\ \dot\beta$ are spinor indices of respectively
positive and negative chiralities.  The nonlinearly realized or
spontaneously broken supersymmetry of the $F^2$ theory acts simply by
\eqn\tugalo{\delta^*\lambda_\alpha ={\epsilon^2\over 4\pi 
\alpha'}\eta_\alpha^*,
~~\delta^*\bar\lambda_{\dot\alpha}={\epsilon^2\over 4\pi
\alpha'}\bar\eta^* _{\dot\alpha}.}
Here $\eta$ and $\bar\eta$ are constants, and we have chosen a
convenient normalization.

One of the many special properties of the DBI theory is that 
\ref\baga{J.~Bagger and A.~Galperin, ``A New Goldstone multiplet for
partially broken supersymmetry,'' Phys. Rev. {\bf D55} (1997) 1091,
hep-th/9608177.} 
it has in four dimensions a supersymmetric extension that preserves
not only the linearly realized supersymmetry -- many bosonic theories
have such a supersymmetric version -- but also, what is much more
special, the nonlinearly realized supersymmetry.  The transformation
law of the photino under the linearly realized supersymmetry is
unchanged from \linsusy\ in going to the DBI theory.  The nonlinearly
realized supersymmetries, however, become much more complicated.  The
generalization of \tugalo\ is
\eqn\nonli{\eqalign{
\delta^*\lambda_\alpha &={1 \over 4\pi\alpha'}\left[\epsilon^2- \pf M
+\sqrt{\det_{ij}(\epsilon\delta_{ij} + M_{ij})}\right]\eta^*_\alpha\cr
&={1\over 4\pi\alpha'}\left[\epsilon^2- \pf M
+\sqrt{\epsilon^4-{\epsilon^2\over 2} \Tr M^2 +(\pf
M)^2}\right]\eta^*_\alpha \cr 
&= {1\over 4\pi\alpha'}\left[-\pf M + |\pf M|+\epsilon^2 {4|\pf
M|-\Tr M^2 \over 4 |\pf M|} + \CO(\epsilon^4)\right]\eta^*_\alpha ,}}
\eqn\nonlio{\eqalign{
\delta^*\bar \lambda_{\dot\alpha} &={1 \over
4\pi\alpha'}\left[\epsilon^2 +\pf M
+\sqrt{\det_{ij}(\epsilon\delta_{ij} + M_{ij})}\right]\bar
\eta^*_{\dot \alpha} \cr  
&={1\over 4\pi\alpha'}\left[\epsilon^2 +\pf M
+\sqrt{\epsilon^4-{\epsilon^2\over 2} \Tr M^2 +(\pf
M)^2}\right]\bar \eta^*_{\dot \alpha} \cr 
&= {1\over 4\pi\alpha'}\left[\pf M + |\pf M|+\epsilon^2 {4|\pf M|-\Tr
M^2 \over 4 |\pf M|} + \CO(\epsilon^4)\right]\bar
\eta^*_{\dot\alpha}.}}
In both \nonli\ and \nonlio, the first two formulas for $\delta^*$ are
taken directly from \baga, and the last is an expansion in small
$\epsilon$ aimed at taking the by now familiar $\alpha'\to 0$ limit.
 
When $B=0$ and we expand around $F=0$, the supersymmetry \linsusy,
\linsusyo\ is realized linearly, while the supersymmetry of \nonli\
and \nonlio\ is spontaneously broken and is realized nonlinearly.  In
expanding around any constant $B$, there is always a linear
combination of the $\delta$ and $\delta^*$ supersymmetries that is
unbroken.  However, this combination depends on $B$.  To see why,
consider an open string ending on the threebrane, and let $\bar\psi$
and $\psi$ be the left and right-moving worldsheet fermions.  In
reflection from the end of the string, we get $\bar \psi = R(B)\psi$
where $R(B)$ is a rotation matrix that can be found from \bouconf\ to
be
\eqn\kilm{R(B)=\left(1-2\pi \alpha'g^{-1}B\right)^{-1}
\left(1+2\pi\alpha'g^{-1}B\right).}
Let $Q_L$ and $Q_R$ be the spacetime supersymmetries carried by left
and right-moving worldsheet degrees of freedom in Type II superstring
theory.  (Because we are mainly focusing now on threebranes, we are
in Type IIB, and $Q_L$ and $Q_R$ both have the same chirality.)  A
general supersymmetry of the closed string theory is generated by
$\epsilon^\alpha_R Q_{R,\alpha} +\epsilon_L^\beta Q_{L,\beta}$ with
constants $\epsilon_L$, $\epsilon_R$.  Reflection at the end of the
open string breaks this down to a subgroup with
\eqn\locon{\epsilon_L = R(B)\epsilon_R,}
where now of course the rotation matrix $R(B)$ must be taken in the
spinor representation.  Here we see explicitly that which
supersymmetries are unbroken depends on $B$, though the number of
unbroken supersymmetries is independent of $B$.\foot{We can also see
explicitly that a system of a three-brane with a $B$-field and a
separated $-1$-brane is only supersymmetric if $B^+=0$, as asserted in
the introduction.  The supersymmetry left unbroken by the $-1$-brane
obeys $\epsilon_L=\Gamma_{0123}\epsilon_R$, where $\Gamma_{0123}$ is
the four-dimensional chirality operator.  At $B=0$, this is compatible
with \locon\ for spinors of $\Gamma_{0123}=1$, the expected result
that the $-1$-brane (like an instanton) preserves the supersymmetry of
positive chirality.  If we now turn on $B\not=0$, compatibility with
\locon\ fails unless $R(B)=1$ for states of $\Gamma_{0123}=1$, that is,
unless $B^+=0$.}

We can easily match the stringy parameters $\epsilon_L$, $\epsilon_R$
with the parameters $\eta$, $\eta^*$ of the supersymmetrized DBI
action.  We have (up to possible inessential constants)
\eqn\jumpo{\eta=\epsilon_L+\epsilon_R,~~\eta^*=\epsilon_L-\epsilon_R.}
In fact, $\eta$ can be identified with $\epsilon_L+\epsilon_R$ as the
generator of a supersymmetry that is unbroken at $B=0$. And $\eta^*$
can be identified with $\epsilon_L-\epsilon_R$ as being odd under a
$\Z_2$ symmetry that acts by $\lambda\to -\lambda$, $F\to -F$ in the
field theory, and by reversal of worldsheet orientation in the string
theory.

Now, specializing again to four dimensions, we want to identify the
unbroken supersymmetry in the $\alpha'\to 0$ limit, which we
temporarily think of as the limit with $g$ fixed and $B\to\infty$.
Here we meet the interesting fact that there are {\it two}
inequivalent zero slope limits in four dimensions.  For nondegenerate
$B$ with all eigenvalues becoming large, we get from \kilm\ that
$R(B)\to -1$ in the vector representation for $B\to \infty$.  But the
element $-1$ of the vector representation can be lifted to spinors in
two different ways: as a group element that is $-1$ on positive
chirality spinors and $1$ on negative chirality spinors, or
vice-versa.  Starting from $R(B)=1$ (on both vectors and spinors) at
$B=0$, what limit we get for $B\to \infty$ depends on the sign of
$\pf(B)$.  In fact, the limit of $R(B)$ is, in acting on spinors,
\eqn\iko{R(B)\to -\Gamma_0\Gamma_1\Gamma_2\Gamma_3 \cdot {\rm
sign}(\pf(B)).} 
To prove this, it is enough to consider the special case that $B$ is
selfdual or anti-selfdual.  If $B$ is selfdual, then from \kilm,
$R(B)\in SU(2)_+$ and is identically $1$ on negative chirality
spinors, and hence approaches $-1$ for $B\to \infty$ on positive
chirality spinors.  For $B$ anti-selfdual, we get the opposite result,
leading to \iko.

There is no loss of essential generality in assuming that we want to
study instantons, or anti-selfdual gauge fields, for which the
unbroken supersymmetry generators are of positive chirality.  In this
case, \iko\ reduces to $R(B)=-{\rm sign}(\pf(B))$.  \jumpo\ tells us
that the unbroken supersymmetry is $\eta$ when $R(B)=1$ and $\eta^*$
when $R(B)=-1$.  We conclude, then, that for instantons, when $B\to\infty$ with
$\pf(B)>0$,  the unbroken supersymmetry 
is the one that at $B=0$ is
nonlinearly realized, while for $B\to\infty$ with
$\pf(B)<0$, the unbroken supersymmetry
is the one that is linearly realized at $B=0$.  For anti-instantons,
the two cases are reversed.  We shall see this difference both in the
discussion below, based on the supersymmetric DBI action of
\baga, and in section 5 in the analysis of the $D0/D4$ system, which will
also lead to an alternative intuitive explanation of the difference.

\bigskip\noindent{\it BPS States Of Supersymmetric Born-Infeld}

We assume that we are in four dimensions with $B$ of rank four.
Instead of taking $B\to\infty$, it is equivalent, of course, to take
the metric to be $g_{ij}=\epsilon\delta_{ij}$ and take $\epsilon\to 0$
with the two-form $B$-fixed.  Using $\Tr M^2=\Tr (M^-)^2+\Tr (M^+)^2$
and $4\pf M=\Tr (M^-)^2-\Tr (M^+)^2$, the equations \nonli\ and
\nonlio\ become in this limit
\eqn\nonlia{
\delta^*\lambda_\alpha =- {1 \over 2\pi\alpha'}\eta^*_\alpha \cases{
\pf M +\CO(\epsilon^2 ) & for $\pf M <0$ \cr
\epsilon^2 {\Tr (M^+)^2 \over 4 \pf M} + \CO(\epsilon^4)& for $\pf
M >0$}}
\eqn\nonliao{
\delta^*\bar\lambda_{\dot\alpha} ={1 \over
2\pi\alpha'}\bar\eta^*_{\dot\alpha}  \cases{
\pf M +\CO(\epsilon^2 ) & for $\pf M >0$ \cr
\epsilon^2 {\Tr (M^-)^2 \over 4 \pf M} + \CO(\epsilon^4)& for $\pf
M <0$ .}}

For $F=0$ and constant $B$, the unbroken supersymmetries are a linear
combination of the original unbroken supersymmetries \linsusy\ and
\linsusyo\ with the supersymmetries described in the last paragraph.
The parameters of the unbroken supersymmetries obey
\eqn\unbroken{\eta^*_\alpha=C_{ij}^+
\sigma^{ij\beta}_{\alpha}\eta_\beta}
\eqn\unbrokena{\bar \eta^*_{\dot\alpha}=C_{ij}^- \bar
\sigma^{ij\dot\beta}_{\dot\alpha} \bar\eta_{\dot\beta}}
\eqn\cpluss{C^+=\cases{
{1\over 2\pi \alpha' \pf B}B^+& for $\pf B<0 $\cr
{8\pi \alpha' \pf B \over\epsilon^2 \Tr (B^+)^2}B^+ & for $\pf
B>0$}}
\eqn\cplussu{C^-=\cases{
-{1\over 2\pi \alpha' \pf B}B^-& for $\pf B>0 $\cr
-{8\pi \alpha' \pf B \over\epsilon^2 \Tr (B^-)^2}B^- & for $\pf
B<0$.}}
The dichotomy between the two cases described above is visible here in
the dependence of the $\epsilon\to 0$ limit on the sign of $\pf(B)$.
\foot{The relation to the discussion surrounding \iko\ is somewhat
obscured by the fact that we have done the scaling with $\epsilon\to
0$ rather than $B\to \infty$, as assumed in the discussion of \iko.
Taking $B$ to infinity with $g_{ij}$, $\eta$, and $\eta^*$ fixed is
equivalent to $\epsilon\to 0$ with $B$-fixed and a nontrivial scaling
of $\eta$ and $\eta^*$.  To compare most directly to the discussion
surrounding \iko, if one takes $B\to\infty$ with fixed $\epsilon$,
then according to \cpluss\ and \cplussu, in the limit one has
$C^{\pm}$ tending to zero or infinity depending on the sign of
$\pf(B)$.  Here $C\to 0$ means that the unbroken supersymmetry is
generated by $\eta$ (the generator of the original linearly realized
supersymmetry) and $C\to\infty$ means that the unbroken supersymmetry
is generated instead by $\eta^*$.}  Now we want to consider BPS
configurations on $\R^4$, with constant $B$ and $F$ approaching zero
at infinity.  Since $F\to 0$ at infinity, an unbroken supersymmetry
must be of the form \cpluss\ or \cplussu.  Without loss of generality,
we examine the condition for instantons, configurations that leave
invariant the positive chirality supersymmetry \cpluss.  The condition
for this to be so is
\eqn\bpsconba{M^+=\cases{
C^+\pf M  & for $\pf B<0 $\cr
C^+ \epsilon^2 {\Tr (M^+)^2 \over 4\pf M} & for $\pf B>0 $ \cr}}
with the appropriate constants $C^+$ in the two cases \cpluss.  This
choice of $C^+$ guarantees that the equations are satisfied at
infinity, where $F$ vanishes.  It turns out that these two equations
are the same, and we get one condition regardless of the sign of $\pf
B$, namely
\eqn\bpsconb{M^+= {B^+ \over 2\pi \alpha' \pf B}\pf M}
or equivalently, since $M=2\pi\alpha'(B+F)$,
\eqn\fincond{F^+={B^+\over 8 \pf B}\epsilon^{ijkl}(2B_{ij} F_{kl} +
F_{ij} F_{kl}).}
The reduction of \bpsconba\ to \bpsconb\ is clear enough, using
\cpluss, if $\pf(B)<0$.  For $\pf(B)>0$, one must work a bit harder.
By taking the trace squared of 
\bpsconba\ in the case $\pf(B)>0$, one gets $\Tr (M^+)^2=\Tr(C^+)^2
\epsilon^4(\Tr(M^+)^2)^2/16\pf(M)^2$, or 
$\Tr(M^+)^2=16\pf(M)^2/\epsilon^4 \Tr(C^+)^2$.  Using this together
with the formula in \cpluss\ for $C^+$ when $\pf(B)>0$, one finally
arrives at \bpsconb\ also for $\pf(B)>0$.

The fact that the BPS condition ends up being the same (even though
the unbroken supersymmetry is completely different) for $\pf(B)>0$ or
$\pf(B)<0$ looks rather miraculous from the point of view of the
supersymmetric DBI theory.  However, in noncommutative gauge theory,
the BPS condition is that $\hat F$ should be anti-selfdual in the open
string metric; this condition, when mapped to commutative gauge theory
by \changran, is manifestly independent of the sign of the Pfaffian.

To compare \fincond\ to $\hat F^+=0$ in greater detail, it is
convenient to consider the vierbein $E$ of \vierde.  Since here we set
$g_{ij}=\epsilon \delta_{ij}$, and we are only interested in the limit
$\epsilon \rightarrow 0$, we have
\eqn\vierdea{E=  -{\sqrt \epsilon \over 2\pi\alpha'} {1\over B},}
which is finite in the limit.  Given any antisymmetric tensor
$\Lambda$, such as $B$ or $F$, denote its selfdual projection relative
to the open string metric $G= (E E^t)^{-1} $ by $\Lambda^+_G$ (as
before, we write simply $\Lambda^+$ and $B^+$ for the selfdual
projections of $\Lambda$ in the closed string metric $g$).  We have,
for example, $F^+_G=(E^t)^{-1}(E^tFE)^+E^{-1}$, the idea being that to
compute $F^+_G$, we first map $F$ to a local orthonormal frame by
$F\to E^tFE$, then take the ordinary selfdual projection, and then map
back to the original frame using the inverse vierbein.  Antisymmetric
bi-vectors like $1\over B$ are treated similarly except that we use
$(E^t)^{-1}$ instead of $E$, so $\left({1\over
B}\right)^+_G=E\left(E^{-1}{1\over B}(E^t)^{-1}\right)^+E^t$.  Since
the vierbein is proportional to $1\over B$, we first derive a useful
identity.  We do that in a frame where $B$ has special form:
\eqn\bspefor{B=\left(\matrix{0&b_1&0&0\cr -b_1&0&0&0 \cr 0&0&0&b_2 \cr
0&0&-b_2&0}\right).}
Then, for any 
\eqn\anyfa{F=\left(\matrix{0&F_{12}&F_{13}&F_{14}\cr
-F_{12}&0&F_{23}&F_{24} \cr -F_{13}&-F_{23}&0&F_{34} \cr
-F_{14}&-F_{24}&-F_{34}&0}\right)} we find
\eqn\usebfb{\eqalign{
\left({1\over B^t}F{1\over B}\right)^+&=\left(\matrix{
0&{1\over b_1^2}F_{12}&{1 \over b_1b_2}F_{24}&-{1\over
b_1b_2}F_{23}\cr
-{1\over b_1^2}F_{12}&0&-{1 \over b_1b_2}F_{14}&{1\over
b_1b_2}F_{13}\cr  
-{1\over b_1b_2}F_{24}&{1 \over b_1b_2}F_{14}&0&{1\over
b_2^2}F_{34}\cr 
{1\over b_1b_2}F_{23}&-{1 \over b_1b_2}F_{13}&-{1\over b_2^2}F_{34}&0}
\right)^+ \cr
&= {1\over 2}\left(\matrix{
0&{1\over b_1^2}F_{12}+{1\over b_2^2}F_{34}&{1 \over
b_1b_2}(F_{24}-F_{13})&-{1\over b_1b_2}(F_{23} +F_{14})\cr 
-{1\over b_1^2}F_{12}-{1\over b_2^2}F_{34}&0& -{1 \over
b_1b_2}(F_{14}+F_{23})&{1\over b_1b_2}(F_{13}-F_{24})\cr 
-{1\over b_1b_2}(F_{24}-F_{13})&{1 \over
b_1b_2}(F_{14}+F_{23})&0&{1\over b_2^2}F_{34}+{1 \over
b_1^2}F_{12}\cr  
{1\over b_1b_2}(F_{23}+F_{14})&-{1 \over
b_1b_2}(F_{13}-F_{24})&-{1\over b_2^2}F_{34}-{1 \over b_1^2}F_{12}
&0}\right)\cr
&= -{1\over b_1b_2}F^++ {b_2F_{12}+b_1F_{34}\over
b_1^2b_2^2}\left({b_1+b_2\over 2}\right)  \left(\matrix{ 0&1&0&0\cr
-1&0&0&0 \cr 0&0&0&1 \cr 0&0&-1&0}\right)\cr
&= -{1\over \pf B}F^++ {\epsilon^{ijkl}B_{ij}F_{kl}
\over 4(\pf B)^2} B^+.}}
We conclude that
\eqn\bfconce{\left({1\over B^t}F{1\over B}\right)^+=-{1\over \pf
B}F^++ {\epsilon^{ijkl}B_{ij}F_{kl} \over 4(\pf B)^2} B^+}
for any antisymmetric $B$, not necessarily of the form \bspefor.
Using the explicit form of $E$ in the present case \vierdea\ and
equation \bfconce, one computes 
\eqn\fplusloc{\eqalign{
&F^+_G=(E^t)^{-1} \left(E^tFE\right)^+ E^{-1}
=\pf E (E^t)^{-1}\left( -F^+ +{ B^+ \over 4 \pf B}
\epsilon^{ijkl}B_{ij} F_{kl}\right) E^{-1},\cr
&B^+_G=(E^t)^{-1} \left(E^tBE\right)^+ E^{-1}
=\pf E (E^t)^{-1}B^+E^{-1} , \cr
&G\left({1\over B}\right)^+_G G=(E^t)^{-1}\left(E^{-1}{1 \over
B}(E^t)^{-1}\right)^+E^{-1}= -{\sqrt{\det G}\over \pf B }B^+_G,}}
where the second equation is derived by substituting $B$ for $F$ in
the first equation, and the third equation is derived by writing $E$
and $G$ in terms of $B$ and using the second equation.
Using \fplusloc\ equation \fincond\ becomes
\eqn\finconda{F^+_G =- {1\over 8 \pf B} B^+_G \epsilon^{ijkl}F_{ij}
F_{kl}= {1\over 8 \sqrt{\det G}} G\left({1 \over B}\right)^+_G G
\epsilon^{ijkl}F_{ij} F_{kl}={1\over 4} G\theta^+_G G F \tilde F,}
which is the same as \modins.

We note that the $\epsilon \rightarrow 0$ limit of the
Dirac-Born-Infeld action \biacm\ has an infinite power series
expansion in $F$, but the condition for a BPS configuration \fincond\
or \finconda\ is polynomial in $F$ and in $\theta$.  Since the action
\biagm\ and the supersymmetry transformation laws \linsusy, \nonli\
are exact when derivatives of $F$ are neglected, the same is true in
the limit $\epsilon \rightarrow 0$.  Therefore, the action \biacm\ and
the BPS condition \fincond\ or \finconda\ are exact when such
derivatives are neglected.  We matched the action \biacm\ with the
noncommutative action \effact\ and the BPS condition \finconda\ with
the noncommutative BPS condition \modins\ at leading order in $F$, but
it is now clear that \finconda\ should agree with
\modins\ to all orders in $F$, if derivatives of $F$ are
neglected.

\bigskip\noindent{\it Classical Solution Of \finconda}

We conclude this section with a computation that is really only
offered for entertainment, as there is no solid basis for interpreting
it.  Nekrasov and Schwarz \nekrasov\ found, using the noncommutative
ADHM equations, an explicit rank one solution (with instanton number
one) of the noncommutative instanton equation $\hat F^+=0$ on $\R^4$.
What makes this interesting is that the corresponding equation $F^+=0$
in ordinary rank one (Abelian) gauge theory has no such smooth, finite
action solution on $\R^4$.

It is amusing to ask whether one can find the noncommutative rank one
solution as a classical solution of DBI theory.  There is no reason to
expect this to work, since the fields in the rank one instanton are
not slowly varying, and on the contrary it cannot work, since a
nonsingular Abelian gauge field with $F=0$ at infinity cannot have a
nonzero value of the instanton number
\eqn\huoco{\int_{{\bf R}^4}d^4x F\tilde F.}
It is nonetheless interesting to see how far we can get.  We will see
that the result is as good as possible: the solution we will find of
\finconda\ has the mildest possible singularity compatible with a
nonzero value of \huoco, and in particular has a milder singularity
than an analogous solution of the linear equation $F^+=0$.

The equation we wish to solve is
\eqn\tuggo{F_{ij}^+=\omega_{ij}^+(F\tilde F)}
for an Abelian gauge field in four dimensions, where $\omega$ is given
in \finconda\ in terms of $B$.  We take $F\tilde
F=4(F_{12}F_{34}+F_{14}F_{23}+F_{13}F_{42})$.  There is no loss of
essential generality in assuming that the nonzero components of
$\omega$ are $\omega_{12}=\omega_{34}=1$.

The noncommutative solution obtained in \nekrasov\ is invariant under
the $U(2)$ subgroup of $SO(4)$ that leaves fixed $\omega$ (or
$B^+$). So we look for a solution of $\tuggo$ with the same symmetry.
Up to a gauge transformation, we can take
\eqn\muggo{A_i=\omega_{ij}x^jh(r)}
where $r=\sqrt{\sum_i(x^i)^2}$.  Since this is the most general
$U(2)$-invariant ansatz, it must be compatible with the equations.

We compute
\eqn\compfi{\eqalign{
&F_{12}=-2h-(x_1^2+x_2^2)\cdot h'/r \cr
&F_{34}=-2h-(x_3^2+x_4^2)\cdot h'/r \cr
&F_{13}=F_{24}=(x_1x_4-x_2x_3)\cdot h'/r \cr
&F_{23}=-F_{14}=(x_2x_4+x_1x_3)\cdot h'/r }}
with $h'=dh/dr$.  It follows that
\eqn\fftie{F\tilde F=8(2h^2+hh'r).}
The only nontrivial component of the equation is the 1-2 component,
which becomes
\eqn\nonleo{-4h-rh'=16(2h^2+hh'r)}
or
\eqn\nonlet{r{d\over dr}(h+8h^2)=-4(h+8h^2).}
Hence the solution is
\eqn\soleno{h+8h^2=Cr^{-4}}
for some constant $C$.  Note that $h\sim r^{-4}$ for $r\to \infty$,
which is the correct behavior for this partial wave (``dipole'') in
Abelian gauge theory, while $h\sim r^{-2}$ for $r\to 0$, which is the
singularity of the solution.  Since
\eqn\ucis{\eqalign{
\int d^4x F\tilde F=&16\pi^2\int_0^\infty r^3
dr\left(2h^2+hh'r\right) = 8\pi^2\int_0^\infty dr {d\over
dr}\left(r^4h^2\right)\cr
=& -8\pi^2\lim_{r\to 0}(r^4h^2)=-\pi^2C,}}
the behavior near $r=0$ is exactly what is needed to give the solution
a finite and nonzero instanton number.  This contrasts with the
solution of the linear equation $F^+=0$ with the same symmetry; in
that case, $h\sim 1/r^4$ near $r=0$, and the instanton number is
divergent.

\newsec{$D$-Branes And Small Instantons In The Presence Of Constant
$B$ Field} 

As we mentioned in the introduction, one of the most interesting
applications of noncommutative Yang-Mills theory has been to
instantons, especially small instantons \refs{\nekrasov,\berkooz}.
Noncommutative Yang-Mills has been related to the possibility of
adding a Fayet-Iliopoulos term to the ADHM equations, a possibility
also seen \abs\ in the study of small instantons via $D$-branes.  In
this section, we will reexamine the $D$-brane approach to small
instantons in the context of the $\alpha'\to 0$ limit \limfi\ with
fixed open string parameters.

\nref\one{T.~Filk, ``Divergencies in a Field Theory on Quantum
Space,'' Phys. Lett. {\bf B376} (1996) 53.}%
\nref\onevgb{J.C.~Varilly and J.M.~Gracia-Bondia, ``On the ultraviolet
behaviour of quantum fields over noncommutative manifolds,''
Int.\ J.\ Mod.\ Phys.\ {\bf A14} (1999) 1305, hep-th/9804001.}
\nref\two{M.~Chaichian, A.~Demichev and P.~Presnajder, ``Quantum Field
Theory on Noncommutative Space-times and the Persistence of
Ultraviolet Divergences,'' hep-th/9812180;  ``Quantum Field Theory on
the Noncommutative Plane with E(q)(2) Symmetry,'' hep-th/9904132.}%
\nref\twopo{C.P.~Martin and D.~Sanchez-Ruiz, ``The One Loop UV
Divergent Structure of U(1) Yang-Mills Theory on Noncommutative
${\bf R}^4$,'' Phys. Rev. Lett. {\bf 83} (1999) 476, hep-th/9903077.}%
\nref\twopt{M.M.~Sheikh-Jabbari, ``Renormalizability of the
Supersymmetric Yang-Mills Theories on the  Noncommutative Torus,''
JHEP {\bf 06} (1999) 015, hep-th/9903107.}%
\nref\twopth{T.~Krajewski and R.~Wulkenhaar, ``Perturbative Quantum
Gauge Fields on the Noncommutative Torus,'' hep-th/9903187.}%
\nref\twopfo{S.~Cho, R.~Hinterding, J.~Madore and H.~Steinacker,
``Finite Field Theory on Noncommutative Geometries,''
hep-th/9903239.}%
\nref\three{E.~Hawkins, ``Noncommutative Regularization for the
Practical Man,'' hep-th/9908052.}%
For this, we have to study the $D(p-4)$-$Dp$ system for some $p\geq
3$.  The first case, $p=3$, has the advantage that  quantum
noncommutative super Yang-Mills is apparently well-defined in four dimensions,
since it seems that in general the deformation to nonzero $\theta$
does not change the renormalization properties of Yang-Mills theory
\refs{\one-\three}.  If so, the structure we will find in the small
instanton problem for $\pf(B)<0$ must be already contained in the
$\hat F^2$ theory in the small instanton regime.  However, the
$D0$-$D4$ system is richer, because one has the chance to study the
quantum mechanics on instanton moduli space, so we will focus on this.
(It takes in any case only relatively minor modifications of the
formulas to convert to $D(p-4)$-$Dp$ for other $p$.)

For simplicity, we set $g_{ij}=\epsilon \delta_{ij}$.  We consider a
$D0$-brane embedded in the $D4$-brane.  The effective Hamiltonian for
this case governs possible deformations to a $D0$-brane outside the
$D4$-brane or a non-small instanton in the $D4$-brane.  The boundary
conditions on the 0-4 open strings are
\eqn\bczf{\partial_\sigma x^0\big|_{\sigma=0,\pi}=
\partial_\tau x^{1\dots 9}\big|_{\sigma=0}=
\partial_\tau x^{5\dots 9}\big|_{\sigma=\pi}=
g_{ij}\partial_\sigma x^j + 2\pi i\alpha' B_{ij}\partial_\tau
x^j\big|_{\sigma=\pi}=0}
for $i=1\dots 4$.  We bring $B$ to a canonical form
\eqn\bform{B= {\epsilon \over 2\pi\alpha'} \left(\matrix{
0&b_1&0&0\cr
-b_1&0&0&0\cr
0&0&0&b_2\cr
0&0&-b_2&0\cr}\right).}
We will eventually be interested in the limit \limfi\ with finite $B$
and hence with $|b_I|\sim \epsilon^{-1/2}$.  Because of the mass shell
condition $\alpha' p^2/2=N$, where $p$ is the momentum in the ``time''
direction common to the $D0$ and $D4$, and $N$ is the worldsheet
Hamiltonian for the string oscillators, we see that states of finite
(spacetime) energy must have
\eqn\juducu{N\sim\alpha'\sim {1\over |b|}}
for $\epsilon\to 0$.  Excitations with higher energy than this are not
part of the limiting theory obtained in the zero slope limit.

As in section 4.2, the qualitative properties of the $\epsilon\to 0$
limit will depend on the sign of $\pf(B)$.  For large $b$, the
boundary conditions \bczf\ become Dirichlet boundary conditions at
both ends.  Those are the boundary conditions of both a supersymmetric
$D0$-$D0$ system, and a nonsupersymmetric $D0$-$\overline{D0}$ system.
We will see that for $\pf(B)>0$, the $D0$-$D4$ system behaves like
$D0$-$\overline{D0}$, and for $\pf(B)<0$, it behaves like
$D0$-$D0$. Intuitively, this is because the induced instanton number
on a $D4$-brane with a $B$-field is proportional to $-\half\int
\pf(B)$, so the $D4$-brane carries $D0$ charge if $\pf(B)<0$, and
$\overline{D0}$ charge if $\pf(B)>0$.

We define $z_1=x_1+ix_2$, $z_2=x_3+ix_4$ and express the boundary
conditions on $x^i$ as
\eqn\zbcz{\eqalign{
&\partial_\tau z_I\big|_{\sigma=0}=\partial_\sigma z_I +
b_I\partial_\tau z_I\big|_{\sigma=\pi}=0 \cr
&\partial_\tau \bar z_I\big|_{\sigma=0}=\partial_\sigma \bar z_I -
b_I\partial_\tau \bar z_I\big|_{\sigma=\pi}=0 .}}
A boson $z$ with boundary conditions \zbcz\ has a mode expansion
\eqn\zmode{\eqalign{
&z=i\sum_n \left( e^{(n+\nu)(\tau+i\sigma)} -
e^{(n+\nu)(\tau-i\sigma)}\right) { \alpha_{n+\nu}  \over n+\nu} \cr
&\bar z=i\sum_n \left( e^{(n-\nu)(\tau+i\sigma)} -
e^{(n - \nu)(\tau-i\sigma)}\right) {\bar \alpha_{-n+\nu}  \over
n - \nu} \cr}}
with 
\eqn\nubrel{e^{2\pi i \nu}=-{1+ib\over 1-ib}, \qquad 0\le \nu < 1.}
With a Euclidean worldsheet, $\bar z$ is not the complex conjugate of
$z$ because the boundary conditions \boucon\ are not compatible with
real $x^{1,2,3,4}$.  With Lorentzian signature, $\tau=it$ and $\bar
\alpha$ is the complex conjugate of $\alpha$ in \zmode.

We find
\eqn\approxnu{\nu \approx \cases{
-{1 \over \pi b} & for $b\rightarrow -\infty $\cr
{1 \over 2} +{b\over \pi} & for $b \approx 0$ \cr
1- {1\over \pi b} & for $b \rightarrow + \infty$ \cr}}
so $\nu$ changes from $0$ to $1$ as $b$ changes from minus infinity to
infinity.  The complex boson ground state energy is
\eqn\gsen{E(\nu)= {1 \over 24} - {1\over 2}\left(\nu- {1\over
2}\right)^2.}  
It is invariant under $\nu \rightarrow 1-\nu$, as it should be because
this exchanges the oscillators of $z$ with those of $\bar z$ in
\zmode. 

We are interested in the spectrum of the 0-4 strings.  In the R sector
the ground state energy is zero, and we find massless fermions.  In
the NS sector the ground state energy is
\eqn\gsenes{E_t=3E(0) +E(\nu_1)+E(\nu_2)-E(0)
-3E(1/2)-E(|\nu_1-1/2|) -E(|\nu_2-1/2|) +E(1/2),}
where the first term is from $x^{0,5\dots 9}$, the second and third
{}from $z_{1,2}$, and the fourth from the bosonic ghosts.  The other
terms arise from the corresponding fermions, whose energies differ by
$\half$ from the corresponding bosons -- they are $n\pm|\nu-\half|$.
{}From \gsenes\ we find
\eqn\gsenese{E_t=-\half\left(\left|\nu_1-\half\right|+\left|\nu_2-\half
\right|\right).}

The four lowest energy states are the ground state and three states
obtained by acting with the fermion creation operators with energies
$|\nu_I-\half|$ on the ground state.  Their energies are
\eqn\gsenese{\pm \half\left(\nu_1-\half\right) \pm \half \left(\nu_2-\half
\right).}
Of these the states with energies 
\eqn\stateene{E^+_\pm=\pm\half\left(\nu_1+\nu_2 -1\right)}
have one sign of $(-1)^F$, while those with energies
\eqn\stateenea{E^-_\pm=\pm\half\left(\nu_1-\nu_2\right),}
have the opposite sign.  The GSO projection projects out one of these
pairs.  Which pair is being projected out depends on whether we study
D0-branes or anti-D0-branes. (This is a general feature of $D$-brane
physics, exploited for instance in 
\ref\sen{A. Sen, ``Tachyon Condensation On The Brane-Antibrane
System,'' hep-th/9805170.}: 
strings ending on $D0$-branes or $\overline{D0}$-branes have the same
boundary condition but opposite GSO projection.) We use the
conventions that for D0-branes we keep the states with energies
$E^+_\pm$ and for anti-D0-branes the states with energies $E^-_\pm$.
With this choice, a D0-brane has instanton number $+1$.

For D0-branes 
\eqn\evariousli{E^+_\pm =\pm\half(\nu_1+\nu_2 -1) \approx \pm {1\over
2\pi}\cases {
b_1+b_2 & for $b_1,b_2\approx 0$ \cr
\pi-\big|{1\over b_1}+{1\over b_2}\big| & for $|b_I| \rightarrow
\infty$ with $\pf (B)>0$ \cr
{1\over b_1}+{1\over b_2} & for $|b_I| \rightarrow \infty$ with $\pf
(B)<0$ .\cr}}

For small $b_I$, a tachyon appears, meaning that turning on a small
$B\not= 0$ perturbs the small instanton problem, adding a
Fayet-Iliopoulos term to the ADHM equations.  This is in keeping, at
least qualitatively, with the proposal in \abs\ that the $B$-field
should add this term to the low energy physics.  The tachyon mass
squared is small for small $b_I$, so the $D0$-$D4$ system is almost on
shell and gives a reliable description of small instanton behavior in
this range.  For generic $b_I$, there is still a tachyon in the
$D0$-$D4$ system, suggesting that the instanton moduli space has no
small instanton singularity, but since the tachyon mass squared is not
small, the $D0$-$D4$ system is significantly off shell, and it is hard
to use it for a quantitative study of instantons.

Now let us consider the $\alpha'\to 0$ limit, or more precisely the
two cases of $|b_I|\to \infty$, with $\pf(B)>0$ or $\pf(B)<0$.  Here
we may hope for a more precise description.

For $\pf(B)>0$, the $D0$-$D4$ system is tachyonic.  The tachyon mass
squared, in units of $1/\alpha'$, is of order 1, since $E^+_{\pm}$ is
of order 1, so the tachyon ``mass'' is of order $1/\sqrt{\alpha'}$.
The tachyon mass squared is actually that of the standard
$D0$-$\overline {D0}$ tachyon, and we interpret the tachyon to mean
that the $D0$ can annihilate one of the induced $\overline
{D0}$-branes in the $D4$, along the lines of \sen. Thus, the $D0$-$D4$
system is an excitation of the $D4$ system.  Its excitation energy is
much too big to obey \juducu.  In fact, the tachyon mass squared gives
an estimate of the excitation energy required to deform a $D4$-brane
to a $D0$-$D4$ system (with an extra induced $\overline{D0}$-brane in
the $D4$), so this excitation energy corresponds to $N=|E^+_\pm|\sim
1$.  Thus, although the $D0$-$D4$ system is a possible excitation of
string theory, it is not one of the excitations of the $D4$-brane that
survives in our favorite $\alpha'\to 0$ limit \limfi\ of the open
string theory.  Hence, in particular, the $D0$-$D4$ system is not part
of the physics that will be described by the $\hat F^2$ theory if
$\pf(B)>0$.  And instanton moduli space of the $\hat F^2$ theory
should have no small instanton singularity (which would be governed
presumably by a $D0$-$D4$ system) if $\pf(B)>0$.  This last statement
is in agreement with the analysis in \nekrasov, where $\theta$ (and
hence $B$) was assumed to be self-dual, and the instanton moduli space
was found, using a noncommutative ADHM transform, to have no small
instanton singularity.

For $\pf(B)<0$, the situation is completely different.  There is still
a tachyon for generic $B$, but $N=|E^+_\pm|\sim 1/|b|\sim \alpha'$,
which means that \juducu\ is obeyed.  Hence, for $\pf(B)<0$, the
$D0$-$D4$ still represents (for generic $B$) an excitation of the
$D4$-brane with a positive excitation energy, but the energy of this
excitation scales correctly so that it survives as part of the
limiting theory for $\alpha'\to 0$.  Hence, the $D0$-$D4$ system is
part of the physics that the limiting $\hat F^2$ theory should
describe.  Moreover, if $\pf(B)<0$, it is possible to have $B^+=0$, in
which case, since $b_1=-b_2$, the tachyon mass squared vanishes.  In
this particular case, the $D0$-$D4$ system is supersymmetric and BPS
and should represent a point on noncommutative instanton moduli space.
Thus, the moduli space of noncommutative instantons should have a
small instanton singularity precisely if $B^+=0$.  The small
instantons at or very near $B^+=0$ should be described by the
$D0$-$D4$ system and the associated ADHM equations; in this
description, the Fayet-Iliopoulos term vanishes, and the small
instanton singularity appears, if $B^+=0$.

Let us now examine the excitation spectrum of the $D0$-$D4$ system for
negative $\pf (B)$ as $|b_I|\rightarrow \infty$. As we will see, this
system has, in addition to the ground state, excited states that are
also part of the limiting theory for $\alpha'\to 0$.  We assume,
without loss of generality, that $b_1 >0$ and $b_2<0$ (their product
is negative because $\pf (B)$ is negative).  Therefore $\nu_1 \approx
1-{1 \over \pi b_1}$ and $\nu_2\approx -{1\over \pi b_2}$.

Before the GSO projection, the lowest energy state $|0\rangle$ has
energy
\eqn\reallow{E^-_-=- {1\over 2} (\nu_1-\nu_2) \approx -{1\over
2} + {1\over 2\pi b_1 }- {1\over 2\pi b_2}.}
Four low energy states that survive the GSO projection are obtained by
acting on $|0\rangle$ with the lowest oscillators of the fermionic
partners of $z_I$ and $\bar z_I$, whose energies are $\nu_1-{1\over
2}\approx {1\over 2} -{1 \over \pi b_1}$, ${3\over 2}-\nu_1 \approx
{1\over 2} +{1 \over \pi b_1}$, $\nu_2+{1\over 2} \approx {1\over
2}-{1\over \pi b_2}$, ${1\over 2} -\nu_2 \approx {1\over 2}+ {1\over
\pi b_2}$.  The resulting four states have energies
\eqn\fourlowstates{\eqalign{
&E^+_{\pm} \approx \pm\left({1 \over 2\pi b_1} + {1 \over 2\pi
b_2}\right) \cr 
&E^+_1\approx {3 \over 2\pi b_1} - {1 \over 2\pi b_2} \cr
&E^+_2\approx {1 \over 2\pi b_1} - {3 \over 2\pi b_2} .\cr}}
Six more states are obtained by acting on $|0\rangle$ with the lowest
oscillators of the fermionic partners of $x^{0,5\dots 9}$, whose
energy is $1\over 2$.  These states have energy
\eqn\fourlowstates{E^+_i\approx {1\over 2\pi b_1} - {1\over 2\pi b_2}
\qquad {\rm for}~i=0,5\dots 9.}
Of these ten states the lowest are the first two, which have already
been mentioned above.  The other eight states have larger energies,
but since they have $N=E\sim 1/|b|\sim \alpha'$, they obey \juducu\
and survive as part of the limiting quantum mechanics for $\alpha'\to
0$.

We can also act on any one of these states with an arbitrary
polynomial in the lowest bosonic creation operator in $z_1$ and in
$\bar z_2$, whose energies are $1- \nu_1 \approx {1 \over \pi b_1}$
and $\nu_2\approx -{1\over \pi b_2}$.  These states again have low
enough energy to survive in the $\alpha'\to 0$ limit.  Of the states
just described, some will obey the physical state conditions, and some
will be projected out of the spectrum.

The existence of this large number of light states can be simply
understood as follows.  For infinite $|b_I|$ with negative $\pf (B)$,
our string is effectively stretched between two D0-branes.  One
$D0$-brane has a fixed position at the origin, but the second
$D0$-brane can be anywhere in the $D4$ (since a $D4$ with a strong $B$
field of $\pf(B)<0$ contains a continuous distribution of $D0$'s).
The fluctuation in position of the second $D0$-brane are given by the
$n=0$ bosonic oscillators in \zmode.  Note that for $n=0$, as $|b|\to
\infty$ or $\nu\to 0$, the exponential factor
$\left(\exp((n+\nu)(\tau+i\sigma))
-\exp((n+\nu)(\tau-i\sigma))\right)/(n+\nu)$ in \zmode\ becomes a
simple linear function of $\sigma$, describing a straight string
connecting the $D0$-brane at the origin with an arbitrary point in the
$D4$.  The fluctuation in the free endpoint of the straight string is
governed by an effective Hamiltonian which is that of a charged
particle in a magnetic field in its lowest Landau level.  In section
6, we revisit these low-lying excitations (and their analogs in other
cases) and use them to construct modules for the ring of functions on
a noncommutative space.

We would like to interpret the low lying $D0$-$D4$ excitation spectrum
that we have found as corresponding to small fluctuations around a
point-like instanton.  In ordinary Yang-Mills theory, the fluctuations
about an instanton solution are given by eigenfunctions of a small
fluctuation operator that one might describe as a generalized
Laplacian. The eigenfunctions depend on all four coordinates of
$\R^4$.  In noncommutative Yang-Mills theory, the fluctuations about
an instanton should be described by a noncommutative analog of a
Laplace operator.  We do not know how to explicitly describe the
appropriate operator directly, especially in the small instanton
limit. But we believe that near $B^+=0$, for perturbing around a small
instanton, its spectrum is given by the states we have just described
in the $D0$-$D4$ system.  These states are naturally regarded as
functions of just two of the four spacetime coordinates, something
which at least intuitively is compatible with noncommutativity of the
spatial coordinates.  For instance, charged particles in a constant
magnetic field in the first Landau level have wave functions that are
functions of just half of the coordinates.  In fact, their wave
functions are the functions of two bosons we found above.

\bigskip\noindent{\it Quantitative Analysis Of FI Coupling}

Let us consider the more general case of $k$ D4-branes and $N$
D0-branes.  The effective theory of $D0$-branes has eight
supersymmetries.  From the 0-0 strings, we get a $U(N)$ gauge group
and (in a language making manifest half of the supersymmetry) two
chiral superfields $X$ and $Y$ in the adjoint representation of
$U(N)$.  Quantization of the 0-4 strings gives two chiral superfields
$q$ and $p$ in the fundamental of $U(N)$. Their Lagrangian includes a
potential proportional to
\eqn\dzbranef{\Tr \left\{
\left([X,X^\dagger]+[Y,Y^\dagger] +qq^\dagger- p^\dagger p
-\zeta\right)^2 +\big|[X ,Y]+qp\big|^2 \right\},}
where $\zeta$ is an FI term, determined by the existence of a tachyon
mass in quantizing the 0-4 strings.  In fact, for nonzero $\zeta$ the
spectrum of the theory at the origin includes a tachyon, which should
match the tachyon found in the 0-4 spectrum (at least when $\zeta$ is
small and the $D0$-$D4$ system is almost stable).  The space of zeros
of the potential \dzbranef\ does not include the origin in this case.
This means that the pointlike instanton, corresponding to $X=Y=q=p=0$,
does not exist when $\zeta\not=0$.

Comparing \dzbranef\ at $X=Y=q=p=0$ with the analysis above of the
energy of 0-4 strings, we see that for $b_I\approx 0$ the spectrum of
the theory is consistent with
\eqn\zetabnzero{\zeta^2 \sim B^+_{ij} B^{+ij}.}
We are more interested in the limit $|b_I| \sim \epsilon^{-\half}
\rightarrow \infty$.  For positive $\pf (B)$, the tachyon mass squared
$m^2=-{E^+\over \alpha'} = -{1 \over 2\alpha'} \sim \epsilon^{-\half}$
diverges, signaling a strong instability and leading us to propose
that in the $\alpha'\to 0$ limit, for $\pf(B)=0$, the small instanton
is not part of the physics described by the $\hat F^2$ theory.  For
negative $\pf (B)$ the tachyon mass squared
\eqn\tachyonmass{m^2=-{E^+\over \alpha'} = -{1 \over 2\alpha'}
|\nu_1+\nu_2-1| \approx -{1 \over 2\pi\alpha'}\big| {1\over b_1}+{1
\over b_2}\big|= -{(\det g)^{1\over 4} \over (2\pi\alpha')^2 (\det
G)^{1\over 4} }\big|\theta_G^+\big|}
scales correctly so that this tachyon, and the $D0$-$D4$ system whose
instability it describes, can be part of the physics described by the
$\hat F^2$ theory.  In \tachyonmass, we used \fplusloc\ to express
$m^2$ in terms of $|\theta_G^+|$, where
\eqn\thetaps{\big|\theta_G^+|^2=-\Tr
G\theta_G^+G\theta_G^+.} 
This relation leads us to identify the FI term
\eqn\zetad{\zeta \sim \big|\theta_G^+|.} 

The space of zeros of \dzbranef\ is, according to \nekrasov, the
moduli space of instantons on noncommutative $\R^4$.  Our
identification of $\zeta$ in terms of the $B$-field and hence as the
noncommutativity parameter $\theta$ gives an independent derivation of
this fact.

The moduli space of instantons depends only on $\zeta$, which in our
limit is proportional to $\big|\theta_G^+|$.  This means that if
$\theta_G^+ =0$, the moduli space of noncommutative instantons is not
deformed from its value at $\theta=0$, and includes the singularity of
point-like instantons.  When $\theta_G^-\not =0$ but $\theta_G^+ =0$,
the underlying space is noncommutative, and the instanton solutions
depends on $\theta$, but the moduli space of instantons is independent
of $\theta$.

The removal of the small instanton singularity for generic $\theta$
seems, intuitively, to be in accord with the idea that the spacetime
coordinates do not commute for $\theta\not= 0$, and hence an instanton
cannot be completely localized.  However, this line of thought cannot
be pushed too far, since the small instanton singularity is present
for $\theta^+=0$.

Since $B^+$ induces negative $D0$-brane charge and instantons are
$D0$-branes, we found an instability when the $D0$-branes are
point-like that prevents them from separating from the $D4$-branes.
This suggests a relation between the problem of instantons on a
noncommutative space and $K$-theory.

\bigskip\noindent{\it Instanton Moduli Space Depends Only On
$\theta^+$} 

By now we have seen several indications that the moduli space of
noncommutative instantons depends only of $\theta^+$, not $\theta^-$.
One indication is that the explicit equation \modins\ describing the
first noncommutative correction to the instanton equation (for small
$\theta$ or $F$) depends only on $\theta^+$.  Also, by determining the
FI term from the $D0$-$D4$ system, we have just obtained a more
general argument that the moduli space depends only on $\theta_G^+$.
We now want to show that this conclusion can actually be obtained
using the methods in \nekrasov\ for direct analysis of noncommutative
instantons via the ADHM construction.  In effect, this gives a direct
mathematical argument for identifying the FI term with $\theta_G^+$.
(In our terminology, the analysis in \nekrasov\ is done in the open
string metric $G$ -- the right metric for the $\hat F^2$ action that
they work with -- and in extending their reasoning below, we work
entirely in this metric.  We will work in coordinates in which the
metric $G$ is $\delta_{ij}$.)

Given any $\theta^{ij}$ on ${\bf R}^4$, we can always pick a complex
structure on ${\bf R}^4$, with complex coordinates $z_0$ and $z_1$,
such that the nonzero commutators are
\eqn\kopo{\eqalign{ [z_0,\bar z_0] &= -{\zeta_0}\cr
                    [z_1,\bar z_1] & =-{\zeta_1}.\cr}}
This can be done in such a way that the instanton equation says that
$\widehat F^{2,0}=\widehat F^{0,2}=0$ (that is, the $(2,0)$ and
$(0,2)$ parts of $\widehat F$ vanish) and $\widehat F_{0\bar
0}+\widehat F_{1\bar 1}=0$.  (The complex structure with these properties
is the one
for which the metric $G$ is Kahler, and $\theta$ is of type $(1,1)$.)

In the study of noncommutative instantons in \nekrasov, only the case
$\zeta_0=\zeta_1=\zeta/2$ (or in other words $\theta^-=0$) was
considered.  But it is straightforward to repeat the computation in
greater generality and show that the moduli space only depends on the
sum $\zeta_0+\zeta_1$, or in other words it only depends on
$\theta^+$.  The key step in \nekrasov\ is the verification of the
second part of eqn. (3.6) of that paper (namely
$\tau_z\tau_z^\dagger=\sigma_z^\dagger\sigma_z$), where the $\tau$'s
and $\sigma$'s were defined earlier in their eqn. (2.2).  The
classical ADHM construction is formulated in eqn (2.1) of \nekrasov,
with $\mu_r=\mu_c=0$, and assumes that the $z$'s commute.  If one
wants to turn on nonzero $\mu$'s, one can (by a rotation) assume that
$\mu_c=0$ and turn on only $\mu_r$.  The basic idea in \nekrasov\ is
that if $\mu_r\not= 0$, one can compensate for this by letting the
$z$'s and $\bar z$'s no longer commute.  It is shown in this paper
that in verifying the key equation
$\tau_z\tau_z^\dagger=\sigma_z^\dagger\sigma_z$, a $c$-number term
coming from the commutator of a $z$ and a $\bar z$ can cancel the
contribution of $\mu_r$.  Indeed, the term coming from the $[z,\bar
z]$ commutators is in general $[z_0,\bar z_0]+[z_1,\bar z_1]$, so the
value of $\mu_r$ that one needs depends only on the sum
$\zeta_0+\zeta_1$.  In particular, the moduli space of noncommutative
instantons, as determined from the ADHM construction, only depends on
the sum $\zeta=\zeta_0+\zeta_1$, as we wished to show.

Although the moduli space only depends on $\zeta$, the instanton
solutions themselves (whose construction is explained in \nekrasov)
depend on both $\zeta_0$ and $\zeta_1$.  Indeed, the components of the
instanton connection take values in an algebra that depends on both
$\zeta_0$ and $\zeta_1$, so it is hard to compare gauge fields for
different $\zeta$'s.  (There is a notion of background independence
for noncommutative gauge fields, which we discuss in
section 3.2, but it involves a transformation that adds a constant to
the curvature measured at infinity, so it is not directly relevant to
comparing instantons on ${\bf R}^4$ with different values of the
$\zeta$'s.)

\newsec{Noncommutative Gauge Theory on a Torus}

In this section we consider $D$-branes compactified on a $p$-torus
$\T^p$ in our usual limit of $\alpha'\to 0$ with the open string
parameters $G$, $\theta$ fixed.  Surprisingly, the effective field
theory based on the $\hat F^2$ action inherits the $T$-duality symmetry
of the underlying string theory.  This is surprising because without
the $B$-field, the effective theory based on the $F^2$ action is not
invariant under $T$-duality.

This $T$-duality appeared in the mathematical literature as Morita
equivalence of different modules.  In the physics literature it was
explored in \refs{\connes,\schwarz-\piosch} using the DLCQ description
of M-theory.  We will devote section 6.1 to  deriving this duality
using our point of view.  It will not involve the DLCQ description of
M-theory, but instead, will use the zero slope limit.  (In section 7
we will explain how these two approaches are related.)  A crucial
element of our discussion will be the difference between the closed
string parameters $g$, $B$ and $g_s$ and the open string parameters
$G$, $\theta$ and $G_s$.  

It is important that unlike $T$-duality of string theory, which is the
reason for $T$-duality of these theories, here we do not relate a theory
on a torus to a theory on the dual torus.  In particular, we do not
have the standard exchange of momentum modes and winding modes.  Since
we are studying open strings there are no winding modes.  Instead, in
open string theory $T$-duality has the effect of changing the
dimensionality of the $D$-branes.  Therefore, this $T$-duality acts on the 
$D$-brane charges.  It changes the rank of the gauge fields and their
topological charges.

In subsections 6.2-6.4, we will interpret  many mathematical
results about construction of modules over a noncommutative torus
and Morita equivalences between them in terms of standard techniques of 
quantizing open strings.

\subsec{$T$-duality}

We start by deriving the $T$-duality of the theories using our point of
view.  We consider Dp-branes on $T^p$ parametrized by $x^i \sim
x^i+2\pi r$ with the closed string metric $g$.  The periods of the $B$
field are $(2\pi r)^2B$ and this motivates us to express the
noncommutativity in terms of the dimensionless matrix $\Theta = {1
\over 2\pi r^2} \theta$.

The $SO(p,p,\Z)$ $T$-duality group is represented by the matrices
\eqn\deftm{T=\left(\matrix{
a&b \cr
c&d}\right),}
where $a$, $b$, $c$ and $d$ are $p\times p$ matrices with integer
entries.  $T$ satisfies
\eqn\soppr{T^t \left(\matrix{0&1 \cr
1&0}\right) T=\left(\matrix{0&1 \cr
1&0}\right),}
and therefore
\eqn\varrela{\eqalign{
&c^ta+a^tc=0\cr
&d^tb+b^td=0\cr
&c^tb+a^td=1.\cr}}
$T$ acts on
\eqn\defea{E={r^2 \over \alpha'} (g + 2\pi \alpha' B)}
and the string coupling $g_s$ as
\ref\tdualityaction{A.~Giveon, M.~Porrati and E.~Rabinovici,
``Target space duality in string theory,''
Phys.\ Rept.\ {\bf 244} (1994) 77, hep-th/9401139.}
\eqn\tactione{E'=(aE+b) {1\over cE+d}}
\eqn\tactiong{g_s'=g_s\left({\det g' \over \det g} \right)^{1 \over
4}.}
Using \varrela, \defea\ and \tactione\
\eqn\gtransf{g'= {\alpha' \over 2r^2}(E'+ (E')^t)=
{1 \over (cE+d)^t}g{1 \over cE+d},}
and \tactiong\ becomes
\eqn\taction{g_s'={g_s \over \det(cE+d)^\half}.}

We are interested in the action of $T$ on the open string parameters
$G$, $\Theta$ and $G_s$.  For simplicity we ignore the more general
variables \changevarg\ including $\Phi$.  The role of $\Phi$ in
$T$-duality was elucidated in \piosch. Using
\eqn\gthetaerel{{1 \over E}={\alpha'\over r^2} G^{-1}+ \Theta}
and \varrela\ we find
\eqn\biggpr{G'= {2\alpha' \over r^2} \left({1 \over E'}+ {1 \over
(E')^t}\right)^{-1} =  (a+bE^{-1})G (a+bE^{-1})^t.} 
Similarly
\eqn\thetaprel{\Theta'={1\over 2}  \left({1 \over E'}- {1 \over
(E')^t}\right)= \left[(c+dE^{-1}) {1 \over a+bE^{-1}}\right]_A,}  
where $[~]_A$ denotes the antisymmetric part.  Finally,
\eqn\biggsdef{G_s= g_s \left({ \det \left({\alpha'\over r^2} E\right)
\over \det g }\right)^{1\over 2}}
transforms to
\eqn\biggstrt{G_s'=G_s\left[\det(a+bE^{-1})\right]^{1\over 2},}
and therefore ${ \sqrt {\det G} \over G_s^2}$ is $T$-duality invariant.
{}From \biggstrt\ we find the transformation law of the Yang-Mills
gauge coupling
\eqn\biggstr{g_{YM}'=g_{YM}\left[\det(a+bE^{-1})\right]^{1\over 4},}

In the zero slope limit with finite $r$ we have $E^{-1}\approx \Theta$
and \biggpr, \thetaprel, \biggstr\ become
\eqn\tdualze{\eqalign{
&G'= (a+b\Theta ) G (a+b\Theta )^t \cr 
&\Theta'= (c+d\Theta) {1 \over
a+b\Theta} \cr 
&g_{YM}'=g_{YM}\left[\det(a+b\Theta)\right]^{1\over 4}}} 
(it is easy to check using the antisymmetry of $\Theta$ and the
relations \varrela\ that $\Theta'$ is antisymmetric).  These
transformation rules have already appeared in the mathematical
literature.  In the physics literature they appeared in \refs{\connes,
\schwarz-\piosch}.  Our expressions \tdualze\ are similar to those in
these papers but differ by $a\leftrightarrow d$ and
$b\leftrightarrow c$; i.e. by conjugation by $T=\left(\matrix{0&1\cr
1&0}\right)$.  In section 7 we will explain the origin for this
difference. 

We note that the transformation of the metric $G$ is unlike the
typical $T$-duality transformation of a metric (like \gtransf).  Since
it is linear in $G$, for every $T$-duality transformation the transformed
metric $G'$ scales like the original metric $G$.  If one of them
scales to zero, so does the other one; for example, there is no
transformation which maps $G$ to $G^{-1}$.  Although for closed
strings there are transformations which map momentum modes to winding
modes, this is not true for the open strings we consider; yet the
theory is $T$-duality invariant!

If some of the components of $\Theta$ are rational, they can be
transformed to zero.  For example, if $\Theta= -d^{-1}c$, where $c$
and $d$ have integer entries, there is a T dual description with
$\Theta'=0$.  It is given by $T=\left(\matrix{a&b\cr c&d}\right)$ with
appropriate $a$ and $b$.  $\Theta'$ vanishes because (using \varrela)
$a+b\Theta=a-bd^{-1}c=(d^t)^{-1}$ is invertible.  This also guarantees
that the transformed metric $G'=d^{-1}G(d^{-1})^t$ and Yang-Mills
coupling constant $g_{YM}'=g_{YM}'(\det d)^{-{1\over 4}}$ are finite.
In this case the noncommutative theory with nonzero $\Theta$ is T dual
to a commutative theory.  The volume of the torus of this dual
commutative theory is smaller by a factor of $\det d$ relative to the
original torus.  More generally, if only some of the components of
$\Theta$ are rational some of the coordinates could be transformed to
commuting coordinates.

There is another point we should mention about the case with rational
$\Theta$.  Then, there exist $T$-duality transformations for which
\tdualze\ is singular.  In particular, for $a+b\Theta=0$ the
transformed $G$ vanishes and the transformed $\Theta$ diverges.  One
way to understand it is to first use a $T$-duality transformation, as
above, to transform to $\Theta=0$.  Then, all the transformations with
$a=0$ are singular.  They include the transformation with $a=d=0$,
$b=c=1$ which can be referred to as ``$T$-duality on all sides of the
torus.''  More generally, if only some of the components of $\Theta$
are rational, there exist $T$-duality transformations to $\Theta$ with
infinite entries and to $G$ with vanishing eigenvalues.  We conclude
that when $\Theta$ is rational not all the $SO(p,p;\Z)$ duality group
acts.

This discussion becomes more clear for the simplest case of the two
torus.  Then, the $T$-duality group is $SO(4,4;\Z) \cong SL(2,\Z)
\times SL(2,\Z)$.  One $SL(2,\Z)$ factor acts geometrically on $G$
leaving $\Theta$, $g_{YM}$ and the volume $V$ of the torus with the
metric $G$ unchanged.  The other $SL(2,\Z)$ acts as
\eqn\tdualzet{\eqalign{
&V'=  V(a+b\Theta )^2 \cr 
&\Theta'= {c+d\Theta \over a+b\Theta} \cr 
&g_{YM}'=g_{YM}(a+b\Theta)^\half,}}
where now $\Theta$, $a$, $b$, $c$ and $d$ are numbers, rather than
matrices as above, and $\left(\matrix{a&b\cr c&d}\right)$ is an
$SL(2,\Z)$ matrix.  We again see that when $\Theta=-c/d$ is rational it
can be transformed to $\Theta'=0$ by an $SL(2,\Z)$ transformation
$\left(\matrix{a&b\cr c&d}\right)$  with
appropriate $a$ and $b$.  For such a transformation $a+b\Theta= 1/d$,
and therefore $V'=V/d^2$ and $g_{YM}'=g_{YM}/d^\half$ are finite.
However, the transformation $\left(\matrix{c&d\cr -a&-b}\right)$
does not act regularly.

The fact that for rational $\Theta=-c/d$ the theory is equivalent to
another theory on a commutative torus whose volume is smaller by a
factor of $d^2$ can be understood as follows.  Before the $T$-duality
transformation the torus is parametrized by $x^i \sim x^i+2\pi r$.
The algebra of functions on the torus is generated by $U_i=e^{i
x^i/r}$, which satisfy $U_1U_2=e^{-2\pi i\Theta}U_2U_1$.  The
$T$-duality transformation does not act on the complex structure of the
torus but affects its volume.  This can be achieved by rescaling the
two coordinates $x^i$ such that the identification is $x^i \sim
x^i+2\pi r/d$, thus reducing the volume by a factor of $d^2$.  Now the
algebra of functions on the torus is generated by $\tilde U_i=e^{i
x^id/r}= U_i^d$.  They satisfy $\tilde U_1 \tilde U_2=e^{-2\pi id^2
\Theta}\tilde U_2 \tilde U_1=
\tilde U_2 \tilde U_1$; i.e.\ the new torus is commutative.

We stress that this transformation to an ordinary theory on another
torus is unrelated to the transformation to ordinary gauge fields we
exhibited in section 3, which exists also on $\R^n$ and on a torus for
any $\Theta$, not necessarily rational.  The transformation in section
3 does not act on the space the theory is formulated on -- if the
theory is formulated on a torus, this transformation maps us to another
theory formulated on the same torus.  This transformation also  does
not change the rank of the gauge group, but it converts the simple
$\hat F^2$ action to a complicated non-polynomial action.  The
transformation we discuss here for rational $\Theta$ acts on the
torus, changes the rank of the gauge group and maps the simple $\hat
F^2$ to a simple $F^2$ action.

One might be concerned that because of the special properties of the
zero slope limit with rational $\Theta$ the noncommutative theory
behaves in a discontinuous fashion as a function of $\Theta$.
Furthermore, for generic $g$ and $B$ their map into the fundamental
domain varies ergodically in the zero slope limit\foot{This fact was
mentioned in
\ref\moorefin{G.~Moore, ``Finite in All Directions,''
hep-th/9305139.}, 
where the role of noncommutative geometry was anticipated.}.  However,
these discontinuities affect only the closed strings, which decouple
in the limit.  The open string parameters $G$, $\Theta$ and $G_s$
transform smoothly under $T$ duality.  Correspondingly, the open
string dynamics varies smoothly with $\Theta$ for fixed $G$ in the
zero slope limit.

\subsec{Modules Over A Noncommutative Torus}

\nref\concon{A. Connes, ``${\bf C}^*$-alg\'ebres et G\'eom\'etrie
Diff\'erentielle,'' C. R. Acad. Sc. Paris {\bf 290} (1980) 559.}%
\nref\rieffel{M. Rieffel, `` `Vector Bundles' Over Higher Dimensional
`Noncommutative Tori,' '' Proc. Conference Operator Algebras,
Connections With Topology And Ergodic Theory, Lecture Notes In Math. 
{\bf 1132} (Springer-Verlag, Berlin, Heidelberg, 1985) 456.}%
Our goal in the rest of this section is to
construct and analyze directly from
quantization of open strings a natural class of representations for
the algebra ${\cal A}$ of functions on a noncommutative torus.
Mathematically, representations of a ring are usually called modules.
We will aim to understand in physical terms the usual
mathematical constructions \refs{\concon,\rieffel,\connes,\schwarz}
of projective modules over a noncommutative
torus, and the ``Morita equivalences'' between them, which \refs{\connes,
\schwarz-\everl} are intimately related to $T$-duality.

For simplicity, we will discuss bosonic strings.  (Incorporating
supersymmetry does not affect the ring ${\cal A}$ and so does not
alter the modules.)  Consider a $p$-brane wrapped on the torus.  The
ground states of the $p$-$p$ open strings are tachyons.  For every
function $f$ on $\T^p$, there is a corresponding tachyon vertex
operator $\O_f$.  In the limit $\alpha'\to 0$, the dimensions of the
$\O_f$'s vanish.  The operator product algebra of the $\O_f$'s
reduces, as we have seen in section 2, to the $*$-product of
functions on $\T^p$, essentially
\eqn\ruggo{\O_f(t)\cdot \O_g(t')\to \O_{f*g}(t'),~~{\rm
for}\,\,\,t>t'.} 

A $p$-$p$ open string has a world-sheet with topology $\Sigma= I\times
\R$, where $\R$ is a copy of the real line parametrizing the proper
time, and $I=[0,\pi]$ is an interval that parametrizes the string at
fixed time.  The algebra ${\cal A}$ of tachyon vertex operators can be
taken to act at either end of the open string.  Operators acting on
the left of the string obviously commute with those acting on the
right, and the open string states form a bimodule for ${\cal A}\times
{\cal A}$.  By a bimodule for a product of rings $\A\times \A'$, we
mean a space that is a left module for the first factor ${\cal A}$,
and a right module for the second factor $\A'$, with the two actions
commuting\foot{A left module for a ring ${\cal A}$ is a set $M$ on
which ${\cal A}$ acts, obeying a condition that will be stated
momentarily.  For $a\in {\cal A}$ and $m\in M$, we write $am$ for the
product of $a$ with $m$.  The defining property of a left module is
that for $a,b\in {\cal A}$ and $m\in M$, one has $(ab)m=a(bm)$.  In
the case that $M$ is a right module, the action of $\A$ on $M$ is
usually written on the right: the product of $a\in \A$ with $m\in M$
is written $ma$.  The defining property of a right module is that
$m(ba)=(mb)a$.}.  That the strings are a left module for the first
factor and a right module for the second can be seen as follows.  The
interaction of open strings with the $B$-field comes from a term
\eqn\toho{-{i\over 2}\int_\Sigma \epsilon^{ab}B_{ij}\partial_ax^i
\partial_bx^j=-{i\over 2}\int_{\{0\}\times \R}B_{ij}x^i{dx^j\over dt}
+{i\over 2}\int_{\{\pi\}\times \R}B_{ij}x^i{d x^j\over dt}.}
The relative minus sign between the two boundary terms in \toho\ means
that if for vertex operators inserted on $\{0\}\times \R$, \ruggo\
holds with some given $\theta$, then for operators inserted on
$\{\pi\}\times \R$, the same OPE holds with $\theta$ replaced by
$-\theta$.  Changing the sign of $\theta$ is equivalent to reversing
the order of multiplication of functions, so if the conventions are
such that \ruggo\ holds as written for vertex operators inserted at
the left end of the string, then for operators inserted at the right
end we have
\eqn\noho{{\cal O}_f(t)\cdot {\cal O}_g(t')\to {\cal O}_{g*f}(t'),~~
{\rm if}~t>t'.}
Comparing to the definition in the footnote, we see that the open
string states form a left module for ${\cal A}$ acting on the left end
of the open string, and a right module for ${\cal A}$ acting on the
right end of the open string.

In the limit $\alpha'\to 0$, the excited string states decouple, and
we can get an $\A\times \A$ bimodule $M$ by just taking the open
string ground states.  In fact, this module is simply a free module,
that is $M=\A$, since the open string ground states are a copy of
$\A$.

We can construct many other left modules for $\A$ as follows.
Consider an arbitrary boundary condition $\gamma$ for open strings on
$\T^p$ with the given open string parameters $G$ and $\theta$.
(Physically, $\gamma$ is determined by a collection of $Dq$-branes for
$q=p,p-2,\dots$.)  Then consider the $p$-$\gamma$ open strings, that
is the strings whose left end is on a fixed $p$-brane and whose right
end obeys the boundary condition $\gamma$.  The $p$-$p$ algebra $\A$
acts on the $p$-$\gamma$ open string ground states for any given
$\gamma$, giving a left module $M_\gamma$ for $\A$.  In section 6.3,
we will construct the usual projective left modules for $\A$ in this
way.  In section 6.4, we examine theoretical issues connected with
this construction.

\subsec{Construction Of Modules}

We turn now to detailed description of the modules.
For brevity of exposition, we will concentrate on modules over a
noncommutative $\T^2$.  The generalization to $\T^p$ does not involve
essential novelty for the type of modules we will construct, which are
obtained from constant curvature connections over an ordinary torus.
(For $p\geq 4$, there should be additional modules constructed from
instantons, but we are not able to describe them very concretely.)

The constructions all start with ordinary actions for open strings on
an ordinary $\T^2$, in the presence of a $B$-field, with twobrane
boundary conditions on the left end of the string and varying boundary
conditions on the right.  We consider four cases for the boundary
conditions on the right of the open string: {\it (i)} twobrane
boundary conditions; {\it (ii)} zerobrane boundary conditions; {\it
(iii)} open strings ending on a twobrane with $m$ units of magnetic
flux; and finally {\it (iv)} the general case of open strings ending
on a system of $n$ twobranes with $m$ units of magnetic flux.
Quantization of the open strings will give standard modules over a
noncommutative $\T^2$.  These modules have been described and used in
section 3.2 of \connes, and reconsidered in \refs{\zumino-\piosch}. 

\bigskip\noindent{\it (i) Twobrane Boundary Conditions}

Naively, the limit $\alpha'\to 0$ can be taken by simply dropping the
kinetic energy term from the open string action.  This leaves only
boundary terms, which are determined by the interaction with the
$B$-field.  We describe the torus by angular coordinates $x^i$,
$i=1,2$, with $0\leq x^i\leq 2\pi$, and we describe the open string
worldsheet by functions $x^i(\sigma,\tau)$, where $\tau$ is the proper
time and $\sigma$, which ranges from 0 to $\pi$, parametrizes the
position along the open string.  If we set $x^i=x^i(0,\tau)$, $\tilde
x^i=x^i(\pi,\tau)$, then the boundary terms in the action become
\eqn\kixno{L_B=-{i\over 2}\int dt B_{ij}x^i{d x^j\over dt} +{i\over
2}\int dt B_{ij}\tilde x^i{d{\tilde x^j}\over dt}.}
If these were the only variables and $L_B$ the full action, then $x^1$
would be the canonical conjugate of $x^2$ -- and similarly $\tilde
x^1$ and $\tilde x^2$ would be canonically paired -- so the classical
phase space would be a copy of $\T^2\times \T^2$.  This cannot be the
full answer since ($\T^2$ being compact) the quantization would be
inconsistent unless $\int_\Sigma B$ is an integral multiple of $2\pi$.
In fact, as we will now see, the phase space is $\T^2\times \R^2$.

We must remember the string connecting the two endpoints.  The
ordinary kinetic energy of the string is
\eqn\nixno{L_K={1\over 4\pi \alpha'}\int_\Sigma
g_{ij}\partial_ax^i\partial^a x^j.}
Here we recall that in the $\alpha'\to 0$ limit with fixed open string
metric, $g_{ij}$ is of order $(\alpha')^2$, so $L_K$ formally
vanishes.  For a given set of endpoints, the contribution to the
energy coming from $L_K$ is minimized by a string that is a geodesic
{}from $x^i$ to $\tilde x^i$.  In each homotopy class of paths from
$x^i$ to $\tilde x^i$, there is a unique geodesic.  The fluctuations
around the minimum involve modes with mass squared of order $1/\alpha'$.
Since we are not interested in such high energy excitations, we can
ignore the fluctuations and identify the phase space as consisting of
a pair of points $x^i$ and $\tilde x^i$ together with a straight line
(or geodesic) connecting them.  (We can also reach this conclusion by
just formally setting $\alpha'=0$ and dropping $L_K$ completely.  Then
there is a gauge invariance under arbitrary variations of
$x^i(\sigma,\tau)$ keeping the endpoints fixed.  We can use this gauge
invariance to fix a gauge in which $x^i(\sigma,\tau)$ is a geodesic.)

Now we write
\eqn\kikop{\eqalign{x^i & = y^i+\half s^i \cr
                    \tilde x^i & = y^i -\half s^i.\cr}}
Here $y^i$ is $\T^2$-valued, but $s^i$ is real-valued.  In fact, $y^i$
is the midpoint of the geodesic from $x^i$ to $\tilde x^i$, while the
real-valuedness of $s^i$ enables us to keep track of how many times
this geodesic wraps around $\T^2$.

The symplectic structure derived from $L_B$ is
\eqn\hikop{\omega=Bdx^1\wedge dx^2-Bd\tilde x^1\wedge d\tilde x^2
=B\left(ds^1\wedge dy^2-ds^2\wedge dy^1\right),}
where we set $B_{12}=B$.  In canonical quantization, we can take the
$y^i$ to be multiplication operators, and identify the $s^i$ as the
canonical momenta:
\eqn\jikop{\eqalign{
s^1=&-{i\over B}{\partial\over\partial y^2}=2\pi i{\Theta}{\partial
\over\partial y^2}\cr                     
s^2=&{i\over B}{\partial\over\partial y^1}=-2\pi i{\Theta}
{\partial\over\partial y^1}.\cr}}
Here we have set as in the discussion of $T$-duality $\Theta=-1/2\pi B$.
(Since we are studying a two dimensional situation, $\Theta$ like $B$
is a number.)  The physical Hilbert space thus consists of functions
on an ordinary $\T^2$, with coordinates $y^i$.  The algebra $\A$ of
functions on the noncommutative $\T^2$ is generated by
\eqn\ujju{\eqalign{U_1 & = \exp({i x^1})
=\exp( i y^1-\pi \Theta(\partial /\partial y^2)), \cr
U_2 & = \exp({ i x^2}) =\exp( i y^2+\pi\Theta(\partial/\partial y^1)).
\cr}}
They obey
\eqn\huddu{U_1U_2=e^{-2\pi i\Theta}U_2U_1.}
The commutant of $\A$ is generated by
\eqn\bujju{\eqalign{
\tilde U_1=& \exp({i \tilde x^1}) = \exp(iy^1+\pi\Theta(\partial
/\partial y^2))\cr 
\tilde U_2= &  \exp({i \tilde x^2}) =
\exp(iy^2-\pi\Theta(\partial/\partial y^1)).}}   
These operators obey 
\eqn\bujju{\tilde U_1\tilde U_2=e^{-2\pi i\tilde \Theta}\tilde
U_2\tilde U_1}
with 
\eqn\tiltet{\tilde \Theta=-\Theta.}
The relative minus sign between $\Theta$ and $\tilde \Theta$ means, as
we have explained in section 6.2, that the open strings are an $\A\times \A$
bimodule (left module for the first factor, acting at $\sigma=0$, and
right module for the second factor, acting at $\sigma=\pi$).  The
formulas that we have arrived at are standard formulas for a free
$\A\times \A$ bimodule.

\bigskip\noindent{\it (ii) Zerobrane Boundary Conditions}

Now, without changing the boundary conditions at $\sigma=0$, we
replace the boundary conditions at $\sigma=\pi$ by zerobrane boundary
conditions.  For example, we can place the zerobrane at the origin and
take the boundary condition at $\sigma=\pi$ to be $\tilde x^i=0$.  The
phase space therefore consists now of a point $x^i$ on $\T^2$ together
with a geodesic from that point to $x^i=0$. A shift $x^i\to x^i+2\pi
n^i $ (with integers $n^i$) acts freely on the phase space, changing
the winding number of the geodesic.  We can forget about the geodesic
if we consider the $x^i$ to be real-valued.
 
The phase space is thus a copy of $\R^2$ with symplectic form
\eqn\tovoo{\omega = B\,dx^1\wedge dx^2.}
To quantize, we can take $x^2$ to be a multiplication operator and
\eqn\ovoo{x^1=-i{1\over B}{\partial\over \partial x^2}=2\pi i{\Theta
}{\partial\over \partial x^2}.}
Hence, the algebra $\A$ of functions on the noncommutative $\T^2$
is generated by
\eqn\povoo{\eqalign{U_1 & = \exp(
{i x^1})=\exp(-2\pi \Theta\partial/\partial x^2)\cr
U_2 & = \exp({ i x^2}),}}
again obeying 
\eqn\ciirel{U_1U_2=e^{-2\pi i\Theta}U_2U_1.}

The commutant of the $U_i$ is generated by
\eqn\jovoo{\eqalign{\tilde U_1= &\exp({i x_1/\Theta})= \exp(-2\pi
\partial/\partial x^2) \cr 
\tilde U_2 = &  \exp({i x_2/\Theta}).\cr}}
We note that unlike $U_i$ which are invariant under $x_i\to x_i +2\pi$,
$\tilde U_i$ are not invariant under this shift.  Yet, they are
valid operators on our Hilbert space because $x_i$ are points in
$\R^2$, rather than in $\T^2$.
These operators obey 
\eqn\ciialg{\tilde U_1\tilde U_2 = e^{-2\pi i\tilde \Theta}\tilde
U_2\tilde U_1,}
where
\eqn\ovoo{\tilde \Theta = {1\over \Theta}.}
The formulas are again standard, and the interpretation is as follows
\connes.  With twobrane boundary conditions, the vertex operators at
$\sigma=\pi$ generate, as we saw above, a noncommutative torus with
$\Theta'=-\Theta$.  The $T$-duality transformation that converts
twobranes to zerobranes acts on $\Theta'$ by $\Theta'\to -1/\Theta'$,
giving us a noncommutative torus $\A'$ with noncommutativity parameter
$\tilde\Theta=-1/\Theta'=1/\Theta$.

In physical terms, the algebra $\A'$ is the algebra of ground state
0-0 strings acting on the 2-0 open strings at $\sigma=\pi$.  The
relevant 0-0 strings are open strings that wind on a geodesic around
the torus, starting and ending at the origin.  That these open strings
in the small volume limit generate the algebra of a noncommutative
torus was the starting point \refs{\connes, \douglas,\cheung} in
applications of noncommutative Yang-Mills theory to string theory.
{}From the point of view of the present paper, the statement can be
justified by computing the OPE's of the 0-0 open strings, which are
equivalent by $T$-duality to the OPE's of 2-2 tachyon vertex
operators.  We studied the 2-2 OPE's in section 2.1.

\bigskip\noindent{\it (iii) $(1,m)$ Boundary Conditions}

Now we consider a more general case: an open string that at $\sigma=0$
has the same twobrane boundary conditions as before while at
$\sigma=\pi$ it terminates on a twobrane that carries $m$ units of
zerobrane charge.  The zerobrane charge can be incorporated by placing
on the twobrane in question a magnetic field of constant curvature
$m/2\pi$.  The boundary terms in the action become now
\eqn\kikxno{L_B=-{i\over 2}\int dt B_{ij}x^i{d x^j\over dt} +{i\over
2}\int dt \left(B_{ij}+{m\over 2\pi}\epsilon_{ij}\right)\tilde
x^i{d{\tilde x^j}\over dt} .}
(Here $\epsilon_{ij}$ is an antisymmetric tensor with
$\epsilon_{12}=1$.)  The symplectic structure is
\eqn\hygo{\omega=Bdx^1\wedge dx^2-\left(B+{m\over 2\pi}\right)
d\tilde x^1\wedge d\tilde x^2.}
While the algebra acting at $\sigma=0$ is a noncommutative torus
with $\Theta=-1/2\pi B$, the algebra acting at $\sigma=\pi$ is
now a noncommutative torus $\A'$ with
\eqn\gulgo{\tilde \Theta = {1\over 2\pi B+m}.}
No essentially new computation is required here; the algebra at
$\sigma=\pi$ is determined in the usual way in terms of the boundary
conditions at $\sigma =\pi$.  (At $m=0$, we had in {\it (i)} above
$\tilde\Theta=-\Theta =1/2\pi B$; replacing $B$ by $B+m/2\pi$ gives
\gulgo.)  We can write this as
\eqn\gulgo{\tilde\Theta={\Theta'\over 1+m\Theta'}}
where $\Theta'=-\Theta$ is the noncommutativity parameter we found in
{\it (i)} above for the algebra at $\sigma=\pi$.  $\tilde\Theta$ is
obtained from $\Theta'$ by the $T$-duality transformation that maps
$(1,0)$ boundary conditions (twobrane charge 1 and zerobrane charge 0)
to $(1,m)$.  In fact, in the zero area limit, the modular parameter
$\tau= 2\pi B +i({\rm Area})$ reduces to $\tau=2\pi B$.  The modular
transformation that maps $(1,0)$ to $(1,m)$ is
\eqn\hulgo{2\pi B\to 2\pi B+m,}
which in terms of $\Theta' = 1/2\pi B$ is 
\eqn\julgo{\Theta'\to {\Theta'\over 1+m\Theta'}.}

Of course, we can also construct explicitly the $\A\times \A'$
bimodule by quantizing the open strings. (The details are a bit
lengthy and might be omitted on first reading.) For this, we set
\eqn\kulgo{\eqalign{ x^i& = y^i +\lambda s^i \cr
                     \tilde x^i& = y^i+(1+\lambda)s^i,\cr}}
$y^i\in \T^2$ and $s^i$ real-valued.   If
\eqn\ilgo{{m\over 2\pi}\lambda^2+(2\lambda+1)\left(B+{m\over
2\pi}\right) =0,}
then the symplectic form in these variables has no $ds^1\wedge ds^2$
term, and reads
\eqn\tomega{\omega= -\left({m\over 2\pi}\lambda+(B+{m\over
2\pi})\right) \left(ds^1\wedge dy^2-ds^2\wedge dy^1\right)-{m\over
2\pi}dy^1\wedge dy^2.} 
A further rescaling
\eqn\oflego{s^i= -w t^i} with \eqn\ryton{w={1\over {m\over 2\pi}\lambda+
  \left(B+{m\over 2\pi}\right)}}
gives
\eqn\romega{\omega=dt^1\wedge dy^2-dt^2\wedge dy^1-{m\over
2\pi}dy^1\wedge dy^2 .}
Because there is no $dt^1\wedge dt^2$ term,  the $y^i$ commute
and can be represented by multiplication operators.  The remaining
commutators can be represented by taking $t^1=-iD/Dy^2$, $t^2=iD/Dy^1$,
where 
\eqn\tomega{\left[{D\over Dy^1},{D\over Dy^2}\right] = i{m\over 2\pi}.}
Hence, the Hilbert space is the space of sections of a line bundle
over $\T^2$ with first Chern class $m$.  The algebra $\A$ that acts at
$\sigma=0$ is generated by
\eqn\typop{\eqalign{U_1&=\exp(ix^1)= \exp\left(iy^1-w\lambda{D\over 
Dy^2}\right) \cr   
U_2&=\exp(ix^2)= \exp\left(iy^2+w\lambda{D\over Dy^1}\right), \cr }}
and using the above formulas, one can verify that $U_1U_2=\exp(-2\pi
i\Theta) U_2U_1$, with as usual $\Theta=-1/2\pi B$.  One can similarly
describe the algebra $\A'$ that acts at $\sigma=\pi$. 

We can pick a gauge in which
\eqn\bursto{\eqalign{{D\over Dt^1} & = {\partial\over\partial t^1}\cr
                      {D\over Dt^2} & = {\partial\over\partial t^2}
                      +i{m\over 2\pi}t^1,\cr}}
with wave functions obeying $\psi(t^1,t^2+2\pi)=\psi(t^1,t^2)$,
$\psi(t^1+2\pi,t^2)=e^{-imt^2}\psi(t^1,t^2)$.  In this gauge, we
expand $\psi(t^1,t^2)=\sum_{k\in{\bf Z}}f_k(t^1)e^{ikt^2}$, where
$f_k(t^1+2\pi)=f_{k+m}(t^1)$.  The $f_k$'s can thus be grouped
together into $m$ functions $f_0(t^1),\dots, f_{m-1}(t^1)$ of a {\it
real} variable $t^1$.  This leads to the description of the module
used in \connes.
   
\bigskip\noindent{\it (iv) $(n,m)$ Boundary Conditions}

The general case of this type is to consider an open string that
terminates at $\sigma=0$ on a twobrane, and at $\sigma=\pi$ on a
cluster of $n$ twobranes with zerobrane charge $m$. We call this a
cluster of charges $(n,m)$.  For simplicity, we suppose that $m$ and
$n$ are relatively prime.

Such a cluster can be described as a system of $n$ twobranes that bear
a $U(n)$ gauge bundle $E$ with a connection of constant curvature
$m/2\pi n$.  The center of $U(n)$ is $U(1)$, and the curvature of $E$
lies in this $U(1)$.  Nonetheless, $E$ cannot be obtained by tensoring
a $U(1)$ bundle with a trivial $U(n)$ bundle, because the first Chern
class of the $U(1)$ bundle would have to be $m/n$, not an integer.
This leads to some complications in the direct description \zumino\ of
the module derived from $E$.

However, $E$, and the open string Hilbert space that comes with it, is
naturally described by an ``orbifolding'' procedure.  We let $\widehat
\T^2$ be obtained from $\T^2$ by an $n^2$-fold cover, obtained by
taking an $n$-fold cover in each direction. (So $\widehat \T^2$ is
described with the same coordinates $x^1,x^2$ as $\T^2$, but they have
period $2\pi n$.)  When pulled back to $\widehat \T^2$, $E$ has a
central curvature with $mn$ Dirac flux units.  In particular, on
$\widehat \T^2$, we can write $E={\cal L}\otimes V$, where ${\cal L}$
is a $U(1)$ line bundle with first Chern class $mn$, and $V$ is a
trivial $U(n)$ bundle, with trivial connection.

To get from $\widehat \T^2$ to $\T^2$, we must divide by the
symmetries $T_i:x^i\to x^i+2\pi$, $i=1,2$.  These symmetries act on
the bundle ${\cal L}$, but in their action on ${\cal L}$ they do not
commute.  If they commuted in acting on ${\cal L}$, then after
dividing by the group generated by the $T_i$, ${\cal L}$ would descend
to a line bundle over the original $\T^2$ of first Chern class $m/n$,
a contradiction as this is not an integer.  Rather than $T_1$ and
$T_2$ commuting in their action on ${\cal L}$, we have
\eqn\uncu{T_1T_2=T_2T_1e^{-2\pi im/n}.}
To get translation operators that do commute, we let $W_1$ and $W_2$
be elements of $U(n)$, regarded as constant gauge transformations of
$V$, that obey
\eqn\buncu{W_1W_2=W_2W_1e^{2\pi im/n}.}
Then the operators
\eqn\juncu{\TT_1=T_1W_1,~\TT_2=T_2W_2}
do commute.  By imposing invariance under the $\TT_i$, the bundle
${\cal L}\otimes V$ on $\widehat \T^2$ descends to the desired $U(n)$
bundle $E$ over $\T^2$.

Now, let us construct the algebras that act at the two ends of the
string.  At $\sigma=0$, the boundary coupling is just
\eqn\ucnub{-{i\over 2}\int d\tau B\left(x^1{dx^2\over d\tau}-x^2{dx^1\over
d\tau}\right).}
There are no $n\times n$ matrices to consider, so after descending
to $\T^2$, the algebra $\A$  of tachyon operators at $\sigma=0$ is
generated simply by $U_i = \exp(ix^i)$, with the usual algebra
$U_1U_2=e^{-2\pi i\Theta}U_2U_1$, with $\Theta=-1/2\pi B$.

Life is more interesting at $\sigma=\pi$.  Before orbifolding, with
the target space understood as $\widehat \T^2$ so that the $x^i$ have
periods $2\pi n$, the boundary couplings are
\eqn\hucub{{i\over 2}\int d\tau \left(B+{m\over 2\pi n}\right)
\left(\tilde x^1{d\tilde x^2\over d\tau}-\tilde x^2{d\tilde x^1\over
d\tau}\right).}
We have included the central curvature of ${\cal L}$.  The algebra of
functions of the $\tilde x^i$ at $\sigma=\pi$ is generated, before
orbifolding, by $Y_i=\exp(i\tilde x^i/n)$ with
\eqn\nufus{Y_1Y_2=Y_2Y_1\exp(-2\pi i/n^2(2\pi B+m/n)).}
Here we have shifted $B\to B+m/2\pi n$ in the usual formula (which at
$\sigma=\pi$ has a phase $\exp(2\pi i \Theta)=\exp(-i/B)$), and also
taken account of the fact that the exponent of $Y_i$ is $n$ times
smaller than usual.  Since we assume that $m$ and $n$ are relatively
prime, the algebra of $n\times n$ matrices is generated by $W_1$ and
$W_2$.  Hence, the $Y_i$ and the $W_i$ together generate the algebra
of operators that can act on the open string ground states at
$\sigma=\pi$.  However, for the orbifolding, we want to consider the
subalgebra of operators that commute with the $\TT_i$.  It is
generated by
\eqn\komiko{\eqalign{\tilde U_1=&Y_1W_2^{b} \cr
                     \tilde U_2=&Y_2W_1^{-b}\cr}  }
where $b$ is an integer such that $mb$ is congruent to $-1$ mod $n$,
or in other words there exists an integer $a$ with
\eqn\homiko{1=an-mb.}  
Equivalently, 
\eqn\konon{P=\left(\matrix{ n & m \cr b & a\cr}\right)}
is an element of $SL(2,\Z)$.  The $\tilde U_i$ obey $\tilde U_1\tilde
U_2=\exp(-2\pi i\tilde\Theta) \tilde U_2\tilde U_1$, with
\eqn\onono{\tilde\Theta={1\over n^2}{1\over 2\pi B+(m/n)}-{mb^2\over
n}.} 
Using \homiko\ and $\Theta'=-\Theta=1/2\pi B$, this is
\eqn\ponono{\tilde\Theta= {a\Theta'+b\over m\Theta'+n}~~{\rm
modulo}~{\bf Z}.} 
This shows that the algebra $\A'$ that acts at $\sigma=\pi$ is the
algebra of a noncommutative torus, with $\tilde \Theta$ obtained from
$\Theta'$ by a $T$-duality transformation that maps a brane cluster of
charges $(1,0)$ to a brane cluster of charges $(n,m)$.

The $\A\times \A'$ bimodule can be described explicitly by quantizing
the open strings.  In fact, before orbifolding, it arises from open
strings on $\hat \T^2$ that end at $\sigma=\pi$ on a cluster with
brane charges $(1,nm)$.  The module with this boundary condition was
described in {\it (iii)} above, and the general case follows by
dividing by $\Z_n\times \Z_n$.  We will omit details.

\bigskip\noindent{\it A Few Loose Ends}

We conclude this subsection by clearing up a few loose ends.

The modules we have constructed are all called projective modules
mathematically.  (An $\A$ module $M$ is called projective if there is
another $\A$ module $N$ such that $M\oplus N$ is equivalent to a
direct sum of free modules.)  We have constructed all of our modules
(except the module of 2-0 strings) starting with a complex line bundle
or vector bundle over an ordinary commutative $\T^2$, which determines
the boundary conditions at the $\sigma=\pi$ end.  Given a complex
vector bundle $E$ over $\T^2$, there is a complex bundle $F$ such that
$E\oplus F$ is trivial.  This is the starting point of complex
$K$-theory; it means that the space of sections of $E$ is a projective
module for the ring of functions on the commutative $\T^2$.  From this
it follows that the modules over the noncommutative $\T^2$ obtained by
quantizing open strings are likewise projective.  
>From this point of view, the 2-0 module
needs special treatment, because it is not determined by a vector
bundle at the $\sigma=\pi$ end.  However, the 2-0 module is determined
by a boundary condition at $\sigma=\pi$ that is associated with an
element of $K(\T^2)$, so it still leads to a projective module
\schwarz.

To an $\A$ module $M(E)$ determined by $E$ (which is a bundle over
$\T^2$ or in the exceptional case a $K$-theory element of $\T^2$), we
can associate the Chern classes of $E$ in $H^*(\T^2,{\bf Z})$.  This
natural topological invariant corresponds to $\mu(M)$ in the language
of \schwarz.  It is related to but differs from the $\Theta$-dependent
``Chern character in noncommutative geometry,'' which we will not try
to elucidate in a physical language.

\subsec{Theoretical Issues}

In this subsection, we will make contact with mathematical approaches
\refs{\connes,\schwarz-\piosch} to $T$-duality of
noncommutative Yang-Mills theory on a torus via Morita equivalence of
algebras, and then we will discuss how this language can be extended,
to some extent, to open string field theory.

Let $\alpha$ be any boundary condition for open strings on $\T^p$, in
the zero slope limit with fixed open string parameters.  The
$\alpha$-$\alpha$ ground states form an algebra $\A_\alpha$.  For any
other boundary condition $\gamma$ that can be introduced in the same
closed string theory, let $M_{\alpha,\gamma}$ denote the open string
tachyon states with $\alpha$ boundary conditions on the left and
$\gamma$ boundary conditions on the right.  It is an $\A_\alpha\times
\A_\gamma$ bimodule. $\A_\gamma$ is the commutant of $\A_\alpha$ in
this module, and vice-versa.  This means that the set of operators on
$M_{\alpha,\gamma}$ that commute with $\A_\alpha$ is precisely
$\A_\gamma$, and vice-versa.  That $\A_\alpha$ and $\A_\gamma$ commute
is clear from the fact that they act at opposite ends of the open
strings.  That they commute only with each other follows from the fact
that, as was clear in the construction of the modules, together they
generate the full algebra of observables in the string ground states.
As usual in quantum mechanics, this full algebra of observables acts
irreducibly on $M_{\alpha,\gamma}$.

There is an interesting analogy between the present open string
discussion and rational conformal field theory.  Consider, for
example, the WZW model in two dimensions on a closed Riemann surface,
with target space a Lie group $G$.  As long as one considers the full
closed string theory, this model has ordinary $G\times G$ symmetry.
Quantum groups arise if one tries to factorize the model into separate
left and right-moving sectors.  Likewise, for open strings, the full
algebra of operators acting on the string ground states, namely
$\A_\alpha\times \A_\gamma$, acts irreducibly on the quantum
mechanical (ground state) Hilbert space $M_{\alpha,\gamma}$, which as
we have seen in the introduction to this section can be naturally
realized in terms of ordinary functions (or sections of ordinary
bundles) over an ordinary torus.  A quantum torus arises if one
attempts to focus attention on just one end of the open string, and to
interpret just $\A_\alpha$ (or $\A_\gamma$) geometrically.

\bigskip\noindent{\it Morita Equivalence}

For any two boundary conditions $\alpha$ and $\beta$, there is a
natural correspondence between $\A_\alpha$ modules that are obtained
by quantizing open strings, and the analogous $\A_\beta$ modules.
Indeed, there is one of each for every boundary condition $\gamma$;
thus, the correspondence is $M_{\alpha,\gamma}\leftrightarrow
M_{\beta,\gamma}$.  This natural one-to-one correspondence between
(projective) modules for two rings is called mathematically a Morita
equivalence.  In fact, the $\alpha$-$\beta$ and $\beta$-$\alpha$ open
strings give bimodules $M_{\alpha,\beta}$ and $M_{\beta,\alpha}$ for
$A_\alpha \times A_\beta$ and $A_\beta\times A_\alpha$.  Using
$M_{\alpha,\beta}$ one can define a map from left $\A_\beta$ modules
to left $\A_\alpha$ modules by
\eqn\kiko{N\to M_{\alpha,\beta}\otimes_{\A_\beta}N,}
for every left $\A_\beta$ module $N$, with the inverse being $L\to
M_{\beta,\alpha}\otimes_{\A_\alpha}L$ for a left $\A_\alpha$ module
$L$.  In physical terms, if $N$ is of the form $M_{\beta,\gamma}$ for
some $\gamma$, then the map from $M_{\alpha,\beta}\times
N=M_{\alpha,\beta}\times M_{\beta,\gamma}$ to $M_{\alpha,\gamma}$
(which coincides with $M_{\alpha,\beta}\otimes_{\cal A_\beta}
M_{\beta,\gamma}$) is just the natural string vertex combining an
$\alpha$-$\beta$ open string and a $\beta$-$\gamma$ open string to
make an $\alpha$-$\gamma$ open string.  In other words, an
$\alpha$-$\beta$ state with vertex operator $\O$ and a
$\beta$-$\gamma$ state with vertex operator $\O'$ is mapped to an
$\alpha$-$\gamma$ state with vertex operator given by the product
$\lim_{\tau\to \tau'}\O(\tau) \O'(\tau')$.  This gives a well-defined
map from $M_{\alpha,\beta}\times M_{\beta,\gamma}$ to
$M_{\alpha,\beta}$ because the dimensions of the operators vanish, and
it can be interpreted as a map from
$M_{\alpha,\beta}\otimes_{\A_\beta}M_{\beta,\gamma} $ to
$M_{\alpha,\gamma}$ because of associativity of the operator product
expansion, which states that $(\O\O'')\O'=\O(\O''\O')$ for any
$\beta$-$\beta$ vertex operator $\O''$.

\bigskip\noindent{\it Relation To $T$-Duality}

Now, we will make a few remarks on the relation between Morita
equivalence and $T$-duality, as exploited in 
\refs{\connes,\schwarz-\piosch}.  First we need to
understand the mathematical notion of ``a connection on a module $M$
over a noncommutative algebra $\A_\alpha$.''  If $\A_\alpha$ is the
algebra of a noncommutative torus, generated by $U_i=\exp( ix^i)$,
with $U_1U_2=e^{-2\pi i \Theta}U_2U_1$, then according to the standard
mathematical definition used in the above-cited papers, such a
connection is supposed to be given by operators
\eqn\iko{D_i={\partial\over\partial x^i}-iA_i,}
where one requires that the $A_i$ commute with the $U_j$.  If $M$ is a
module $M_{\alpha,\gamma}$, constructed from $\alpha$-$\gamma$ open
strings (for some $\gamma$), then this definition means that the $A_i$
are elements of $\A_\gamma$.  Physically, to describe strings
propagating with such a connection, we must perturb the open string
worldsheet action by adding a boundary perturbation at the $\gamma$
end, without modifying the action at the $\alpha$ end\foot{This
mathematical definition differs from a more naive physical notion,
adopted in most of this paper, which is that if the gauge fields are
elements of $\A_\gamma$, then we say we are working on the torus whose
algebra of functions is $\A_\gamma$.}.

Now we can see Morita equivalence of noncommutative gauge fields (a
more precise notion than Morita equivalence of modules), in the sense
defined and exploited in \schwarz.  This asserts that gauge theory
over $\A_\alpha$ in the module $M_{\alpha, \gamma}$ is equivalent to
gauge theory over $\A_\beta$ in the module $M_{\beta,\gamma}$.  In
fact, to do gauge theory over $\A_\alpha$ in the module
$M_{\alpha,\gamma}$, we add a boundary perturbation at the $\gamma$
end; likewise, to do gauge theory over $\A_\beta$ in the module
$M_{\beta,\gamma}$, we add a boundary perturbation at the $\gamma$
end.  By using in the two cases the {\it same} boundary perturbation
at the $\gamma$ end, we get the equivalence between gauge theory over
$\A_\alpha$ and gauge theory over $\A_\beta$ that is claimed
mathematically.  If $\A_\alpha$ and $\A_\beta$ are algebras of
functions on noncommutative tori, then these must be $T$-dual tori (as
proved in \refs{\connes,\schwarz, \zumino,\piosch} and in section
6.3).  Conversely, any pair of $T$-dual tori can be boundary
conditions in the same closed string field theory.  So this gives the
$T$-duality of noncommutative Yang-Mills theory on different tori as
described mathematically.

Let $w$ be a $T$-duality transformation and suppose that
$\beta=w^{-1}\alpha$.  A special case of what was just said is that
gauge theory over $\A_\alpha$ with the module $M_{\alpha,\gamma}$ is
equivalent to gauge theory over $\A_{w^{-1}\alpha}$ with the module
$M_{w^{-1}\alpha,\gamma}$.  Acting with $T$-duality on the boundary
conditions at both ends of the string is certainly a symmetry of the
full string theory (even before taking a zero slope limit).  This
operation transforms $\A_{w^{-1}\alpha}$ back to $\A_\alpha$ while
transforming $M_{w^{-1}\alpha,\gamma}$ to $M_{\alpha,w\gamma}$. So it
follows that gauge theory over $\A_\alpha$ with module
$M_{\alpha,\gamma}$ is equivalent to gauge theory over $\A_\alpha$
with module $M_{\alpha,w\gamma}$.  Thus we can, if we wish, consider
the $T$-duality to leave fixed the torus $\A_\alpha$ and act only on
the commutant $\A_\gamma$.  This alternative formulation of how the
$T$-duality acts is more in line with the naive notion of ``gauge
theory on a noncommutative torus'' mentioned in the last footnote.

Regardless of which approach one takes, the key simplification that
causes the $\hat F^2$ action to be invariant under $T$-duality, while
the conventional $F^2$ Yang-Mills action is not, is that in the zero
slope limit, one can consider independent $T$-duality transformations
at the two ends of the open string, in this way defining a
transformation that leaves the torus fixed and acts only on the
quantum numbers of the gauge bundle.

\bigskip\noindent{\it Relation To Open String Field Theory}

In all of this discussion, we have considered a $*$ product
constructed just from the string ground states, in the limit
$\alpha'\to 0$.  It is natural to ask whether one can define a more
general $*$ product that incorporates all of the string states.  The
only apparent way to do this is to use the $*$ product defined by
gluing open strings in the noncommutative geometry approach to open
string field theory \ewitten.  This $*$ product can be introduced for
open strings defined with any boundary condition in any closed string
conformal field theory.  Consider the case of oriented bosonic
open string theory.  In flat $\R^{26}$, with $B=0$, taking free
(Neumann) boundary conditions for the open strings, one gets a $*$
product (considered in \ewitten) whose $\alpha'\to 0$ limit is the
ordinary commutative multiplication of functions on $\R^{26}$.
What if we repeat
the same exercise with a constant nonzero $B$-field?
Using the relation of the $*$ product of open strings to the operator
product algebra, in this situation the string field theory $*$ product
reduces in the zero slope limit to the $*$ product of noncommutative
Yang-Mills theory on $\R^{26}$.  (If instead of Neumann boundary
conditions for open strings, one takes $p$-brane boundary conditions
for some $p<25$, one gets instead noncommutative Yang-Mills on the
$p$-brane worldvolume.)  Thus, noncommutative Yang-Mills theory can be
regarded as a low energy limit of string field theory.  This gives an
interesting illustration of the open string field theory philosophy,
though it remains that the open string field theory does not seem
particularly useful for computation (being superseded by the $\widehat
F^2$ action).

\nref\moore{R. Minasian and G. Moore, ``$K$ Theory And Ramond-Ramond
Charge,'' JHEP {\bf 9711:002} (1997), hep-th/9710230.}%
\nref\ugwitten{E. Witten, ``$D$-Branes And $K$ Theory,'' 
JHEP {\bf 9812:019} (1998), hep-th/9810188.}%
The discussion of Morita equivalence of algebras makes sense in the
full generality of open string field theory.  Keeping fixed the closed
string conformal field theory, let $\alpha$ be any possible boundary
condition for the open strings.  Then the $*$ product of the
$\alpha$-$\alpha$ open strings (keeping all of the excited open string
states, and without any zero slope limit) gives an algebra
$\A_\alpha$.  For any other boundary condition $\beta$, the
$\alpha$-$\beta$ open strings give an $\A_\alpha\times\A_\beta$
bimodule $M_{\alpha,\beta}$ that by the same construction as above
establishes a Morita equivalence between $\A_\alpha$ and $\A_\beta$.
This Morita equivalence gives a natural equivalence between the
$K$-theory of $\A_\alpha$ and that of $\A_\beta$.  Presumably, in Type
II superstring theory (in an arbitrary closed string background) the
$D$-brane charge takes values in this $K$-theory -- generalizing the
relation of $D$-brane charge to ordinary complex $K$-theory in the
long distance limit \refs{\moore,\ugwitten}.

\newsec{Relation To M-Theory In DLCQ}

Motivated by the spectrum of BPS states, Connes, Douglas and Schwarz
\connes\ proposed that M-theory compactified on a null circle (DLCQ)
\eqn\xmiden{x^- = {1\over 2}(x^0 - x^1) \sim x^- + 2\pi R}
with nonzero $C_{-ij}$ leads to noncommutative geometry.  This
suggestion was further explored in \zumino.  Here, we
will examine this problem in more detail from our perspective.

\nref\hepol{S. Hellerman and J. Polchinski, unpublished.}%
\nref\alllc{N.~Seiberg, ``Why is the Matrix Model Correct?''
Phys. Rev. Lett. {\bf 79} (1997) 3577, hep-th/9710009.}%

The compactification on a null circle needs a careful definition.
Here we will define it as a limit of a compactification on a
space-like circle in the $(x^0,x^1)$ plane which is almost null
\refs{\hepol,\alllc}.  Its invariant radius $\epsilon R$ will
eventually be taken to zero.  In order to study this system we follow
the following steps:
\item{(1)} We boost the system to bring the space-like circle to be
along $x^1$. 
\item{(2)} We scale the energy such that energies of order one
before the boost remain of order one after the boost.
\item{(3)} We scale the transverse directions and momenta in order
to let them affect the energy as before the boost.
\item{(4)} We interpret the system with the small circle as type IIA
string theory.
\item{(5)} If the original system is also compactified on a spatial
torus, we perform $T$-duality on all its sides.

\noindent
After this sequence of steps we find a system which typically has a
smooth limit as $\epsilon$ is taken to zero.  This limit was first
discussed in
\ref\dkps{M. Douglas, D. Kabat, P. Pouliot, and S. Shenker, 
``$D$-branes and Short Distances in String Theory'',
Nucl. Phys. {\bf B485} (1997) 85, hep-th/9608024.}
and later in the context of this definition of the Matrix Model \bfss\
in
\nref\senmat{A.~Sen, ``D0-branes on ${\bf T}^n$ and Matrix Theory,''
Adv. Theor. Math. Phys. {\bf 2} (1998) 51, hep-th/9709220.}%
\refs{\senmat,\alllc}.  We will now follow these steps for the case of
compactification on a null circle with constant nonzero $C_{\pm ij}$.

We consider M-theory compactified on the null circle \xmiden.  We
assume that the metric is $g_{+-}=1$, $g_{\pm i}=0$, but the metric in
the transverse directions $g_{ij}$ can be arbitrary.  If the
transverse space includes a torus, we identify the transverse
coordinates as $x^i \sim x^i + 2\pi r^i$.

It is important that we keep all the parameters of the M-theory
compactification fixed.  We hold the Planck scale $M_p$, the metric
$g$ and the identification parameters $r^i$ fixed of order one as we
let $\epsilon \rightarrow 0$.

Since $x^-$ is compact in the DLCQ, the time coordinate is $x^+$.
Therefore, the Hamiltonian, which is the generator of time translation,
is $P_+=P^-$.  The parameter $R$ in \xmiden\ can be changed by a
longitudinal boost.  Therefore, the dependence of various quantities
on $R$ is easily determined using longitudinal boost invariance.  In
particular, the DLCQ Hamiltonian is of the general form
\eqn\hamm{P^-=P_+=RM_p^2F\left({x^i M_p}, {RC_{-ij}\over M_p^2},
{C_{+ij}\over RM_p^4} \right),}
for some function $F$ (we suppress the transverse momenta $P\sim
1/x$).  The dependence on the Planck scale $M_p$ was determined on
dimensional grounds.  The system is invariant under translation along
the null circle.  The corresponding conserved charge, the longitudinal
momentum, is
\eqn\ongm{P_-=P^+={N \over R}.}
The Hilbert space is split into sectors of fixed $N$.  Since $R$ can
be changed by a boost, the way to describe the decompactification of
the null circle is to consider the limit $N \rightarrow \infty$,
$R\rightarrow \infty$ with fixed $P_-$.

As we said above, we view this system as the $\epsilon \rightarrow 0$
limit of a compactification on a space-like circle of invariant
radius $\epsilon R$.  Since we plan to take $\epsilon $ to zero, we
will expand various expressions below in powers of $\epsilon$ keeping
only the terms which will be of significance.  The Hamiltonian is also
changed to $P_+ +\epsilon P_-$, so that it does not generate
translations along the space-like circle. 

We now perform step (1) above and boost the system such that the
circle is along the $x^1$ direction.  We denote the various quantities
after the boost with the subscript $\epsilon$.  The generator of time
translation, $P_{\epsilon,0}$ receives a large additive contribution
{}from $P_{\epsilon,-}= {N\over \epsilon R}$, which we are not
interested in.  Therefore, we consider the new Hamiltonian
\eqn\hamms{H=P_{\epsilon,0} - {N\over \epsilon R}=\epsilon
RM_p^2F\left(x^i M_p,{\epsilon R C_{\epsilon,-ij}\over
M_p^2},{C_{\epsilon,+ij}\over  \epsilon R M_p^4}\right).}  
$C_{\epsilon,\pm ij}$ are related to $C_{\pm ij}$ by a boost
\eqn\resch{C_{\epsilon,-ij}={1\over \epsilon}C_{-ij}, \qquad
C_{\epsilon,+ij}=\epsilon C_{+ij}.}

One of the consequences of the large boost is that energies which were
originally of order one are now very small.  This is clear from
\hamms\ where there is an explicit factor of $\epsilon$ in front of
the Hamiltonian.  In order to focus on these low energies, we perform
step (2) above and scale the energy by $1\over \epsilon$.  We do that
by replacing the system with Planck scale $M_p$ with a similar system
with Planck scale
\eqn\resmp{\tilde M_p^2= {1\over \epsilon} M_p^2.}
Then, the new Hamiltonian 
\eqn\hammsat{\tilde H=\epsilon R\tilde M_p^2F= R M_p^2F}
is of order one.

In order to keep the dependence on $x$ and $C$ as it was before the
scaling of $M_p$, we should follow step (3) above and also scale them
such that the arguments of $F$ are unchanged:
\eqn\hammsa{\tilde H= R M_p^2F\left(\tilde x^i \tilde
M_p,{\epsilon R\tilde C_{\epsilon,-ij}\over \tilde M_p^2},{\tilde
C_{\epsilon,+ij}\over \epsilon R\tilde M_p^4}\right).}  
That is
\eqn\impresc{\tilde x^i =x^i { M_p \over \tilde M_p} =
x^i\epsilon^\half, \qquad \tilde C_{\epsilon,-ij} =
C_{\epsilon,-ij}{\tilde M_p^2 \over M_p^2} = C_{-ij}{1 \over
\epsilon^2}, \qquad \tilde C_{\epsilon,+ij} = C_{\epsilon,+ij}{\tilde
M_p^4 \over M_p^4} = C_{+ij}{1 \over \epsilon}.}

We now move to step (4) above and interpret this system, M-theory on
a spatial circle of radius $\epsilon R$, as type IIA string theory.
The parameter $N$ is now interpreted as the number of D0-branes.  The
string theory parameters are
\eqn\stringvarir{\tilde \alpha' \sim {1 \over \tilde M_p^3R
\epsilon }={1 \over M_p^3R} \epsilon^{1\over 2}, \qquad
\tilde g_s \sim(\tilde M_p R\epsilon)^{3\over
2}=(M_p^2R^2\epsilon)^{3\over 4},}
and there is a nonzero $B$-field
\eqn\valbf{\tilde B_{\epsilon,ij}= {RC_{-ij} \over \epsilon}}
(the contribution of $C_{+ij}$ to $\tilde B$ is negligible for small
$\epsilon$).  We note as a consistency check of our various changes of
variables that just as the periods of $C$ around a three cycle
including the null circle are of order one, so are the periods of
$\tilde B$ around a two cycle $\tilde \Sigma^{(2)}$
\eqn\intbo{\int_{\tilde \Sigma^{(2)}} \tilde B \sim 1.}

We now move to step (5) above.  If the target space includes a torus
$T^p$, $\tilde x^i$ is identified with $\tilde x^i + 2\pi \tilde r^i$,
where $\tilde r^i = r^i \epsilon^\half$.  The metric on the torus was
not changed by our various changes of variables and remained as it was
in the original M-theory problem of order one, $g_{ij}\sim 1$.
Therefore, the volume of the torus $\tilde V = (\det g)^\half \prod_i
2\pi \tilde r^i \sim \epsilon^{p\over 2}$.

Let the rank of $\tilde B$ be $r\le p$ ($r$ is even).  For simplicity
we assume that $\tilde B_{ij}$ is nonzero only for $i,j=1,\dots,r$.  We
now perform $T$-duality on this torus converting the D0-branes to
Dp-branes.  After the transformation, the new coordinates $\hat x^i$
are identified as 
\eqn\hatr{\hat x^i \sim \hat x^i + 2\pi\hat r^i}
with
\eqn\rhatd{\hat r^i = {\tilde \alpha' \over \tilde r^i} \sim {1 \over
M_p^3R r^i}.} 
Hence, the periods of $\hat x^i$ are of order one.

The new metric and $B$-field are
\eqn\partdu{\eqalign{
&\hat g_{ij}=\left({1\over g + 2\pi \tilde \alpha' \tilde
B}\right)_{Sij} \sim \cases 
{\epsilon & for $i,j=1,\dots,r$ \cr
1& otherwise \cr}\cr
&\hat B_{ij}= {1\over 2\pi\tilde \alpha'}\left({1\over g +2\pi\tilde
\alpha' \tilde B}\right)_{Aij} = \cases{
\left({1 \over (2\pi\tilde \alpha')^2\tilde B}\right)_{ij} \sim 1& for
$i,j=1,\dots,r$ \cr
0 & otherwise.}}}
Therefore, the new volume of the torus is $\hat V= (\det \hat g)^\half
\prod_i 2\pi \hat r^i\sim \epsilon^{r\over 2}$, and the new string
coupling is 
\eqn\hatst{\hat g_s= \tilde g_s  {\hat V^\half \over \tilde V^\half}
\sim \epsilon^{3-p+r\over 4}.} 

This is exactly the limit studied above in \limfi\ and \scalegs, which leads to
quantum Yang-Mills with finite coupling constant on a noncommutative
space with finite volume.  More explicitly, we can determine the
metric $G$, the noncommutativity parameter $\theta$ and the Yang-Mills
coupling using \limmet\ and \ymcoup, and the expressions for $\hat g$, $\hat
B$ and $\hat g_s$ in \partdu\ and \hatst: 
\eqn\finbg{G_{ij} = g^{ij}, \qquad \theta^{ij}= {C_{-ij}\over RM_p^6},
\qquad g_{YM}^2 \sim { M_p^6R^3 \over(M_p^3R)^p V},}
where $V=(\det g)^\half \prod_i 2\pi r^i$ is the volume of the
original torus.  We would like to make a few comments:
\item{(1)}  The reason the indices in the left hand side and the right
hand side of these equations do not seem to match is because the torus
we end up with is dual to the original torus of the underlying M
theory. 
\item{(2)}  It is important that all these quantities are independent
of $\epsilon$ and hence the limit $\epsilon \rightarrow 0$ leads to a
well defined theory.  
\item{(3)}  The expressions for $G$ and $g_{YM}$ are as in the case
without the $C$  field \refs{\mmrev,\senmat,\alllc}.
\item{(4)}  Even though in the zero slope limit of string theory
$\theta ={1 \over B}$, our various changes of variables lead to $\theta
\sim C$ indicating that the behavior of the theory is smooth for $C$
near zero.
\item{(5)}  The DLCQ limit $\epsilon \rightarrow 0$ is clearly smooth
in $C$ in the theory of the D0-branes.  The $T$-duality transformation
to the Dp-branes leads to the metric $\hat g$ and string coupling
$\hat g_s$, which do not depend smoothly on $C$.  However, the
noncommutative Yang-Mills parameters \finbg\ depend smoothly on $C$.
\item{(6)}  The discussion of $T$-duality in
\refs{\connes,\schwarz-\piosch} is in terms of the D0-branes and the
torus with coordinates $\tilde x$ and metric $\tilde g$.  The
discussion in section 6 is in terms of the torus with coordinates
$\hat x$ and metric $\hat g$.  Hence, the expressions for the
$T$-duality transformations on $G$, $\Theta$ and $g_{YM}$ differ by
conjugation by the $T$-duality transformation which relates the two tori
$T= \left(\matrix{0&1\cr 1&0}\right)$; i.e.\ they are related by
$a\leftrightarrow d$ and $b\leftrightarrow c$.
\item{(7)}  For $p\ge 4$ the parameters in \finbg\ including the
Yang-Mills coupling are finite, but the theory is not likely to make
sense, as it is not renormalizable \refs{\one-\three}.  For $C=0$
($r=0$) this follows from $\hat g_s \rightarrow \infty$, which forces
the use of strongly coupled string theory leading to the $(2,0)$
theory for $p=4$
\nref\rozali{M.~Rozali, Phys. Lett. {\bf B400} (1997) 260,
hep-th/9702136.}%
\nref\brse{M.~Berkooz, M.~Rozali and N.~Seiberg,
Phys. Lett. {\bf B408} (1997) 105, hep-th/9704089.}%
\refs{\rozali,\brse} and the little string theory for $p=5$
\refs{\brse,\seilst}.  This suggests that also for nonzero $C$
there exist new consistent theories for which noncommutative
Yang-Mills theory is the low energy approximation.

\newsec{Noncommutative Version Of The Six Dimensional $(2,0)$ Theory} 

So far we have been studying open strings ending on $D$-branes in the
presence of a $B$-field yielding a generalization of the ordinary
gauge theory on $D$-branes.  It is natural to ask whether the discussion
can be extended to M5-branes in M-theory in the presence of a $C$
field, or perhaps even to NS5-branes in string theory with background
RR fields.  This could lead to a generalization of the $(2,0)$ six
dimensional field theory and the enigmatic little string theory
\seilst. 

There are several reasons to expect that such generalizations
exist. First, reference \abs\ proposed a DLCQ description of a
deformation of the $(2,0)$ field theory in terms of quantum mechanics
on the deformed moduli space of instantons.  This was motivated as a
regularization of the $(2,0)$ theory in which the small instanton
singularities are absent.  A similar two-dimensional field theory in
that target space could lead to a deformation of the little string
theory.  The second reason for the existence of these theories is the
fact that the 4+1 and 5+1 dimensional noncommutative Yang-Mills
theories are not renormalizable \refs{\one - \three}.  Yet, they
arise, as in section 7, in the DLCQ description of M-theory on $\T^4$
and $\T^5$.  It is natural to ask then whether they can be embedded in
consistent theories without gravity.  In the commutative case the
supersymmetric Yang-Mills theory in five dimensions can be embedded in
the six-dimensional $(2,0)$ theory and the six-dimensional gauge
theory can be embedded in the little string theory.  Therefore, we are
led to look for generalizations of these theories in which the
noncommutative gauge theories can be embedded.

We want to study M5-branes in a background $C$-field in an appropriate
limit.  In $\R^{11}$ a constant $C$-field can be gauged away.  But in
the presence of M5-branes it leads to a constant $H$-field at
infinity.  This is similar to the fact that a constant $B$-field can
be gauged away but it leads to a constant background $F$ in the
presence of $D$-branes.  There is, however, a crucial difference between
these cases.  The constant background $F$ on $D$-branes is arbitrary.
Here, however the background $H$ satisfies an algebraic constraint
(for example, when $H$ is small it is selfdual), and therefore only
half of its degrees of freedom can be specified.

We start by analyzing the constraints on $H$ on a single M5-brane in
$\R^6$.  There is no completely satisfactory covariant Lagrangian
describing the dynamical twoform $B$ on the M5-brane.  We can use
either a non-covariant Lagrangian
\ref\persch{M.~Perry and J.H.~Schwarz, ``Interacting chiral gauge
fields in six-dimensions and Born-Infeld theory,'' Nucl. Phys. {\bf
B489} (1997) 47, hep-th/9611065.}
or covariant equations of motion
\ref\wessez{P.S.~Howe, E.~Sezgin and P.C.~West, ``Covariant field
equations of the M-theory five-brane,'' Phys. Lett. {\bf B399} (1997)
49, hep-th/9702008; ``The Six-dimensional selfdual tensor,''
Phys. Lett. {\bf B400} (1997) 255, hep-th/9702111.}.  
Here we will follow the latter approach.

The background $H$ field transforms as ${\bf 10}\oplus {\bf 10'}$
under the six-dimensional Lorentz group.  It can be written as
$H=H^++H^-$, where $H^+$ is selfdual (${\bf 10}$ of the Lorentz
group), and $H^-$ is anti-selfdual (${\bf 10'}$ of the Lorentz group).
Out of $H^+$ we can form
\eqn\varitens{\eqalign{
&K_i^j = H^+_{ikl}H^{+jkl}\cr
&U_{ijk}=K_i^lH^+_{ljk}\cr
&D=K_i^jK_j^i.}}
which transform as ${\bf 20'}$ ($K_{ij}$ is a traceless symmetric
tensor), ${\bf 10'}$ ($U_{ijk}$ is an anti-selfdual threeform) and
${\bf 1}$ ($D$ is a scalar).  For slowly varying $H$ the equation of
motion express $H^-$ in terms of $H^+$
\eqn\eomh{H^-={1\over f(D)} U,}
where $U$ is the anti-selfdual form in \varitens\ and $f(D)$ is a
function of the scalar $D$ in \varitens, which is given implicitly by 
\eqn\solq{D=6f(D)^2\left(1-{2\sqrt{2}M_p^3 \over f(D)^\half}\right).}
In \solq, we take the solution which continuously interpolates between
$f(0)=8M_p^6$ and $\lim_{D\rightarrow \infty} f(D) = \sqrt{D\over
6}$. 

We are looking for constant $H$ solutions.  For simplicity, we start
with the metric $g_{ij}=\eta_{ij}=(-1,1,1,1,1,1)$ (we will later scale
it).  Up to a Lorentz transformation, the nonzero components of the
generic\foot{By taking a limit as $h,h_0\to 0$ while boosting,
one can get a nongeneric constant $H$ solution of  \eomh\ which
in a suitable coordinate system takes the form
$H_{012}=-H_{125}=H_{034}=-H_{345}=h$ with any constant $h$.  For this solution
$K_0^0=-K_5^5=K_0^5=-K_5^0=-4h^2$, $U=D=0$.  It is selfdual both in
the subspace spanned by $x^0,x^5$, and in the subspace spanned by
$x^1,x^2,x^3,x^4$.  Therefore, it is possible that this
solution is relevant to the discussion in \abs, which was based on
such a selfdual tensor.} $H$ satisfying \eomh\ are
\eqn\cancho{H_{012}=h_0, \quad H_{345}=h}
with a relation between $h_0$ and $h$.  These nonzero values of $H$
break the $SO(5,1)$ Lorentz group to $SO(2,1) \times SO(3)$.  Using
$\epsilon^{012345}=1$ we find
\eqn\usehkd{\eqalign{
&H^+_{012}=-H^+_{345}=-\half(h-h_0), \cr
&K_0^0=K_1^1=K_2^2= - K_3^3=-K_4^4=-K_5^5=-\half(h-h_0)^2, \cr
& U_{012}=U_{345}={1 \over 4}(h-h_0)^3, \quad D= {3\over
2}(h-h_0)^4.}}
Then equation \eomh\ determines the relation between $h_0$ and $h$
\eqn\hohtr{h_0^2-h^2+M_p^{-6} h_0^2h^2=0.}
For a given $h$, we have to take the solution
\eqn\solhz{h_0=-{h \over (1+M_p^{-6} h^2)^\half}}
(for an anti-M5-brane we have the opposite sign in \solhz).
For small $h$ we have $h_0 \approx -h$ and $H^-\approx 0$, i.e.\ $H$
is selfdual.  For large $h$ we have $h_0 \approx -M_p^3$ and $H$ is
dominated by its spatial components $H_{345}=h$.

We are looking for a limit in which the theory in the bulk of
spacetime decouples.  It should be such that after compactification on
a circle, the low energy theory will be the five-dimensional $\hat F^2$
theory we found in the zero slope limit.  In order to motivate the
appropriate limit, we examine a compactification of the theory on a
spacelike $\S^1$ of invariant radius $r$ to five dimensions.  The
resulting five-dimensional theory includes a gauge field with $F_{ij}
= \oint dx^k H_{ijk}$, whose dynamics is controlled for slowly varying
fields by the DBI action $\CL_{DBI} \sim \sqrt{-\det(g+{1\over 2\pi r
M_p^3} F)} $.  The magnetic dual of $F$ is the threeform $\tilde
F={}^*\left(\partial \CL_{DBI}/\partial F\right)$, where ${}^*(~)$
denotes a dual in five dimensions.  The six-dimensional equation
\eomh\ states that $\tilde F=H$ with indices in the five noncompact
dimensions. 

Let us examine various different compactifications:
\item{(1)}  The compactification is along a circle in the subspace
spanned by $x^{3,4,5}$.  It can be taken, without loss of generality,
to be along $x^5$.  Here the background $H$ we study leads to magnetic
field of rank two $F_{34} = 2\pi r h$ and a three-form $\tilde
F_{012}=h_0$.  The limit $h\rightarrow \infty$ leads to $\tilde F$ of
order one.
\item{(2)}  The compactification is along a circle in the subspace
spanned by $x^{0,1,2}$.  It can be taken without loss of generality to
be along $x^1$.  Here the background $H$ we study leads to electric
field $F_{02} = 2\pi r h_0$ and a three-form $\tilde F_{345}=h$.  In
the limit $h\rightarrow \infty$ the electric field reaches its
critical value.  It is satisfying that the large magnetic field and
the critical electric field limits of the five-dimensional theory are
the same limit in six dimensions.  Also, the six-dimensional equation
of motion \eomh\ guarantees that the five-dimensional electric field
cannot exceed its maximal value.
\item{(3)}  The compactification is along a circle, which can be
brought using a boost along $x^{1,2}$ to the subspace spanned by
$x^{1,2,3,4,5}$.  It can be taken to be along $x^1$ and $x^5$.  This
leads to a background magnetic field of rank two $F_{34} \sim h$ and
an electric field $F_{02} \sim h_0$.  There is also a background three
form.
\item{(4)}  The generic compactification along a spacelike
circle which is not of the form (3) above, can be taken, using the
$SO(2,1)\times SO(3)$ freedom, to be in the $x^{0,5}$ plane,
\eqn\comci{(x^0,x^5) \sim (x^0,x^5) +2\pi r (-a, \sqrt {1+a^2})} 
for some constant $a$.  Or in terms of the boosted and rescaled
coordinates $\tilde x^0= \sqrt{1+a^2}x^0+ax^5 $, $\tilde x^5={1\over
r}(ax^0 + \sqrt{1+a^2}x^5)$, it is
\eqn\comcit{(\tilde x^0, \tilde x^5) \sim (\tilde x^0,\tilde x^5) +(0,
2\pi ).}  
We will use these new coordinates and will drop the tilde over
$x^{0,5}$.  Then, the metric is $g_{ij}=(-1,1,1,1,1,r^2)$ and the
background $H$ field is $H_{012}= h_0\sqrt{1+a^2}$, $H_{034}= -ha$,
$H_{125}= -h_0ar$, $H_{345}= hr\sqrt{1+a^2}$.  The five-dimensional
theory has background magnetic field of rank four $F_{12} =-2\pi r
h_0a$, $F_{34}=2\pi r h\sqrt {1+a^2} $, there is no background
electric field, and there is a background three-form $\tilde F_{034}
=-a h$, $\tilde F_{012}=\sqrt{1+a^2}h_0$.

We are interested in the zero slope limit in which ${1 \over 2\pi r
M_p^3}F \gg g$ and $F$ is held fixed with rank four.  The nonlinearity
of the equation of motion \eomh\ imposes restrictions on possible
scalings.  Since we want $g_{IJ}=\epsilon \delta_{IJ}$ for
$I,J=1,2,3,4$, but keep $g_{00}=-1$, $g_{55}=r^2$, we should scale
$M_p \sim \epsilon^{-1/2}$ and change $H$ to $H_{012}=
h_0\sqrt{1+a^2}\epsilon^{-1/2}$, $H_{034}= -ha\epsilon^{-1/2}$,
$H_{125}= -h_0ar\epsilon^{-1/2}$, $H_{345}=
hr\sqrt{1+a^2}\epsilon^{-1/2}$.  Our desired scaling $M_s^2= 2\pi
rM_p^3 \sim \epsilon^{-1/2}$ and $g_s \sim (rM_p)^{3/2} \sim
\epsilon^{3/4}$ is reproduced with $g_{55} =r^2\sim \epsilon^2$.  In
order to keep $F_{IJ}$ fixed we should also have $a =
\epsilon^{-1/2}$.  Then, the background threeform field $\tilde F$ is
of order $1/\epsilon$. 

We conclude that we analyze the theory with
\eqn\consct{\eqalign{
&g_{IJ}=\epsilon \delta_{IJ}, \quad g_{55}=r^2 \sim \epsilon^2, \quad
g_{00}=-1, \quad M_p \sim \epsilon^{-1/2}, \cr 
&H_{012} \approx h_0\epsilon^{-1}, \quad H_{034}\approx
-h\epsilon^{-1}, \quad H_{125}\approx -h_0r\epsilon^{-1}, \quad
H_{345}\approx  hr\epsilon^{-1}}} 
compactified on $x^5\sim x^5+2\pi$.  In terms of the null coordinates
$x^\pm=\half(x^0\pm rx^5)$, the dominant components of $H$ are
$H_{-12}\approx 2h_0\epsilon^{-1}$, $H_{-34} \approx
-2h\epsilon^{-1}$, while $H_{+IJ}$ are of order one.

We would like to make a few comments:
\item{(1)}  $H_{-IJ} \sim \epsilon^{-1}$, but $\oint dx^-H_{-IJ} \sim
1$ is a finite, generic rank four twoform.
\item{(2)}  $H_{+IJ} \ll H_{-IJ} $ in the limit.  This fact has a
simple reason.  In the the zero slope limit with $F_{0I}=0$ the 
Lagrangian $\sqrt{-\det(g+{1\over 2\pi r M_p^3} F)}\sim \pf F $.
Therefore, $\tilde F_{0IJ} \sim \epsilon^{0IJKL}F_{KL}$.  Hence
$H_{0IJ} \approx -r^{-1} H_{5IJ}$, and $H$ is dominated by its
$H_{-IJ}$ components. 
\item{(3)}  The limit of the five-dimensional theory is the same as a
DLCQ compactification of the six-dimensional theory.  This fact is
consistent with the way this theory was derived in section 7 by
starting with M-theory in DLCQ on $\T^4$.  In the last step we
performed $T$-duality to convert the theory to a five dimensional field
theory.  But because of the $T$-duality symmetry of the noncommutative
gauge theories, we must have the same limit as before the $T$-duality
transformation.

This discussion motivates us to examine the theory in $\R^6$ with
\eqn\consctn{\eqalign{
&g_{IJ}=\epsilon \delta_{IJ}, \quad g_{55}=\epsilon^2,  \quad
g_{00}=-1, \quad  M_p =A \epsilon^{-1/2}, \cr
&H_{012}= h_0\epsilon^{-1}\sqrt{1+\epsilon}, \quad H_{034}=
-h\epsilon^{-1}, \quad H_{125}= -h_0, \quad H_{345}=
h\sqrt{1+\epsilon}\cr
&h_0^2-h^2+A^{-6} h_0^2h^2=0, \quad {\rm for} \quad \epsilon \to 0,
\quad h,h_0,A, x^i = {\rm fixed}}} 
(the subleading terms in $H_{012}$ and $H_{345}$ are needed in order
to satisfy \eomh).  We note that in the limit $H_{012}$ and $H_{034}$
diverge like $\epsilon^{-1}$.  In terms of null coordinates $x^\pm =
\half (x^0\pm \epsilon x^5)$, 
\eqn\nullcoo{H_{-12}\approx 2h_0\epsilon^{-1}, \quad H_{-34}\approx
-2h\epsilon^{-1}, \quad H_{+12}\approx \half h_0, \quad
H_{+34}\approx \half h,}  
i.e.\ $H_{-IJ}$ are taken to infinity and $H_{+IJ} $ are of order
one. 

It is not clear to us whether the limit \consctn\ of M-theory with
M5-branes exists.  What we showed is that this limit satisfies the
equation of motion on the M5-branes \eomh, and that after
compactification on $\S^1$ it leads to D4-branes in the zero slope
limit, i.e.\ to five dimensional noncommutative Yang-Mills with the
${1 \over g_{YM}^2} \hat F^2$ Lagrangian with finite $g_{YM}$.
Therefore, if this theory can be embedded in a consistent theory
within M-theory, it must come from the limit \consctn.

\bigskip
\centerline{\bf Acknowledgements}
We have benefited from discussions with O. Aharony, M. Berkooz,
G. Moore and A. Schwarz.  The remarks in the concluding paragraph of
section 6.4 were stimulated by a discussion with G. Segal.  This work
was supported in part by grant \#DE-FG02-90ER40542 and grant
\#NSF-PHY-9513835.

\listrefs
\end